\documentclass[apj]{emulateapj}
\usepackage{apjfonts, natbib, pslatex, graphicx, gensymb, epstopdf, amsmath, multirow, float, rotating}

\begin{document}

\title{The influence of Red Spiral Galaxies on the Shape of the Local K-Band Luminosity Function}

\author{Nicolas J. Bonne$^{1,2}$}
\author{Michael J. I. Brown$^1$}
\author{Heath Jones$^{1,3,4}$}
\author{Kevin A. Pimbblet$^{1,5,6}$}
\affiliation{$^1$School of Physics, P.O. Box 27 Monash University, Victoria, Australia 3800; nicolas.bonne@monash.edu\\
$^2$Institute of Cosmology and Gravitation, University of Portsmouth, Dennis Sciama Building, Burnaby Road, Portsmouth, PO1 3FX, United Kingdom\\
$^3$Department of Physics and Astronomy, Macquarie University, Sydney NSW 2109, Australia\\
$^4$Research Centre in Astronomy, Astrophysics \& Astrophotonics, Macquarie University, Sydney NSW 2109, Australia\\
$^5$Department of Physics, University of Oxford, Denys Wilkinson Building, Keble Road, Oxford OX1 3RH, U.K.\\
$^6$Department of Physics and Mathematics, University of Hull, Cottingham Road, Kingston-upon-Hull, HU6 7RX, U.K.}

\begin{abstract}

We have determined K-band luminosity functions for 13,325 local Universe galaxies as a function of morphology and color (for $K_{\rm{tot}}$ $\leq$ 10.75). Our sample is drawn from the 2MASS Extended Source Catalog, with all sample galaxies having measured morphologies and distances (including 4,219 archival redshift- independent distances). The luminosity function for our total sample is in good agreement with previous works, but is relatively smooth at faint magnitudes (due to bulk flow distance corrections). We investigated the differences due to morphological and color-selection using 5,417 sample galaxies with NASA Sloan Atlas optical colors and find that red spirals comprise 20 to 50 \% of all spirals with -25 $\leq$ $M_K$ $<$ -20. Fainter than $M_K$ = -24, red spirals are as common as early-types, explaining the different faint end slopes ($\alpha$ = -0.87 and -1.00 for red and early-types, respectively).  While we find red spirals comprise more than 50\% of all $M_{K}$ $<$ -25 spiral galaxies, they do not dominate the bright end of the overall red galaxy luminosity function, which is dominated by early-type galaxies. The brightest red spirals have ongoing star formation and those without are frequently misclassified as early-types. The faintest ones have an appearance and Sersic indices consistent with faded disks, rather than true bulge dominated galaxies. 
\end{abstract}


\maketitle

\section{Introduction}

Understanding how different galaxy types evolve is one of the most important unresolved issues in modern astronomy. 
The galaxies that form in the early Universe do so from the gravitational collapse of dark matter into halos. Baryonic matter falls into these halos, eventually collapsing enough to make stars and finally forming galaxies \citep[e.g.,][]{sprin05, ben10}. However, the behavior of baryonic matter inside these halos is much more difficult to model than the dark matter halos themselves. This is largely due to the fact that baryonic matter interacts via all of the physical forces, rather than the simple gravitational interactions of dark matter particles. The presence of stars, active galactic nuclei, supernovae and other celestial sources can all influence the gas within galaxies in ways that are still not yet fully understood \citep[e.g.,][]{fab03a, croto06}. As a result, theory has made many plausible predictions about aspects of galaxy evolution that are testable. The shape of the galaxy luminosity function has a long history in this regard \citep[e.g.,][]{ben03_2}.

The galaxy luminosity function describes the number of galaxies per unit volume per unit luminosity. By measuring the luminosity function and its evolution, we can better understand what factors contribute to the star formation rate and growth of galaxies. \citep[e.g.,][]{bel04}. Luminosity functions are vital for testing our theories for galaxy evolution \citep[e.g.,][]{ben03} as, to be plausible, any model proposed must match the observed luminosity function. This is highlighted by the fact that luminosity functions have driven the current paradigm of galaxy evolution and feedback \citep[e.g.,][]{bel04, croto06}, as models without feedback were unable to reproduce both ends of the observed luminosity function simultaneously.

The consensus view of galaxy evolution \citep[e.g.,][]{bel04, croto06, hop06} is that primordial, irregularly shaped galaxies in the early Universe can grow via star formation, and may eventually evolve into disk dominated spiral galaxies. Spiral galaxies will grow by forming new stars, but star formation must be truncated above some critical mass \citep{kau03_2}. Red elliptical galaxies form via mergers of smaller galaxies \citep[e.g.][]{tom72}. These mergers would destroy disks and cause the variety of orbital planes of stars observed in elliptical galaxies today.

By constructing luminosity functions as functions of color \citep[e.g.,][]{bel04, brown07, fab07} we can trace the color evolution of galaxy populations. In the past decade, measurements of color-selected luminosity functions have yielded important insights into galaxy evolution. \citet{bel04}, \citet{brown07} and \citet{fab07}, amongst others, have shown that the stellar mass contained within the red galaxy population has roughly doubled since $z$ $\simeq$ 1. As red galaxies should not be producing any new stars, this indicates that stars from the blue galaxy population are being transferred to the red population. This is best explained by the truncation of star formation in blue galaxies, resulting in the transformation of blue galaxies into red galaxies.

However, to equate this to a measure of morphology evolution, we are forced to make certain assumptions about the relationship between galaxy color and morphology. Direct  studies of morphology-selected luminosity functions \citep[e.g.,][]{marzk98, kocha01, dever09} have helped us to better quantify galaxy populations in our current epoch. These have shown that the typical luminosity and mass of early-type galaxies (elliptical galaxies and lenticular galaxies) is higher than that of more numerous late-type galaxies (spiral galaxies).

 Morphology and color give insight into different galaxy properties. Morphology reflects the motion of stars within a galaxy, and provides information on galaxy formation and assembly. If a galaxy is predominantly spheroidal or bulge dominated, it is believed to have formed through either hierarchical assembly \citep[the merging of multiple smaller galaxies over time; e.g.,][]{tom72} and/or monolithic collapse \citep[multiple star forming regions rapidly collapsing in the early Universe; e.g.,][]{eggen62}. The light from spheroidal galaxies is typically dominated by older stars with a variety orbital planes. The stars in disk galaxies are formed over long periods of time from gas that cools and collapses to form a disk.

Dust corrected optical color is a proxy for star formation, with star forming galaxies containing short-lived luminous blue stars and galaxies with little to no star formation containing older red stellar populations. Observationally, color and galaxy shape correlate \citep[e.g.,][]{strat01, hog02, conse06, mig09} and thus, are often used as proxies for one another. However, with more evidence mounting for the existence of objects such as red spirals \citep[e.g.,][]{goto03, wolf09, mas10} and blue pseudo-bluges \citep[e.g.,][]{drive07, gad09, mci14}, it is becoming more apparent that color can no longer be safely used as such a proxy.

Early-type luminosity functions differ in shape from red luminosity functions, as do those of blue compared to late-type galaxies. Most strikingly, red functions typically have a significant turnover and power-law index of $\alpha$ $\simeq$-0.5 to -0.6 at higher z \citep[e.g.,][]{bel04, brown07, fab07} and $\simeq$-0.8 for the local Universe \citep[e.g.,][]{baldr04}, whereas early-type functions have a nearly flat slope \citep[e.g.,][]{marzk98, kocha01, dever09}. The most plausible explanation is that there must either be significant numbers of blue elliptical galaxies \citep[e.g.,][]{drive07, gad09, mci14} and/or red spiral galaxies \citep[e.g.,][]{goto03, wolf09, mas10} in the Universe. If red galaxies are formed via the truncation or ``switching off" of star formation in blue galaxies, one may expect red spirals to be fainter than blue spirals (and we return to this point later in the paper).

In this paper, we aim to test the hypothesis that the observed difference in shape between optical color-selected luminosity functions and morphology-selected luminosity functions can be explained by the presence of a significant population of red late-type galaxies. To this end, we measure the K-band luminosity function for a large sample of bright local Universe galaxies with data taken from a number of sources. We provide luminosity functions for our sample separated by late and early morphological types. We compare these morphology-selected functions to functions that we calculate based on optical blue and red color separation. Through this comparison, we are able to show that using color as a proxy for morphology is extremely unreliable, as well as explaining why the shapes of morphology-selected and color-selected functions differ so markedly. 

The structure of this paper is as follows. In \S 2, we discuss the sources of our data and our sample selection, as well as the criterion we have used to morphologically divide our sample. In \S 3, we discuss the uniformity and completeness of the sample. In \S 4 we outline both the $V/V_{\rm{MAX}}$ and maximum likelihood methods used to derive and fit our luminosity function respectively. We also discuss the methods used to divide our sample by color, how we account for over-densities in our galaxy sample and our final comparison between our morphologically and color defined luminosity functions. In \S 5 we discuss the significance of our results, as well as compare them to results from previous literature. In \S 6 we draw conclusions and provide a summary of our findings. 

For this work we adopt cosmological parameters from \citet{komat11}; $H_0$ = 70.4 $\rm{km~s^{-1}~Mpc^{-1}}$, $\Omega_M$ = 0.27, $\Omega_\Lambda$ = 0 and $\Omega_A$ = 0.73. We also use two different sources of photometry, 2MASS K$_s$-band photometry, which uses Vega based magnitudes, and NASA Sloan Atlas Petrosian u and r-band photometry, which uses AB based magnitudes.   

\section{Sample Selection}

The datasets used for this work are from a variety of sources. Near-infrared K-band photometry was sourced from 2MASS Extended Source Catalog K$_s$-band photometry \citep{jarre00}. Redshifts were taken predominantly from 6dFGS \citep{jon09}, with others from CFA \citep{huchr99}, revised ZCAT \citep{fal99}, RC3 galaxy catalogue \citep{devac91}, \citet{weg03} as well as from the 2MASS Redshift Survey \citep{huch12}. Morphological classifications were sourced from CFA \citep{huchr99}, the RC3 galaxy catalogue \citep{devac91}, PGC \citep{patur03}, \citet{weg03}, and 2MASS Redshift Survey \citep{huch12}. Redshift-independent distances were obtained from a number of sources listed in Table \ref{dl_tab}, magnitudes used to determine galaxy color were taken from the NASA-Sloan Atlas and dust maps from \citet{sch98}. The final numbers of galaxies taken from each of the above sources is listed in Table \ref{maglim_tab}.

\subsection{2MASS Photometry}

The basis of our galaxy sample is the 2MASS Extended Source Catalog \citep{jarre00} and the photometry it provides. Photometry used is the 2MASS $K_{\rm{tot}}$ value, or a total magnitude extrapolated from a fit to the galaxy's radial profile. We choose this over the standard 2MASS isophotal magnitude as it is not truncated and is closer to the true luminosity of the galaxies in the catalog. To maximize the completeness of our sample, we impose a limit of $K_{\rm{tot}} \leq 10.75$. To produce a sample with low foreground extinction and high spectroscopic completeness, we excluded galaxies within $10^\circ$ of the Galactic Plane. To ensure that the sources we are selecting are galaxies, we select only objects with a 2MASS Object Type of 1, a visual verification score indicating that the object is a galaxy. We also exclude sources with a 2MASS XSC confusion and contamination flag. Lastly, as some sources in the 2MASS catalog are duplicates, we use the dup\_src and use\_src parameters to select the best version of each source, that is, the image that produces the most realistic total magnitude and best represents the galaxy center. These selection criteria result in a base sample of 14,170 galaxies with associated K-band photometry.

To further ensure the quality of our sample, we compare the isophotal magnitude and total magnitudes of our sample galaxies to one another. Any galaxy with a magnitude difference of > 1 was visually inspected. These objects were found to be large (tens of arcmin across in some cases), diffuse nearby objects, such as IC 2574, NGC 4861 and NGC 4395. To correct for this, we include 26 K-band galaxy photometry measurements made by \cite{eng08} that better measure the photometry of these large objects.

When comparing data from different catalogs, we sometimes find inconsistencies in the coordinates recorded for individual galaxies (sometimes differences of tens of arcseconds). This is usually in the case of very large galaxies, galaxies with bright features close to their cores, or galaxies with irregular morphologies. This difference is also due in part to the large range in vintage of the various catalogs and the variety of methods used to determine the center of each galaxy. To match recession velocities/distance estimates and morphological classifications with as many of these galaxies as possible, we use an algorithm that pairs galaxy coordinates to within a varying radius. The positional errors for bright galaxies, particularly from catalogs predating the year 2000, can have large positional errors \citep[see][ on arcsec positions of UGC galaxies]{cot99}. To match objects from the different catalogs, we use the smaller of a magnitude dependent matching radius and the 2MASS XSC semi-major axis divided one third. The magnitude dependent matching radius is $40^{\prime\prime}$ at $K=4$, dropping linearly with magnitude to $12^{\prime\prime}$ at $K=11$. 

\subsection{Galaxy Redshifts}

The bulk of our redshifts come from the 6dF Galaxy Survey \citep[Southern Hemisphere; ][]{jon09}, the CFA Redshift Survey \citep[Northern Hemisphere; ][]{huchr99}, the revised ZCAT \citep[Northern Hemisphere; ][]{fal99} the RC3 galaxy catalog \citep[whole sky; ][]{devac91}, those provided by \citet{weg03}, as well as data from the newly updated 2MASS Redshift Survey \citep[whole sky; ][]{huch12}. Our original sample consists of 13,599 galaxies with redshifts from one of the above mentioned sources. We manually inspect galaxies that have large variations between different redshift measurements. We correct for this by selecting the best redshift for each galaxy by favoring measurements taken by the 6dF and CFA redshift surveys which generally have higher signal-to-noise ratio spectra (and more secure redshifts).

\subsubsection{Correcting for Galaxy Peculiar Motions with Redshift-Independent Distances and Flow Models}

As our galaxies have $z \lesssim 0.05$, and are often members of large clusters such as Virgo we need to contend with galaxy peculiar motions. To account for this we have replaced recession velocity distances wherever possible with redshift-independent distances from the sources summarized in Table \ref{dl_tab}. The distance estimation methods are listed in order of our selection preference. That is, if Fundamental Plane distances are available, we select these preferentially. If not, and Tip of the Red Giant Branch (TRGB) distances are available, we select these, followed by Cepheid distances, as listed.
\begin{table*}\centering
\caption{Redshift-Independent Distances}
\scriptsize \begin{tabular}{ll}
\hline
&\\
Distance Estimation Method & Source \\
&\\
\hline
&\\ 
Fundamental Plane & \citet{bla02} ; (LEDA) \\
TRGB & \citet{dalca09} ; \citet{tul06} \\
Cepheid & \citet{kan03} ; \citet{patur02} ; \citet{fre01} ; \citet{nge06} ; \citet{mac01} \\
SNIa & \citet{hic09} ; \citet{kes09} ; \citet{woo07} ; \citet{kowal08} \\
Tully-Fisher & \citet{the07} ; \citet{sprin09} ; \citet{wil97} ; \citet{tul09} ; \citet{rus05} ; (LEDA) \\
D-$\Sigma$ & \citet{wil97} \\
Surface Brightness Fluctuations & \citet{bla01} ; \citet{ton01} \\
&\\
\hline
\label{dl_tab}
\end{tabular}
\end{table*}
We find redshift-independent distances for 4,219 galaxies from our sample. 

For the remainder, we follow the method of \citet{jones06} in applying field flow corrections from J. Huchra for the \textit{Hubble Space Telescope Key Project} \citep[appendix A of ][]{mou00} to their redshifts (\S 2.2). This model accounts for the presence of the Virgo cluster, Shapley Supercluster and Great Attractor, modeling each cluster as a spherical mass concentration and building a linear flow field around them. These are the major mass concentrations influencing galaxy motions in our $z<0.05$ volume. Thus, by using redshift-independent distances when available, and flow corrected redshifts when not, we are able to correct our entire sample for peculiar velocities. 

To avoid any issues associated with galaxies whose flow corrected recession velocities are small enough to still be distorted by peculiar velocities, and to avoid regions of extreme galaxy over density, we impose a lower limit of luminosity distance $D_L$ = 10 {\rm{Mpc}}, and apply this limit using the best available distance derived with the method discussed previously. Galaxies with TRGB distances are excluded from the final sample as they all have distances of $D_L$ < 10 {\rm{Mpc}}.

\subsection{Morphological Classifications}

As one of the aims of this project is to derive a galaxy luminosity function as a function of simple late and early-type morphology, we obtain morphological classifications for as many of our sample galaxies as possible. To this end, we utilize classifications from the RC3 catalog \citep{devac91}, PGC catalog \citep{patur03}, the CFA survey \citep{huchr99}, data published in \citet{weg03} and HYPERLEDA \citep[many of which are taken from the recently updated 2MASS Redshift Survey ][]{huch12}. As some of these sources use alphabetical classification schemes, we convert all such data to the corresponding revised numerical de Vaucouleurs T-type galaxy classification scheme \citep{devac91}. 

For the purposes of our final luminosity function, we set a morphological cut-off such that early-type galaxies have $T \leq -1$ and that late-type galaxies have $T > -1$. The early-type sample contains ellipticals ($-7 \leq T \leq -4$), and lenticulars ($-4 < T \leq -1$) and the late-type contains spirals ($-1 < T \leq 9$) and unclassified spirals ($T = 20$).

Any eyeball morphological classification scheme is bound to have errors, so to verify that there are no major discrepancies in galaxy classifications between different catalogs, we compare each source with any other available source. We find very good agreement, particularly between PGC, RC3 and HYPERLEDA. When we compare all catalogs, we find that (at most) $\sim$6\% of galaxies classified as late-type in one catalog are classified as early-type in another. 13,509 (95\%) of our original 2MASS galaxy sample are matched with morphological classifications. 

As the majority of catalogs mentioned above have made classifications using photographic plates, we have also tested and confirmed the validity of these classifications against modern, digital data-sets such as the SDSS. We compare our best morphological classification against data from GalaxyZoo \citep{lint11} and \citet{nai10}. $\sim$29\% of galaxies in \citet{nai10} exactly match our best morphology estimation, $\sim$61\% are within 1 $\delta T$ and $\sim$77\% are within 2 $\delta T$. We also find that only $\sim11\%$ of galaxies in our sample are misclassified as late or early-types when compared to GalaxyZoo and \citet{nai10}. 

\subsection{Testing for Uniformity and Completeness}

We test the uniformity of the sample with the $V/V_{\rm{MAX}}$ method \citep{sch68}, where $V$ is the survey volume between the galaxy and the observer, and $V_{\rm{MAX}}$ is the maximum volume the same galaxy could have been found, given the magnitude limit of the survey. Thus, by finding the $V/V_{\rm{MAX}}$ for all of our galaxies individually and then averaging, we can determine the uniformity across our sample. A score of 0.5 means a sample is consistent with being uniformly selected. Our limit of $K_{\rm{tot}}\,\leq$ 10.75 was chosen to achieve high spectroscopic and morphological completeness, and in Figure \ref{uniformity_fig} we show that this limit achieves an average $V/V_{\rm{MAX}}$ of 0.51.  For our luminosity functions, we also measured the $V/V_{\rm{MAX}}$ values for each of our bins, and find that values are between 0.38 and 0.62 except for bins containing small numbers of galaxies (< 7).

To verify the uniformity and sample completeness, we visually inspected all galaxies that did not have either a morphological classification or a recorded redshift. We find that the majority of these ``galaxie" are in fact nebulae, planetary nebulae or star clusters that have been erroneously identified as galaxies in the 2MASS classification scheme. In addition, there are several actual galaxies that either have central coordinates that are vastly different from those of other catalogs, or have bright internal features that have resulted in multiple 2MASS designations being recorded (e.g., IC 5052, IC 4362, NGC 3347B). These objects are, in the case of the former, removed from the catalog and in the latter, classified correctly and left in the catalog. 

\begin{figure}[]
\begin{center}
\includegraphics[width = .49\textwidth] {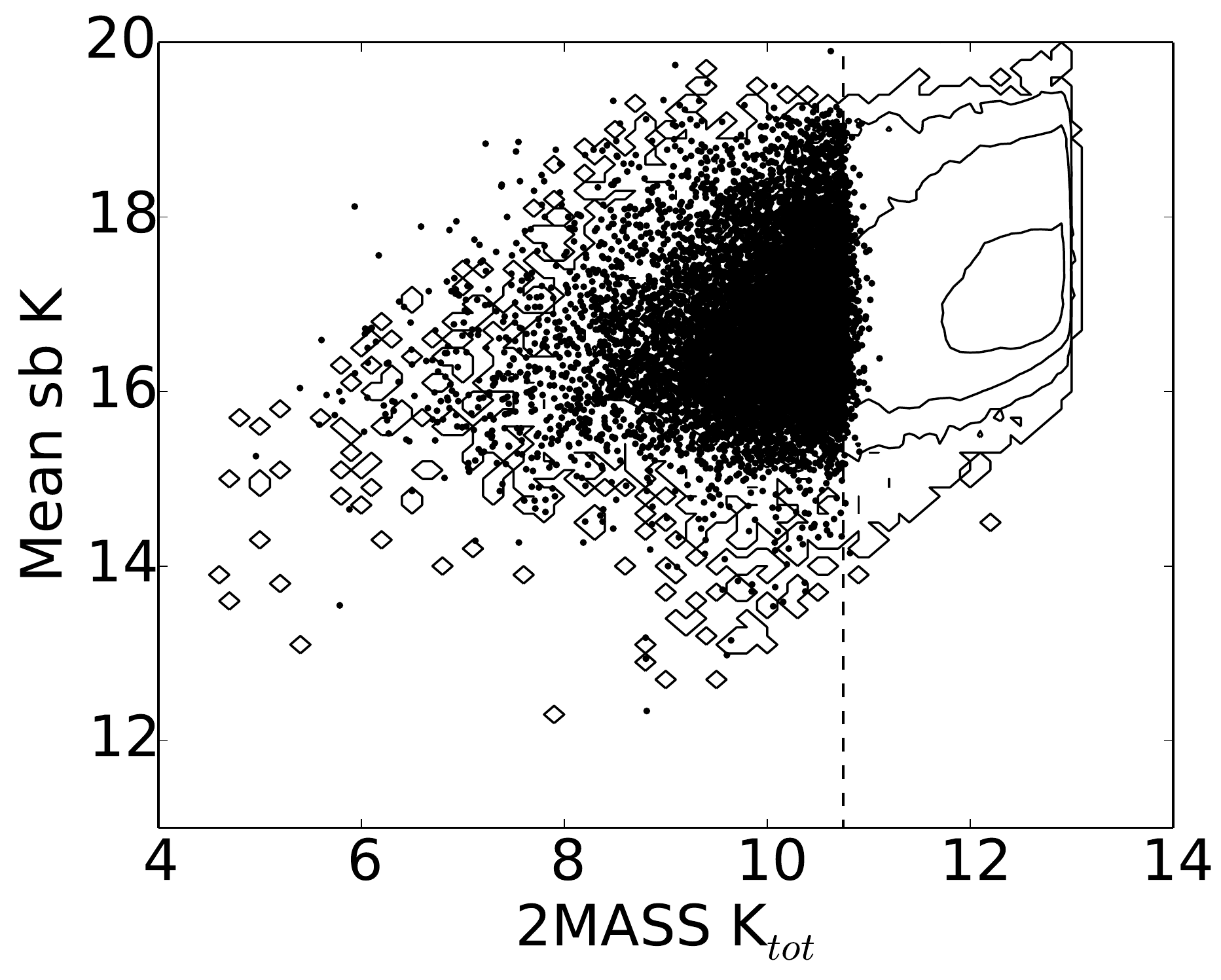}
\caption{A plot of 2MASS XSC K$_{tot}$ against 2MASS XSC mean surface brightness in K. Black points are all K$_{tot}$ $\leq$ 10.75 galaxies and the contours represent densities of 0, 5, 50, 500 and 5000 galaxies in the entire 2MASS XSC sample. 2MASS XSC appears to hit a surface brightness limit of $\simeq$19 mag arcsec$^{-2}$ at K$_{tot}$ magnitudes slightly brighter than our sample limit of K$_{tot}$ $\leq$ 10.75.}
\label{meansb_fig}
\end{center}
\end{figure}

As 2MASS is a relatively shallow survey, even for a bright galaxy sample such as ours it is necessary to test for surface brightness incompleteness. We plot 2MASS XSC K$_{tot}$ $\leq$ 10.75 against 2MASS XSC K mean surface brightness for all 2MASS galaxies in Figure \ref{meansb_fig}. It appears that at fainter K$_{tot}$ magnitudes, 2MASS XSC begins to hit a surface brightness limit of $\simeq$19 mag arcsec$^{-2}$ for galaxies slightly brighter (K$_{tot}$ $\simeq$ 10.5) than our limit of K$_{tot}$ $\leq$ 10.75.

{From Figure \ref{meansb_fig} we can see that most galaxy types in our sample are not affected by surface brightness incompleteness. However, some incompleteness is present, so we return to this issue when exploring morphology and color-selected galaxy samples  in \S 3.6.

We also compared galaxy counts and parameters for a small area of sky (100 square degrees) in both our master catalog and in the UKIDSS DR9 \citep{law07} which probes to much fainter magnitudes in $K$. We find that all 15 galaxies detected in 2MASS up to our faint limit of K$_{tot}$ $\leq$ 10.75 are also found in UKIDSS. As UKIDSS K-band is generally fainter than 2MASS's K$_{tot}$ (by an average of $\simeq$ 0.6 magnitudes), we extend beyond this to UKIDSS $K$ $\leq$ 11.75. For the 64 galaxies present in UKIDSS to this magnitude limit, and in this patch of sky, all but 1 are found in 2MASS. The exception is ARK227, a faint K = 12.39 unidentified elliptical galaxy which is identified by 2MASS as a point source, and is thus not included in our initial sample. Fainter than our magnitude cutoff, it is obvious that the number of galaxies contained in UKIDSS will increase drastically when compared to detections in 2MASS, however these objects are currently not usable in our work as they do not have associated morphological classifications.

Identical conclusions are found by \citet{mci06}. They show that for an SDSS MGS magnitude limited sample of r $\leq$ 15, a matched 2MASS XSC sample, limited to K $\leq$ 13.57, has 96.1\% completeness. As our faintest sample galaxy has a corresponding SDSS magnitude of r = 15, and our extinction corrected faint limit for K is only 10.75, we expect an even higher level of completeness. 

For this magnitude limited sample, and with all erroneously classified objects either removed or correctly classified, 13,649 galaxies remain, and 13,489 have both associated morphological and redshift data. Twenty have associated morphological classifications but no redshift data, 99 have redshift data but no morphological classification and 41 have neither a morphological classification nor recorded redshift. Imposing our distance limit of $D_L$ = 10 {\rm Mpc} further reduces the sample size to 13,325 galaxies, 7,685 of which are late-type and 5,640 of which are early-type. Thus, this sample has an overall completeness of 99\% and does not need to undergo completeness corrections. In addition, by limiting our sample to brighter galaxies, we also increase the reliability our morphological classifications.

Like all galaxy samples of large volumes of space, we are affected by cosmic variance. Using the methodology of \citet{drive10}, we calculate that the cosmic variance of this sample will be of order 6\%. We expect this to dominate over the Poisson uncertainties for our luminosity functions. Figure \ref{radec_fig} shows the distribution on the sky for the new magnitude limited sample.

\begin{figure}[]
\begin{center}
\includegraphics[width = .49\textwidth]{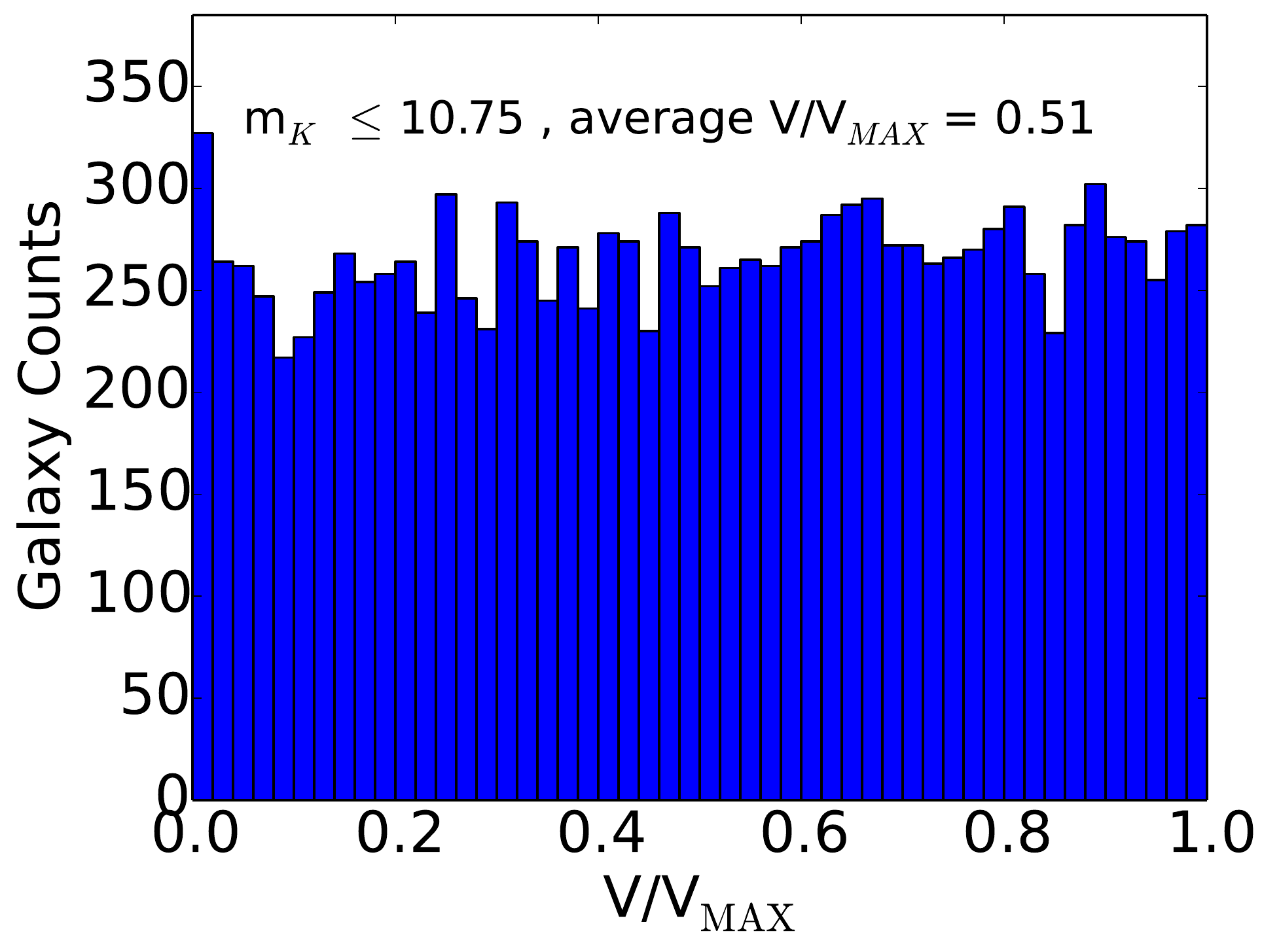}
\caption{A histogram of galaxy counts against $V/V_{\rm{MAX}}$ value. A good completeness is represented by a similar galaxy count in each bin across the full range of possible $V/V_{\rm{MAX}}$ values, and a mean $V/V_{\rm{MAX}}$ of 0.5.}
\label{uniformity_fig}
\end{center}
\end{figure}
\begin{figure}[]
\begin{center}
\includegraphics[width = .49\textwidth]{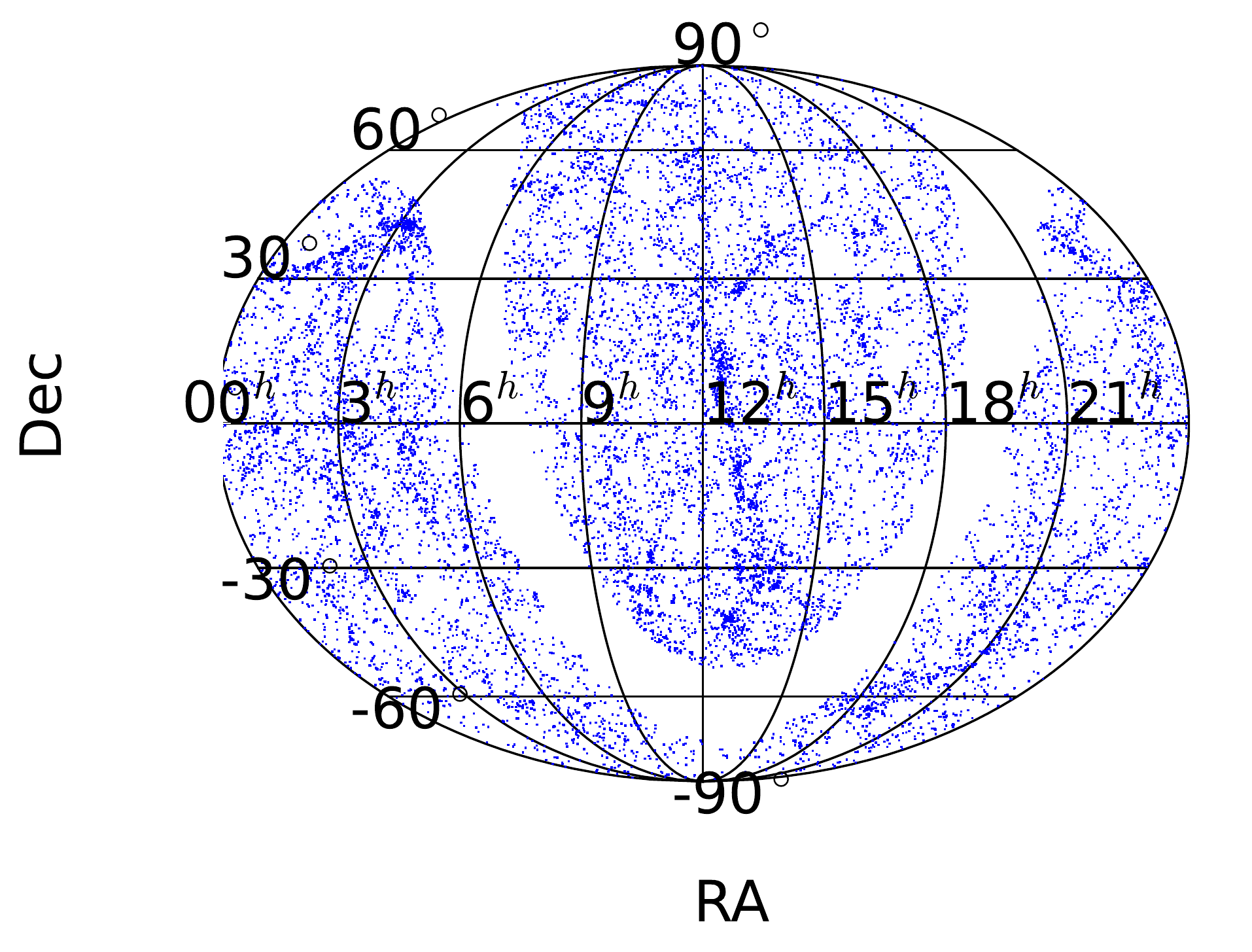}
\caption{Distribution on the sky of the final galaxy sample for $K_{\rm{tot}} \leq 10.75$ There is some obvious structure (portions of large clusters such as Virgo are included in our redshift range) but overall galaxies are evenly distributed across the sky.}
\label{radec_fig}
\end{center}
\end{figure}

\subsection{Final Galaxy Sample}

Our final sample, limited to $K_{\rm{tot}}\,\leq$ 10.75, is comprised of 13,325 galaxies with both redshift or redshift-independent distance estimates and morphological classifications. The sample distribution has some obvious structure, which is seen for both early and late-type galaxies, but there is no evidence of sample incompleteness. In Table \ref{maglim_tab} we show the final counts for data taken from the sources previously discussed in \S 1, for our magnitude limit of $K_{\rm{tot}}\,\leq$ 10.75 (discussed in \S 3). Sources and distance determination methods, sources of morphological classification and sources of redshifts are listed in order of selection preference.
\begin{table}\centering
\caption{Full Data Set}
\scriptsize \begin{tabular}{lll}
\hline
&\\
Data Source & $K_{\rm{tot}}$ $\leq$ 10.75 Counts \\
&\\
\hline
&\\ 
Redshifts \\
&\\
\hline
&\\ 
6dFGS & 4286\\
CFA & 564\\
ZCAT & 2898\\
\citet{weg03} & 19\\
RC3 & 58\\
RC3 (V21) & 20\\
2MRS & 1261\\
&\\
\hline
&\\ 
Redshift-Independent Distances \\
&\\
\hline
&\\ 
Fundamental Plane & 151\\
Cepheid-based & 3 \\
SNIa-based & 94 \\
Tully Fisher & 3847 \\
D-$\Sigma$ method & 76 \\
Surface Brightness Fluctuation & 48 \\
&\\
\hline
&\\ 
Morphological Classifications \\
&\\
\hline
&\\ 
PGC & 10226\\
RC3 & 455\\
\citet{weg03} & 24\\
CFA & 39\\
HYPERLEDA & 2581\\
&\\
\hline
&\\ Total & 13325\\
&\\
\hline
\label{maglim_tab}
\end{tabular}
\end{table}

\section{Luminosity Function}

We determined the K-band luminosity function using the non-parametric $1/V_{\rm{MAX}}$ method \citep{sch68} and by fitting Schechter functions  \citep{sch76} to our sample using the maximum likelihood method \citep[e.g.,][]{mar83}. We choose not to use either of the stepwise maximum likelihood (SWML) method of \citet{efs88} or the STY methods of \citet{san79} as neither offers an independent normalization in the way that the maximum-likelihood method does.
 
In order to derive a luminosity function using either of the above discussed methods, we determine absolute magnitudes using
\begin{equation}
M_{K}\, =\, K_{\rm{tot}}\, -\, 5\, \log\left(\frac{D_{L}}{10{\rm pc}}\right)\, -\, K(z) \,- \,A_K
\label{absmag_eq}
\end{equation}
where $K_{\rm{tot}}$ is our apparent K magnitude, $D_L$ is the luminosity distance for each galaxy, $K(z)$ is our K-correction and $A_K$ is a term for Galactic extinction. For calculations of $D_L$ we preferentially use redshift-independent distances if available, otherwise we use available redshifts. To calculate extinction, we use data from \citet{sch98} dust maps. As our photometric sample and that of \citet{jones06} are both taken from 2MASS and contain many of the same objects, we do not include any corrections for magnitude errors as these were found to be negligible for the 2MASS sample in that paper.

We determined k-corrections by fitting straight lines to the relationship between absolute and apparent K-band magnitudes as a function of observed $H-K$ color and redshift. This relationship was determined using 129 galaxy spectral energy distributions from \citet{brown14}. As these k-corrections are a weak function of $H-K$ color and an almost linear function of redshift ($k(z)\,\sim\,4z-5(H-K)z$ at $z<0.05$), we approximate the $K$-band k-correction with $k(z)$ equal to the values in Table \ref{kcor_tab} by interpolating between values of $z$ to match to values for each individual galaxy.
\begin{table}[]\centering
\caption{Color dependent k-correction values}
\scriptsize \begin{tabular}{ll}
\hline
&\\
z & Color dependent k-corrections \\
&\\
\hline
&\\ 
0.01 & $-$ 0.049 (H - K$_s$) + 0.039\\ 
0.02 & $-$ 0.104 (H - K$_s$) +0.076\\    
0.03 &$-$ 0.157 (H - K$_s$) +0.112 \\    
0.04 &$-$ 0.208 (H - K$_s$) +0.149 \\   
0.05 &$-$ 0.257 (H - K$_s$) +0.187 \\    
0.06 & $-$ 0.304 (H - K$_s$) +0.226\\    
0.07 & $-$ 0.352 (H - K$_s$) +0.267\\    
&\\
\hline
\label{kcor_tab}
\end{tabular}
\end{table}

\subsection{1/V$_{\rm{MAX}}$ Method}

To calculate the galaxy luminosity function $\Phi$, we bin our data by absolute magnitude. We calculate the maximum volume, $V_{\rm{MAX}}$, at which the galaxies in each bin can reside by using the the magnitude limits for bins and apparent magnitude limit for the sample. We then take galaxy number counts for each bin N and divide by $V_{\rm{MAX}}$ to obtain a number density. We use a bin size of 0.25 mag as any size smaller than this is not found to alter the shape of the final luminosity function. The derived full sample K-band luminosity function is presented in Figure \ref{lftotal_fig} and functions, separated into late and early-type galaxies can be seen in Figure \ref{lfearlylate_fig}. Uncertainties in $\Phi$ are Poisson in nature and are derived using the methodology outlined in \citet{geh86}. We note that uncertainties only exceed the expected level of cosmic variance for bins with very small galaxy numbers. Individual data points that make up the $1/V_{\rm{MAX}}$ functions are provided in Table \ref{lfdata_tab}. We provide $\Phi$ values scaled for over and under-densities, which we discuss in more detail in \S3.3 Comparisons between these functions and others from the literature will be discussed in \S 4.3.

\begin{figure}[]
\begin{center}
\includegraphics[width = .49\textwidth]{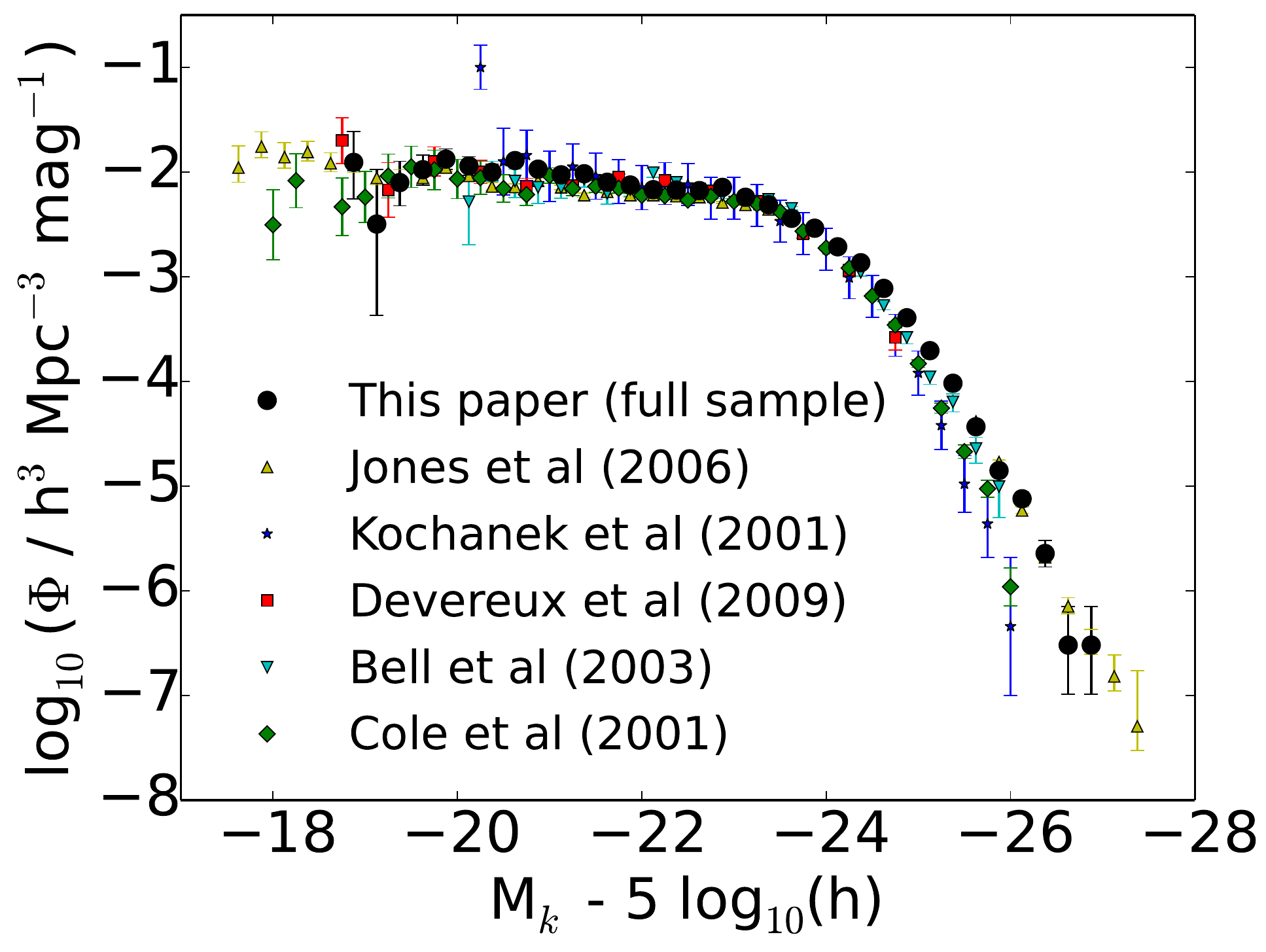}
\caption{1/$V_{\rm{MAX}}$ K-band luminosity function for the total galaxy sample. The faint end of our function is very similar in shape to previous literature, however, some difference is noted at the bright end. This end of the function sits in line with that of \citet{jones06} and \citet{bell03}, but is considerably higher and brighter than that of \citet{kocha01}, \citet{cole01} and \citet{dever09}.}
\label{lftotal_fig}
\end{center}
\end{figure}
\begin{figure}[]
\begin{center}
\includegraphics[width = .49\textwidth]{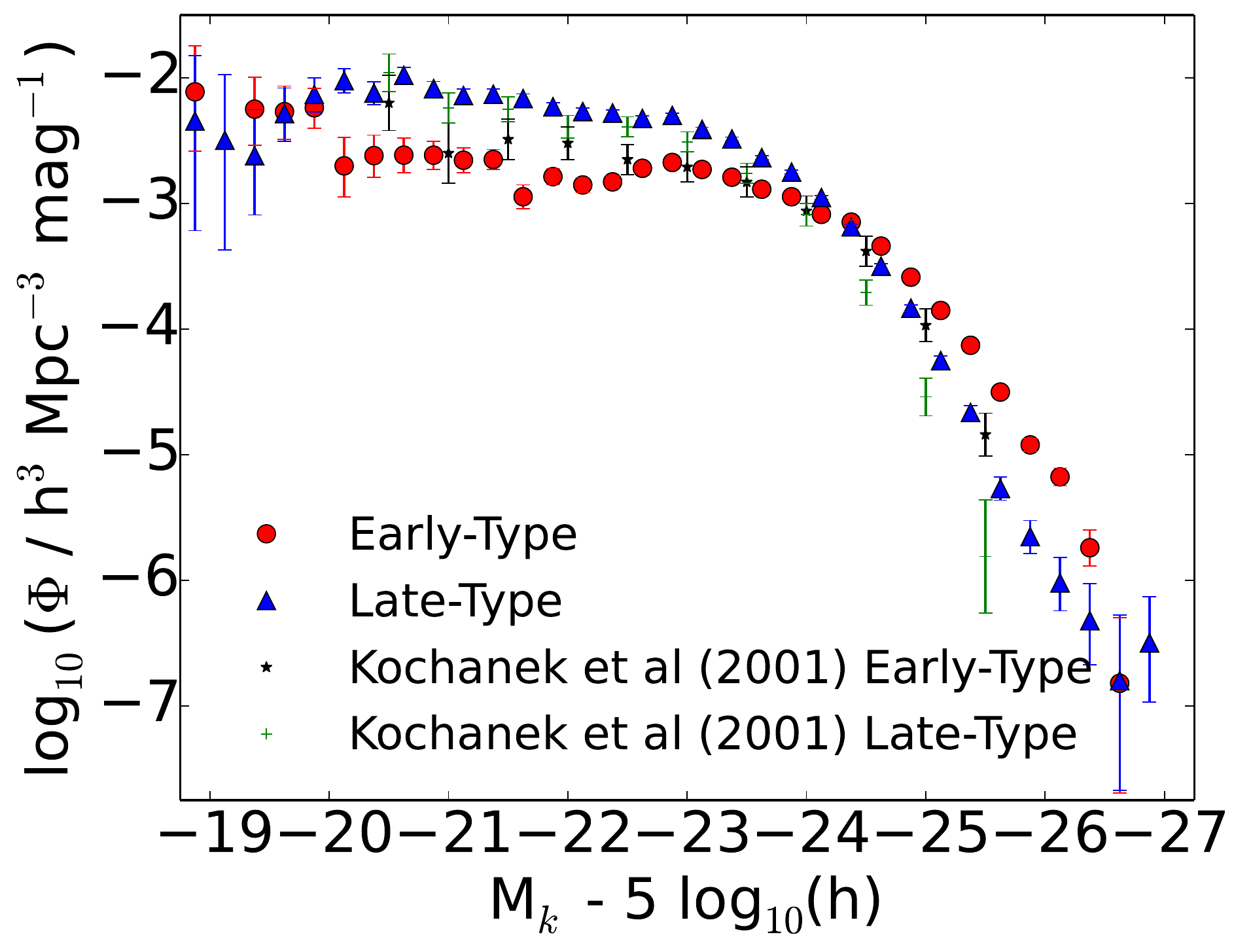}
\caption{1/$V_{\rm{MAX}}$ K-band luminosity functions for the early and late-type galaxy samples. The shapes of our early and late-type luminosity functions differ from those of \citet{kocha01}. Notably, the bright end of both the early and late-type functions from \citet{kocha01} are much lower, as can be expected from the comparison seen in Figure \ref{lftotal_fig}. Though the presented functions data points are similar in shape beyond the bright end, neither of the early-type functions appear to have a faint end slope with a strong turnover. If morphological type was truly a good proxy for color, we would expect an obvious downward slope at the faint end of these curves.}
\label{lfearlylate_fig}
\end{center}
\end{figure}
\begin{table*}[]\centering
\caption{1/V$_{\rm{MAX}}$ Method values (scaled for over-density)}
\scriptsize \begin{tabular}{ccccccccccc}
\hline
&&&&&&&&&&\\
&Total &   &  & & Early-Type &   &  & & Late-Type &  \\[6pt] 
$M_k$ $-$ $5$ $\rm{log_{10}(h)}$ & log$_{10}\,$($\Phi$) & N & & $M_k$ $-$ $5$ $\rm{log_{10}(h)}$ & log$_{10}\,$($\Phi$) & N & & $M_k$ $-$ $5$ $\rm{log_{10}(h)}$ & log$_{10}\,$($\Phi$) & N \\ 
  & (h$^{3}$\, {\rm Mpc}$^{-3}\,$ mag$^{-1}$) & (Galaxy) &   & & (h$^{3}$\, {\rm Mpc}$^{-3}\,$ mag$^{-1}$) & (Galaxy) &   & & (h$^{3}$\, {\rm Mpc}$^{-3}\,$ mag$^{-1}$) & (Galaxy) \\[4pt] 
&&&&&&&&&&\\
\hline
&&&&&&&&&&\\
-26.875 & -6.52$^{\,\, + 0.37}_{\,\,-0.47}$ & 2 &     &   &   &   &     & -26.875 & -6.50$^{\,\, + 0.37}_{\,\,-0.47}$ & 2 \\[4pt] 
-26.625 & -6.52$^{\,\, + 0.37}_{\,\,-0.47}$ & 2 &     & -26.625 & -6.82$^{\,\, + 0.52}_{\,\,-0.87}$ & 1 &    & -26.625 & -6.80$^{\,\, + 0.52}_{\,\,-0.87}$ & 1 \\[4pt] 
-26.375 & -5.64$^{\,\, + 0.12}_{\,\,-0.13}$ & 15 &     & -26.375 & -5.74$^{\,\, + 0.14}_{\,\,-0.15}$ & 12 &    & -26.375 & -6.32$^{\,\, + 0.30}_{\,\,-0.35}$ & 3 \\[4pt] 
-26.125 & -5.12$^{\,\, + 0.07}_{\,\,-0.07}$ & 50 &     & -26.125 & -5.18$^{\,\, + 0.07}_{\,\,-0.07}$ & 44 &    & -26.125 & -6.02$^{\,\, + 0.20}_{\,\,-0.22}$ & 6 \\[4pt] 
-25.875 & -4.85$^{\,\, + 0.05}_{\,\,-0.05}$ & 93 &     & -25.875 & -4.92$^{\,\, + 0.05}_{\,\,-0.05}$ & 79 &    & -25.875 & -5.65$^{\,\, + 0.13}_{\,\,-0.13}$ & 14 \\[4pt] 
-25.625 & -4.43$^{\,\, + 0.03}_{\,\,-0.03}$ & 179 &     & -25.625 & -4.50$^{\,\, + 0.04}_{\,\,-0.04}$ & 153 &    & -25.625 & -5.27$^{\,\, + 0.09}_{\,\,-0.09}$ & 26 \\[4pt] 
-25.375 & -4.02$^{\,\, + 0.02}_{\,\,-0.02}$ & 336 &     & -25.375 & -4.13$^{\,\, + 0.03}_{\,\,-0.03}$ & 260 &    & -25.375 & -4.66$^{\,\, + 0.05}_{\,\,-0.05}$ & 76 \\[4pt] 
-25.125 & -3.71$^{\,\, + 0.02}_{\,\,-0.02}$ & 495 &     & -25.125 & -3.85$^{\,\, + 0.02}_{\,\,-0.02}$ & 354 &    & -25.125 & -4.25$^{\,\, + 0.04}_{\,\,-0.04}$ & 141 \\[4pt] 
-24.875 & -3.39$^{\,\, + 0.02}_{\,\,-0.02}$ & 733 &     & -24.875 & -3.59$^{\,\, + 0.02}_{\,\,-0.02}$ & 468 &    & -24.875 & -3.83$^{\,\, + 0.03}_{\,\,-0.03}$ & 265 \\[4pt] 
-24.625 & -3.11$^{\,\, + 0.01}_{\,\,-0.01}$ & 1003 &     & -24.625 & -3.34$^{\,\, + 0.02}_{\,\,-0.02}$ & 593 &    & -24.625 & -3.50$^{\,\, + 0.02}_{\,\,-0.02}$ & 410 \\[4pt] 
-24.375 & -2.87$^{\,\, + 0.01}_{\,\,-0.01}$ & 1264 &     & -24.375 & -3.15$^{\,\, + 0.02}_{\,\,-0.02}$ & 660 &    & -24.375 & -3.19$^{\,\, + 0.02}_{\,\,-0.02}$ & 604 \\[4pt] 
-24.125 & -2.71$^{\,\, + 0.01}_{\,\,-0.01}$ & 1284 &     & -24.125 & -3.09$^{\,\, + 0.02}_{\,\,-0.02}$ & 544 &    & -24.125 & -2.95$^{\,\, + 0.02}_{\,\,-0.02}$ & 740 \\[4pt] 
-23.875 & -2.53$^{\,\, + 0.01}_{\,\,-0.01}$ & 1380 &     & -23.875 & -2.94$^{\,\, + 0.02}_{\,\,-0.02}$ & 537 &    & -23.875 & -2.75$^{\,\, + 0.02}_{\,\,-0.02}$ & 843 \\[4pt] 
-23.625 & -2.41$^{\,\, + 0.01}_{\,\,-0.01}$ & 1224 &     & -23.625 & -2.86$^{\,\, + 0.02}_{\,\,-0.02}$ & 440 &    & -23.625 & -2.61$^{\,\, + 0.02}_{\,\,-0.02}$ & 784 \\[4pt] 
-23.375 & -2.29$^{\,\, + 0.01}_{\,\,-0.01}$ & 1179 &     & -23.375 & -2.77$^{\,\, + 0.02}_{\,\,-0.02}$ & 391 &    & -23.375 & -2.47$^{\,\, + 0.02}_{\,\,-0.02}$ & 788 \\[4pt] 
-23.125 & -2.24$^{\,\, + 0.01}_{\,\,-0.01}$ & 989 &     & -23.125 & -2.73$^{\,\, + 0.02}_{\,\,-0.02}$ & 321 &    & -23.125 & -2.41$^{\,\, + 0.02}_{\,\,-0.02}$ & 668 \\[4pt] 
-22.875 & -2.14$^{\,\, + 0.01}_{\,\,-0.01}$ & 875 &     & -22.875 & -2.67$^{\,\, + 0.03}_{\,\,-0.03}$ & 260 &    & -22.875 & -2.30$^{\,\, + 0.02}_{\,\,-0.02}$ & 615 \\[4pt] 
-22.625 & -2.14$^{\,\, + 0.02}_{\,\,-0.02}$ & 580 &     & -22.625 & -2.68$^{\,\, + 0.03}_{\,\,-0.04}$ & 166 &    & -22.625 & -2.28$^{\,\, + 0.02}_{\,\,-0.02}$ & 414 \\[4pt] 
-22.375 & -2.12$^{\,\, + 0.02}_{\,\,-0.02}$ & 417 &     & -22.375 & -2.78$^{\,\, + 0.05}_{\,\,-0.05}$ & 92 &    & -22.375 & -2.23$^{\,\, + 0.02}_{\,\,-0.02}$ & 325 \\[4pt] 
-22.125 & -2.13$^{\,\, + 0.03}_{\,\,-0.03}$ & 299 &     & -22.125 & -2.81$^{\,\, + 0.06}_{\,\,-0.06}$ & 62 &    & -22.125 & -2.23$^{\,\, + 0.03}_{\,\,-0.03}$ & 237 \\[4pt] 
-21.875 & -2.10$^{\,\, + 0.03}_{\,\,-0.03}$ & 234 &     & -21.875 & -2.77$^{\,\, + 0.06}_{\,\,-0.07}$ & 51 &    & -21.875 & -2.21$^{\,\, + 0.03}_{\,\,-0.03}$ & 183 \\[4pt] 
-21.625 & -2.09$^{\,\, + 0.03}_{\,\,-0.03}$ & 177 &     & -21.625 & -2.94$^{\,\, + 0.09}_{\,\,-0.10}$ & 25 &    & -21.625 & -2.16$^{\,\, + 0.04}_{\,\,-0.04}$ & 152 \\[4pt] 
-21.375 & -2.04$^{\,\, + 0.04}_{\,\,-0.04}$ & 150 &     & -21.375 & -2.68$^{\,\, + 0.08}_{\,\,-0.08}$ & 35 &    & -21.375 & -2.16$^{\,\, + 0.04}_{\,\,-0.04}$ & 115 \\[4pt] 
-21.125 & -2.10$^{\,\, + 0.04}_{\,\,-0.04}$ & 103 &     & -21.125 & -2.73$^{\,\, + 0.10}_{\,\,-0.10}$ & 24 &    & -21.125 & -2.22$^{\,\, + 0.05}_{\,\,-0.05}$ & 79 \\[4pt] 
-20.875 & -2.09$^{\,\, + 0.05}_{\,\,-0.05}$ & 82 &     & -20.875 & -2.74$^{\,\, + 0.11}_{\,\,-0.11}$ & 19 &    & -20.875 & -2.21$^{\,\, + 0.06}_{\,\,-0.06}$ & 63 \\[4pt] 
-20.625 & -2.06$^{\,\, + 0.06}_{\,\,-0.06}$ & 69 &     & -20.625 & -2.78$^{\,\, + 0.13}_{\,\,-0.14}$ & 13 &    & -20.625 & -2.15$^{\,\, + 0.06}_{\,\,-0.06}$ & 56 \\[4pt] 
-20.375 & -2.18$^{\,\, + 0.08}_{\,\,-0.08}$ & 37 &     & -20.375 & -2.80$^{\,\, + 0.16}_{\,\,-0.17}$ & 9 &    & -20.375 & -2.30$^{\,\, + 0.09}_{\,\,-0.09}$ & 28 \\[4pt] 
-20.125 & -2.13$^{\,\, + 0.09}_{\,\,-0.09}$ & 29 &     & -20.125 & -2.89$^{\,\, + 0.23}_{\,\,-0.25}$ & 5 &    & -20.125 & -2.21$^{\,\, + 0.10}_{\,\,-0.10}$ & 24 \\[4pt] 
-19.875 & -2.05$^{\,\, + 0.10}_{\,\,-0.10}$ & 23 &     & -19.875 & -2.41$^{\,\, + 0.15}_{\,\,-0.16}$ & 10 &    & -19.875 & -2.31$^{\,\, + 0.13}_{\,\,-0.14}$ & 13 \\[4pt] 
-19.625 & -2.13$^{\,\, + 0.14}_{\,\,-0.15}$ & 12 &     & -19.625 & -2.42$^{\,\, + 0.20}_{\,\,-0.22}$ & 6 &    & -19.625 & -2.44$^{\,\, + 0.20}_{\,\,-0.22}$ & 6 \\[4pt] 
-19.375 & -2.24$^{\,\, + 0.20}_{\,\,-0.22}$ & 6 &     & -19.375 & -2.39$^{\,\, + 0.25}_{\,\,-0.29}$ & 4 &    & -19.375 & -2.76$^{\,\, + 0.37}_{\,\,-0.47}$ & 2 \\[4pt] 
-19.125 & -2.63$^{\,\, + 0.52}_{\,\,-0.87}$ & 1 &     &   &   &   &     & -19.125 & -2.64$^{\,\, + 0.52}_{\,\,-0.87}$ & 1 \\[4pt] 
-18.875 & -2.04$^{\,\, + 0.30}_{\,\,-0.35}$ & 3 &     & -18.875 & -2.24$^{\,\, + 0.37}_{\,\,-0.47}$ & 2 &    & -18.875 & -2.48$^{\,\, + 0.52}_{\,\,-0.87}$ & 1 \\[4pt] 
&&&&&&&&&&\\
\hline
\label{lfdata_tab}
\end{tabular}
\end{table*}

\subsection{Modelling the Luminosity Function}

We fit Schechter functions to our sample using the maximum likelihood method described by \citet{mar83}. This particular form of the method works as follows. If we take our sample of galaxies, which have redshift defined volumes and magnitudes, we can calculate the volume within which these galaxies should reside. If we then break the volume up into portions defined by dM (magnitude range) and dV (volume range), which are small enough to only contain 1 or no galaxies, we can use Poisson statistics to determine the probability of each ``box'' containing either 1 or 0 galaxies. This is represented by the likelihood equation
\begin{equation}
\mathcal{L} = \displaystyle\prod_i^N[\lambda(V_i,M_i)dVdM \exp^{-\lambda(V_i,M_i)dVdM}]\displaystyle\prod_j \exp^{-\lambda(V_j,M_j)dVdM}
\end{equation}
the first product relating to boxes containing galaxies and the second for empty boxes. $\lambda(V, M)dVdM = \rho(V, M)\Omega(V, M)dVdM$ is the expected number of galaxies contained in dVdM. Index j relates to boxes where no galaxies are found. 

Next, rather than taking all of our probabilities and multiplying them together to find the best fit, if the likelihood $S = -2\ln(\mathcal{L})$, we effectively transform all of our products into sums and we find that
\begin{equation}
S = -2\displaystyle\sum_i^N \ln[\rho(V_i, M_i)] + 2\displaystyle\iint\rho(V, M)\Omega(V, M)dVdM
\end{equation}
To obtain the best parameters for our fit, we find the minimum of S. This method is convenient as it closely relates to least-squares fitting. 

The Schechter function that the maximum-likelihood method fits to the galaxy distribution is of the form
\begin{equation}
\Phi(M)dM\, =\, \frac{(0.4\, \ln\, 10)\, \Phi^*\, 10^{0.4(\alpha\, +\, 1)\,(M^* \,-\, M)}}{\exp\, (10^{0.4(M^*\, -\, M)})}dM
\end{equation}
The term $\Phi(M)dM$ is the space density of galaxies between magnitudes $M$ and $M + dM$. The parameters $M^*$, $\Phi^*$ and $\alpha$ shape the function and are the 3 parameters that the maximum-likelihood method will have to fit. $M^*$ is the magnitude of an average galaxy, $\Phi^*$ is the average space density of galaxies and $\alpha$ is a term that defines the slope of the faint end of the LF.

A simple method for determining the uncertainties of the ML fit would have been to observe the difference between best fit parameters and fit parameters at S + 1. This was tested and was found to produce smaller than expected uncertainties as this method does not account for large scale structure in the sample and will be dominated by cosmic variance.

Instead, we use a Jackknife approximation \citep[e.g.,][]{que56, efron82} to model the impact of cosmic variance on the errors. By dividing the sky into 20 equally sized regions, we run our maximum likelihood fitting algorithm for the entire sky, minus each subsample separately. Each jackknife replication results in a slightly different value for $\Phi(M^*_i,\Phi^*_i,\alpha_i)$ that is calculated using the maximum likelihood method described in \S 4.1. We then observe the differences between each jackknife run. The new fit parameter will be the mean of the mean of $\Phi(M^*_i,\Phi^*_i,\alpha_i)$ for all runs and errors will be determined using the standard errors of this mean. 

The fits made to the galaxy luminosity distribution can be seen in Figures \ref{schechtertotal_fig} and \ref{schechterearlylate_fig} (total and late/early-type fits respectively). Fits are over-plotted on the $1/V_{\rm{MAX}}$ functions. As can be seen, the luminosity functions are generally well fitted by the Schechter form at the faint end but underestimate the function at the bright end. This was also noted by \citet{jones06} and can be explained by two separate problems. The first is that it is impossible to get the Schechter function to turn over at the knee of the luminosity function as sharply as the real galaxy distribution does. This is simply a limitation of the exponential-power law combination of the Schechter form. The second problem is that the most massive galaxies (such as BCGs and massive spirals) appear as a toe at the end of the bright end. Obviously a function like the Schechter can not accommodate upturns like this either. In the case of the late-type function, the Schechter form also slightly overestimates the function at the far faint end. Parameters for these fits are displayed in Table \ref{lfcompare_tab}.

\begin{figure}[]
\begin{center}
\includegraphics[width = .49\textwidth]{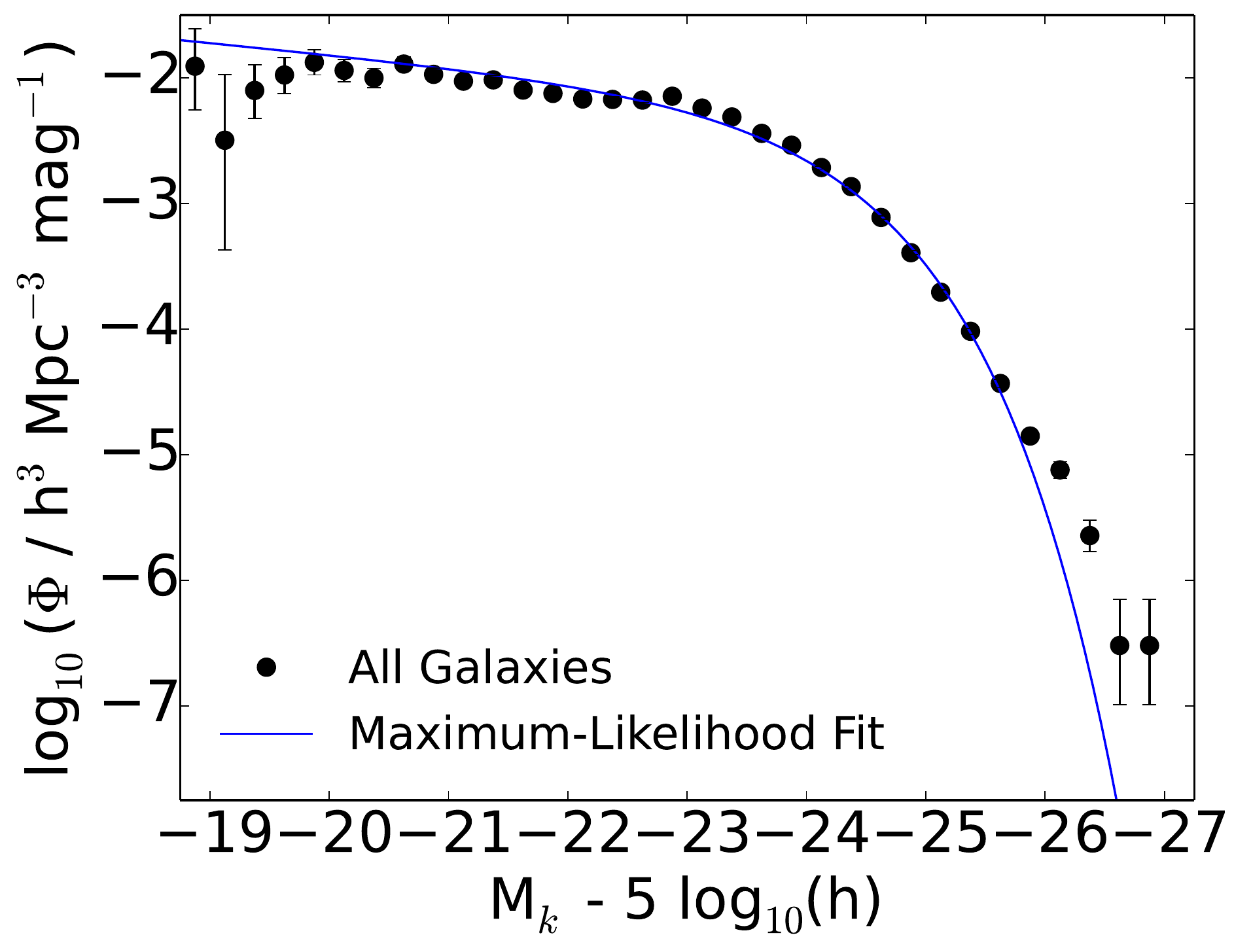}
\caption{1/$V_{\rm{MAX}}$ K-band luminosity function for the total galaxy sample with best maximum-likelihood Schechter function fit. The function is well fitted by the Schechter form at the faint end, bar the faintest points, but does not fit well at the bright end. Parameter fit values are displayed in Table \ref{lfcompare_tab}.}
\label{schechtertotal_fig}
\end{center}
\end{figure}
\begin{figure}[]
\begin{center}
\includegraphics[width = .49\textwidth]{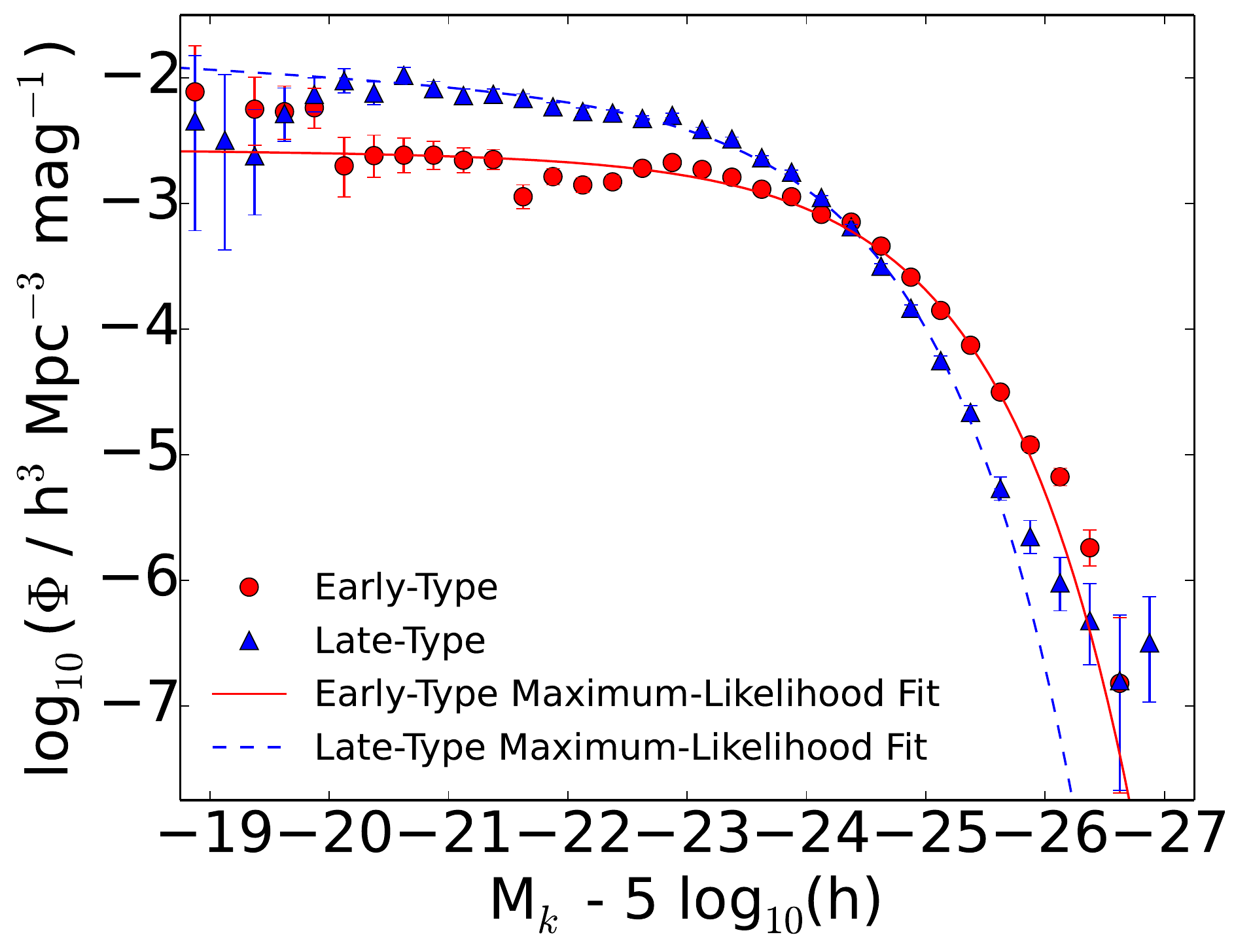}
\caption{1/$V_{\rm{MAX}}$ K-band luminosity functions for early and late-type galaxy samples with best maximum-likelihood Schechter function fits. In this case the Schechter form fits the early-type function extremely well, however, it deviates considerably from the bright end of the late-type function. See Table \ref{lfcompare_tab} for parameter fit values.}
\label{schechterearlylate_fig}
\end{center}
\end{figure}

\subsection{Accounting for Over/Under-Densities in the Sample}

In Figure \ref{densitycorrect_fig} we plot a measure of the over-density of the most luminous galaxies in our sample as a function of luminosity density. For our faintest bins and smallest volumes, it is likely that our sample will be systematically over-dense or under-dense relative to the low redshift Universe. To correct for this, we assume that the largest volume measured by our sample is a representative volume, and then measure how the density of bright galaxies varies from this for smaller volumes. As the number of bright galaxies decreases as we move to smaller volumes, we use progressively fainter galaxies to measure the over-density for the smallest volumes used to measure the 
luminosity functions. We calculate over-densities for our total sample, early-type as well as late-type samples. There is little difference noted between the early and late-type over-densities at corresponding values of $D_L$ so we apply the total sample scaling values to all other samples. Our corrections make use of galaxies with magnitudes of $\simeq$M* and dimmer, which are not strongly biased relative to the overall galaxy population.

\begin{figure}[]
\begin{center}
\includegraphics[width = .49\textwidth]{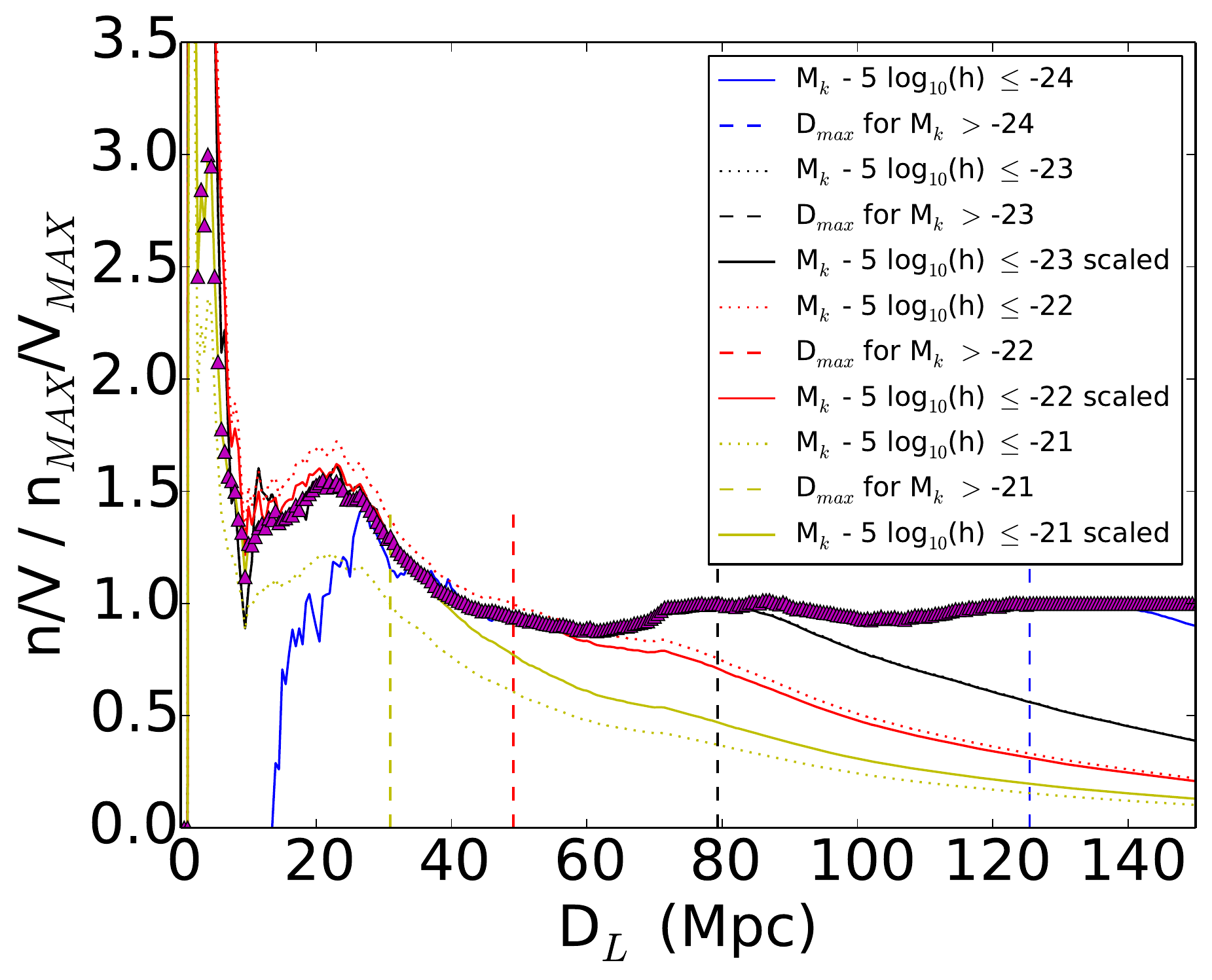}
\caption{A plot of over-density of bright galaxies in the total sample as a function of $D_L$. The over-density was determined by comparing the density of very luminous galaxies within a given distance and volume with the density of comparable galaxies over the entire survey. As the space density of very luminous galaxies is small, we used progressively fainter galaxies for smaller distances and volumes. The plot shows a significant over-density of $\sim$ 1.5 at small $D_L$ in the local Universe, and considerably higher below our limit of $D_L$ = 10 Mpc. There are several clusters at $D_L$ $\simeq$ 20 Mpc that could account for the observed rise in density (e.g. the Virgo Southern Extension, the Fornax Cluster, the Fornax Wall).}
\label{densitycorrect_fig}
\end{center}
\end{figure}

To then apply this to the luminosity function (1/V$_{\rm{MAX}}$ and maximum-likelihood methods), we calculate the absolute magnitude values corresponding to each luminosity distance at which we have calculated over-densities. This factor is then applied to the appropriate $\Phi$ values to scale the LF. Figure \ref{lfscaledtotal_fig} demonstrates an example of this applied to the luminosity function of the total sample. This method is similar, but not identical to that used by \citet{baldr12}. Note that all luminosity functions displayed from this point onwards will have the above discussed density correction applied.

\begin{figure}[]
\begin{center}
\includegraphics[width = .49\textwidth]{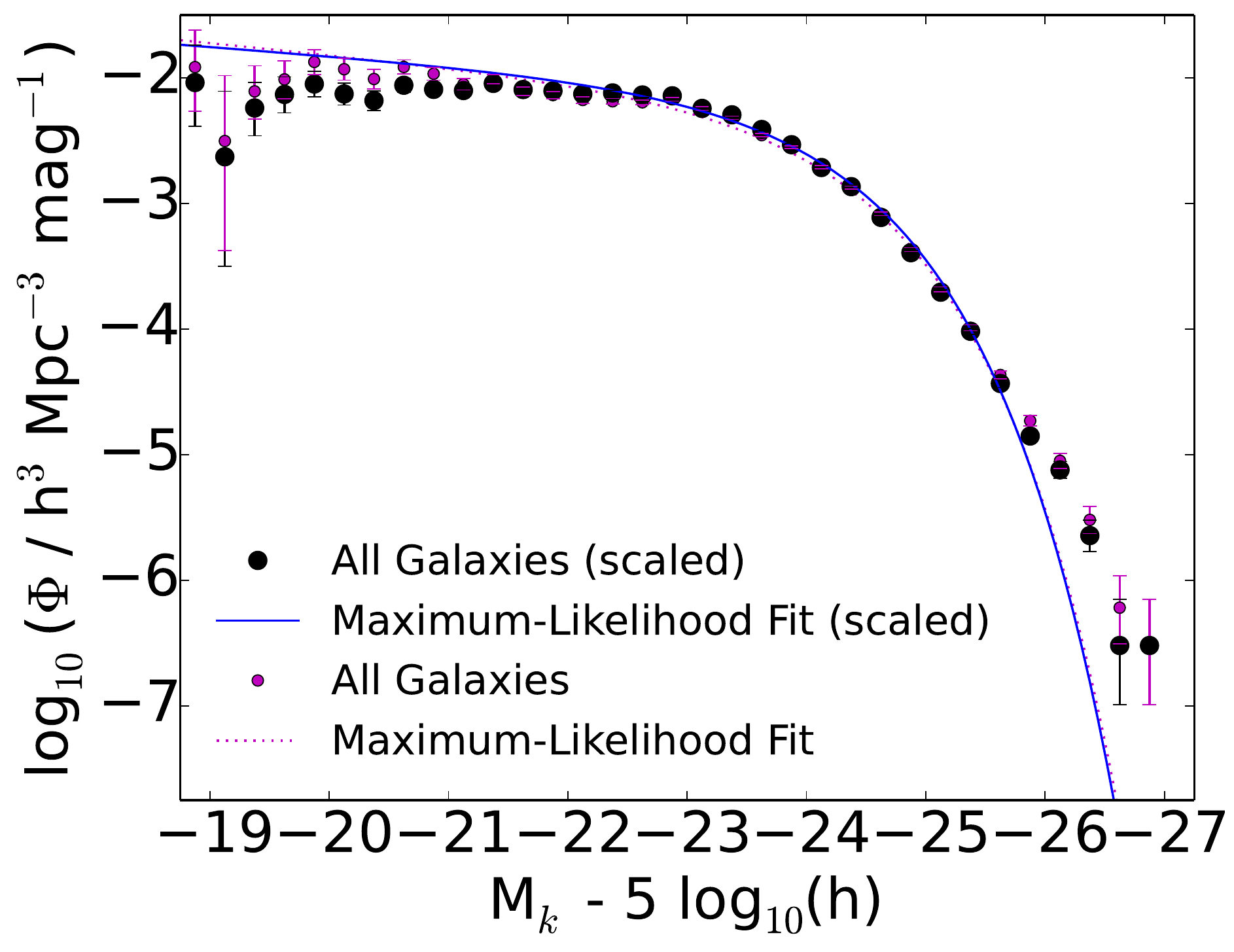}
\caption{The total galaxy sample LF with density corrections applied. Note the change to the shape of the function, particularly at the faint end. Applying this correction slightly lowers the faint end of the function.}
\label{lfscaledtotal_fig}
\end{center}
\end{figure}

\subsection{Effects of Varying T-Type Cut-Off Value for Late-Type/Early-Type Classification}

When applying morphological classifications to galaxies, there will always be some fraction of galaxies that are misclassified. This is very true for the case of S0 galaxies, e.g., a disk-like galaxy with no discernible star-forming regions could be an S0 or a spiral, or alternatively, a featureless elliptical galaxy could also be an S0 at low inclination. To validate our choice of a late-type/early-type boundary at T = -1, we have made new late-type and early-type luminosity functions with the boundary value set to T = 0 and T = -2. By performing a maximum likelihood fit to these new functions, we can determine whether this change significantly alters the shape of the functions. The resulting fit parameters are presented in Table \ref{varymorph_tab}.
\begin{table}[]\centering
\caption{Maximum Likelihood fit parameters for varying T-type cut-off}
\scriptsize \begin{tabular}{ccccc}
\hline
sample & objects & M$^*$ $-$ $5$ $\rm{log_{10}(h)}$ & $\alpha$ & log$_{10}$($\Phi^*$)\\ 
 & & (mag) & & ({h$^{3}\,$ \rm Mpc}$^{-3}\,$ mag$^{-1}$) \\
 & & & &\\ 
\hline
 & & & & \\ 
 T $>$ -2 & 7,857 & -23.50 $\pm$ 0.07 & -1.13 $\pm$ 0.09 & -2.13 $\pm$ 0.04  \\
T $\le$ -2 & 5,468 & -24.04 $\pm$ 0.06 & -1.03 $\pm$ 0.10 & -2.60 $\pm$ 0.05 \\
  & & & & \\
 T $>$ -1 & 7,685 & -23.49 $\pm$ 0.06 & -1.13 $\pm$ 0.10 & -2.13 $\pm$ 0.04 \\
 T $\le$ -1 & 5,640  & -24.03 $\pm$  0.06 &  -1.02 $\pm$ 0.10 & -2.58 $\pm$ 0.05  \\
 & & & & \\
T $>$ 0 & 6,429 & -23.46 $\pm$ 0.07 & -1.14 $\pm$ 0.10 & -2.19 $\pm$ 0.04 \\
T $\le$ 0 & 6,854 & -24.00 $\pm$ 0.06 & -1.08 $\pm$ 0.10 & -2.48 $\pm$ 0.05 \\
 & & & & \\ 
\hline
\label{varymorph_tab}
\end{tabular}
\end{table}

Though the overall shape of the fits do not change significantly, it is obvious that moving the early-type and late-type cutoff value up and down by 1 changes some parameters more than others. The largest difference between all fits is in the $\Phi^*$ parameter, with some change in $\alpha$ and nearly no change for the M$^*$ parameter which seems relatively insensitive to the change in T-type cutoff. From T = -2 to T = -1, for the late-type sample, $\Phi^*$ remains the same, $\alpha$ r M$^*$ changing by 0.01 magnitudes. Alternatively, for the early-type sample, $\Phi^*$ changes by 5\%, $\alpha$ changes by 0.01 and M$^*$ by 0.01 magnitudes. The largest change is seen in the T = 0 functions, with an 15\% change from the original $\Phi^*$ value, a difference of 0.01 between $\alpha$ values and 0.03 magnitudes from M$^*$ for the late-type sample and a 31\% change from the original $\Phi^*$ value, 0.06 from $\alpha$ and 0.03 magnitudes from M$^*$ for the early-type sample.

The large change in $\Phi^*$ between the T = 0 and T = -1 fits is expected and is simply a reflection of the fact that $\simeq$8\% of galaxies in the total sample have a T-type of 0. As the M$^*$ and $\alpha$ parameters remain virtually unchanged, the overall shape of the fits to the luminosity functions also remain relatively similar. 

\subsection{Galaxy Colors}

As an aim of this paper is to demonstrate that the difference in shape between morphology-defined luminosity functions and color-defined luminosity functions can be explained by the presence of a large population of red spirals, we investigate the optical blue/red color distribution of galaxies in our sample using NASA Sloan Atlas u and r-band Petrosian flux photometry. From our total sample, 5,417 galaxies have NASA Sloan Atlas counterparts, 3,155 of which are late-type and 2,262 of which are early-type. 

Unlike the Petrosian fluxes in the original SDSS galaxy catalog, the radius for the fits for each galaxy in the NASA Sloan Atlas does not differ between bands, rather, the radius for the Petrosian fit to a given galaxy is initially fixed for the r-band and then used for all other fits in subsequent bands. This makes Petrosian magnitudes a far better choice for determining color than was previously the case. While photographic photometry from SuperCOSMOS \citep{ham01b, ham01a} was examined for galaxies outside the SDSS area, it had insufficient accuracy to reliably differentiate optically red and blue galaxies. In the case of NASA Sloan Atlas, using Petrosian photometry from g and r filters to determine galaxy color is more difficult as separation between optically blue and red galaxies is less obvious, and it is once again too difficult to easily differentiate between these two populations using a simple color cut. As the u and r filters are significantly separated in wavelength they provide a better indication of color.

Our choice of u-r color over g-r color is also justified by comparing galaxies with and without H$_\alpha$ emission in the \citet{brown14} sample. We find that when using matched aperture photometry of bright galaxies, u-r is more effective than g-r for selecting galaxies with and without H$_\alpha$ emission. Particular examples of galaxies from \citet{brown14} with H$_\alpha$ emission and red g-r colours are NGC 3351, NGC 7591 and NGC 7771, all of which are identified as blue using the u-r color cut in this paper. 

We use the Petrosian fluxes rather than the Sersic fluxes from the NASA Sloan Atlas as we have found that the Sersic fluxes tend to attribute too much light to the bulges of galaxies, causing some blue galaxies with bright red bulges to have a color redder than the Petrosian photometry and at odds with visual inspections. An illustrative example is the photometry of spiral (T = 4) NGC 0151, where the Sersic and Petrosian u-r colors are 2.57 and 1.77 respectively. The Sersic fit photometry is consistent with a red sequence galaxy, but this galaxy has nebular emission lines and a blue star forming disk. This point is further demonstrated by studying NASA Sloan Atlas Sersic indices vs. T-type morphological classifications. There are a number of early-type galaxies with low Sersic indices and late-type galaxies with high Sersic indices. Respectively, these galaxies are generally large, diffuse red objects and blue spirals with large red bulges and diffuse blue disks. We might expect to see examples of blue pseudo-bulges in the population of early-types with low Sersic indices but we do not.

We opt for a simple red/blue color cut of,
\begin{equation}
u-r = -0.05 \ M_K - 5 \rm{log_{10}(h)} \ + 1.1
\end{equation}
with a slope defined by the shape of the red sequence. As an alternative, we also tested the u-r color cut defined in \citet{baldr04} on our data. No color cut is perfect, but we find that upon visual inspection of the red galaxy populations resulting from both cuts, there are a larger number of galaxies with blue, star forming, disks included in the case of \citet{baldr04} ($\simeq$3\% for our cut as apposed to $\simeq$9\% for the \citet{baldr04} cut).

A color-magnitude diagram for the NASA Sloan Atlas sub-sample is displayed in Figure \ref{color_fig} with the above mentioned simple red/blue color cut. Color values are plotted against our calculated K-band magnitudes to provide a better comparison with our derived luminosity functions. As can be seen in Figure \ref{color_fig}, most lenticular and elliptical galaxies reside above the color cut line, however, a large fraction of spirals, between K-band magnitude range of -23 < $M_K$ < -26 also sit above this line.

\begin{figure}[]
\begin{center}
\includegraphics[width = .49\textwidth]{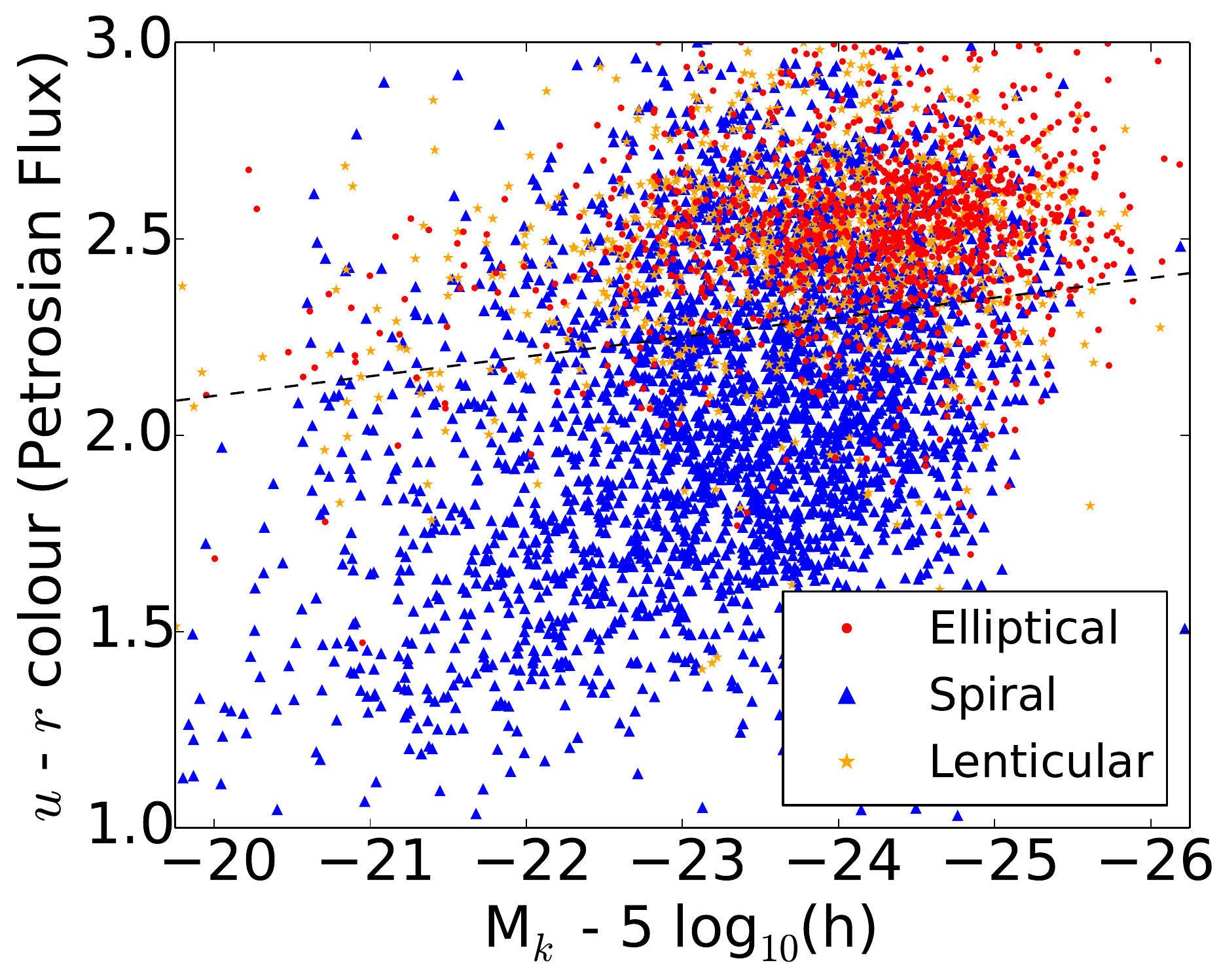}
\caption{Color-Magnitude for NASA Sloan Atlas Petrosian u-r colors for all galaxy types. We have applied a simple sloping color cut for this data set to better fit the shape of the red-sequence. A large number of late-type galaxies can be seen to sit above the cut-off and are thus classified as red.}
\label{color_fig}
\end{center}
\end{figure}

To confirm this observation, in Figure \ref{spiralfrac_fig} we show the fraction of red spiral galaxies to all spiral galaxies against K-band absolute magnitudes, accompanied by plots of galaxy number counts against K-band magnitudes. This shows us that at brighter K-band absolute magnitudes, $\simeq$50$\%$ of spirals are red rather than blue. It is also important to note that, even though these galaxies have comparable magnitudes, the red spirals will have slightly higher stellar masses due to their older stellar populations \citep[][shows that, for galaxies with the same K-band absolute magnitude and u-r colors of 3 and 1, K-band M/L will only differ by $\simeq$0.3 mag]{bell03}. As luminosity decreases, the fraction of red to blue spirals decreases but even at the faint end of our distribution, $\sim$20$\%$ of the overall spiral population are still red.

\begin{figure*}[]
\begin{center}
\plottwo{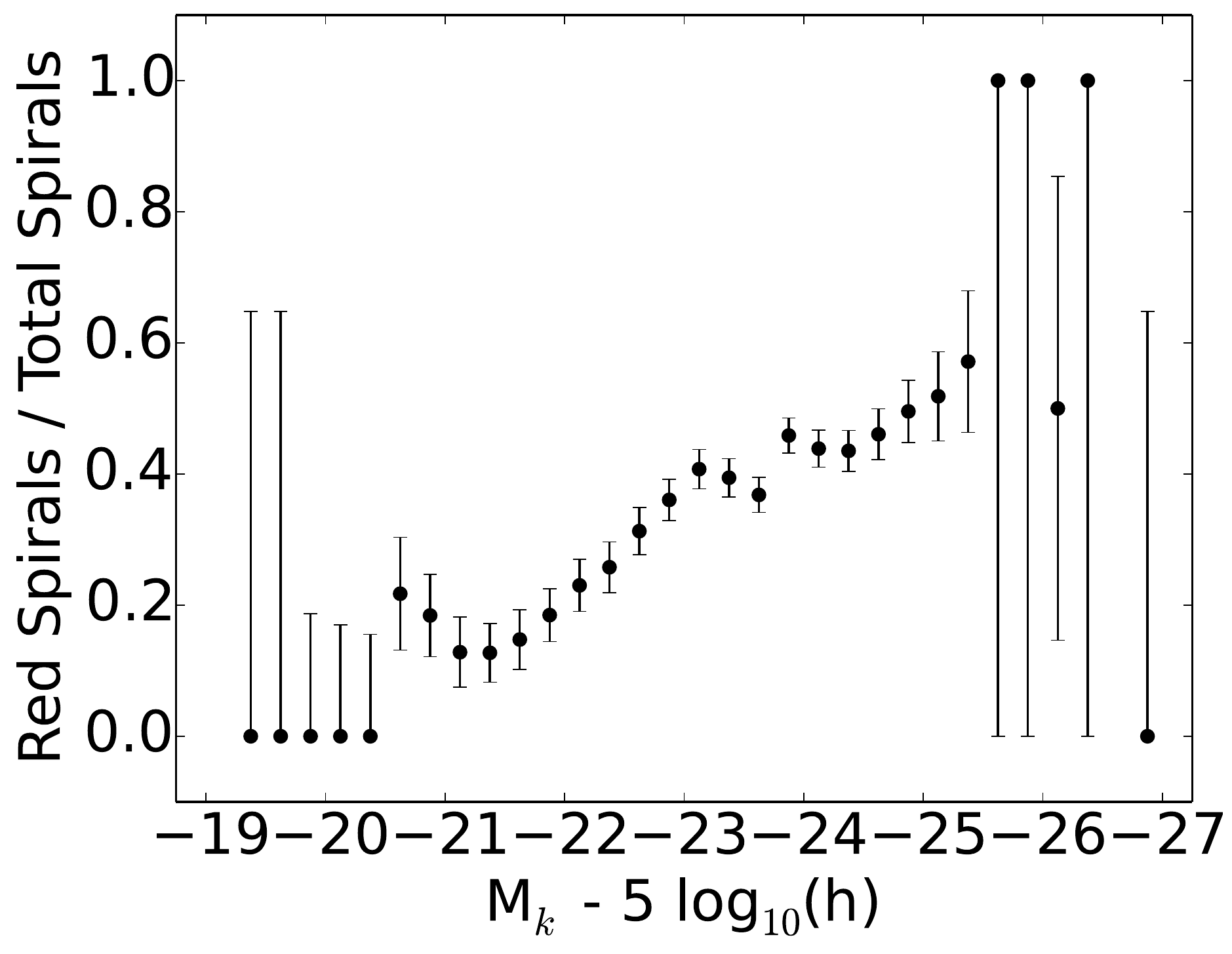}{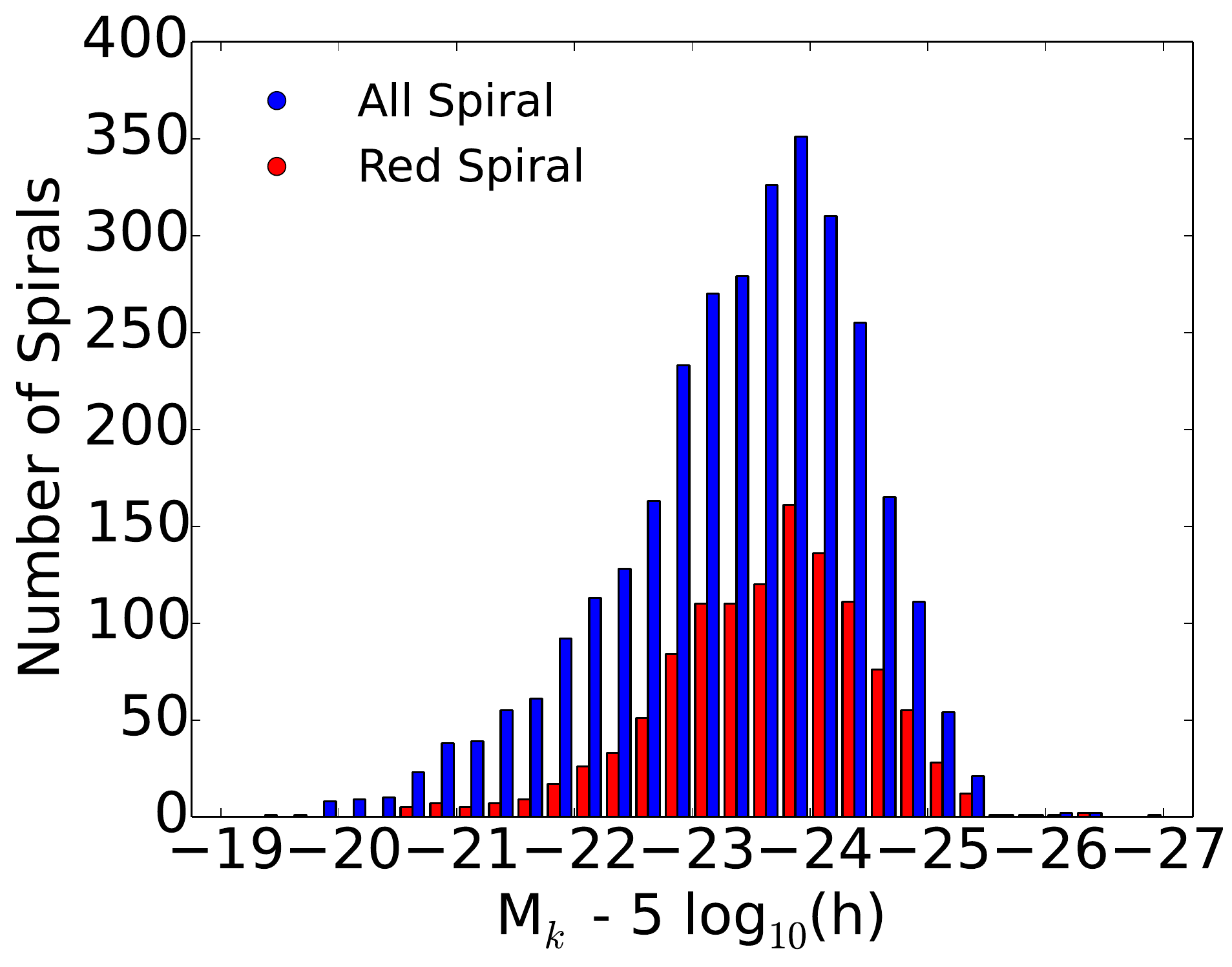}
\caption{Fraction of red to total spiral population (\textit{left}), and galaxy number counts  (\textit{right}) as a function of absolute K magnitude for NASA Sloan Atlas u-r colors. These plots show the increase in the fraction of red spirals to blue spirals in the overall spiral population with increasing stellar mass and absolute K magnitude.}
\label{spiralfrac_fig}
\end{center}
\end{figure*}

To account for the possibility of inclination affecting the colors of our galaxy sample, we plot axis ratios, taken from 2MASS K-band data. If all of our spirals were highly inclined, this would naturally make them appear redder in color, though, as can be seen in Figure \ref{spiralaxis_fig}, there is a trend towards higher inclinations but this is not the case for the entire red spiral population.

\begin{figure}[]
\begin{center}
\includegraphics[width = .49\textwidth]{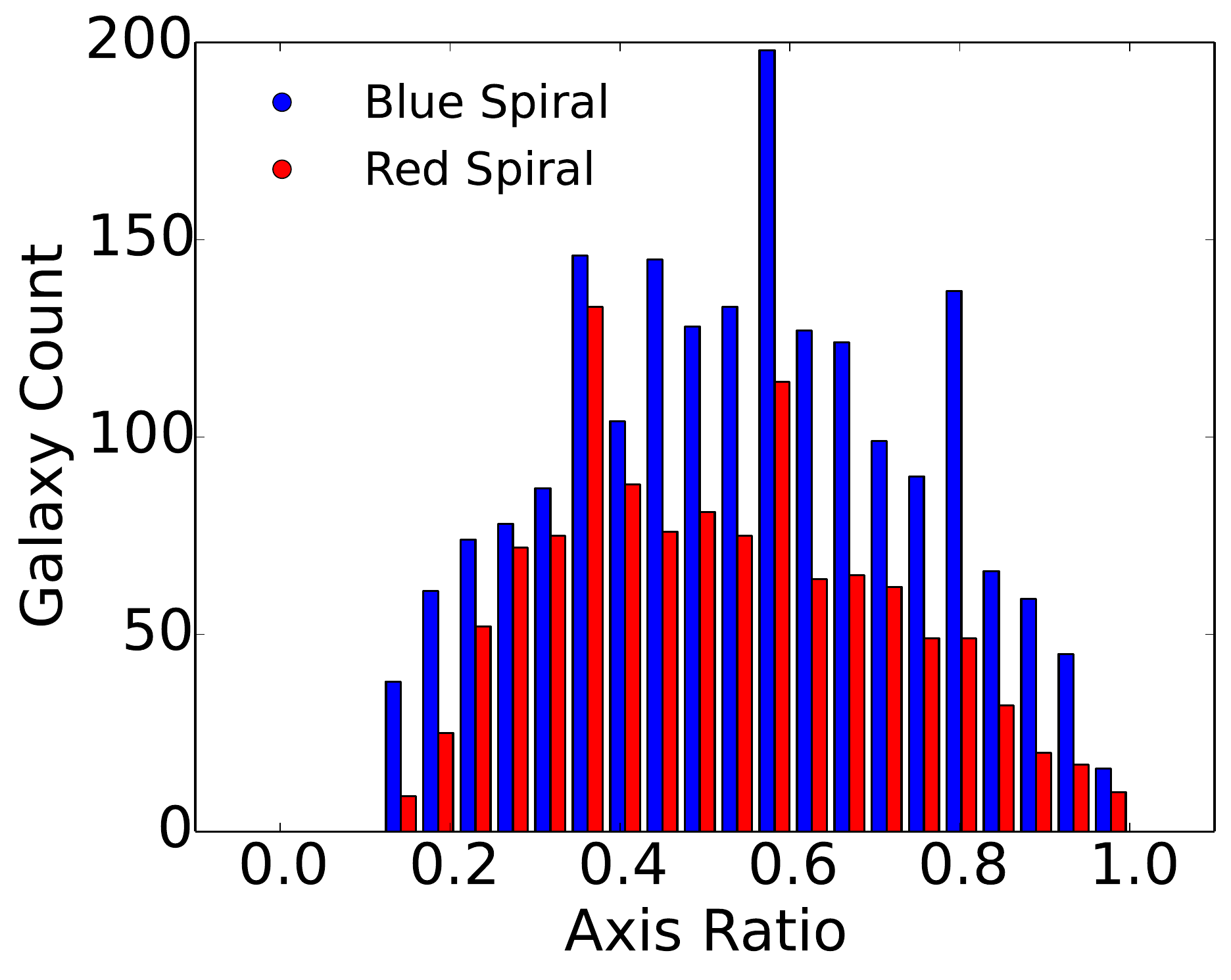}
\caption{Axis ratio comparison for red/blue spirals. There is a slight bias for red spirals to be edge-on, but this does not explain the distribution of galaxy colors.}
\label{spiralaxis_fig}
\end{center}
\end{figure}

To further investigate the effects of dust reddening on our sample, and on the relationship between stellar mass and red spiral fraction, we use the same inclination limits as \citet{mas10} and examine only face-on spiral galaxies with a 2MASS axis-ratio $b/a >$ 0.6. Figure \ref{spiralfracface_fig} shows the results of this. This subsample is considerably smaller than the full spiral sample but still shows the same upward trend in fraction of red/total spirals with increasing stellar mass, though it is less extreme than in the case of the inclination independent sample. This trend was also noted by \citet{mas10} and is comparable in magnitude to these results.

\begin{figure*}[]
\begin{center}
\plottwo{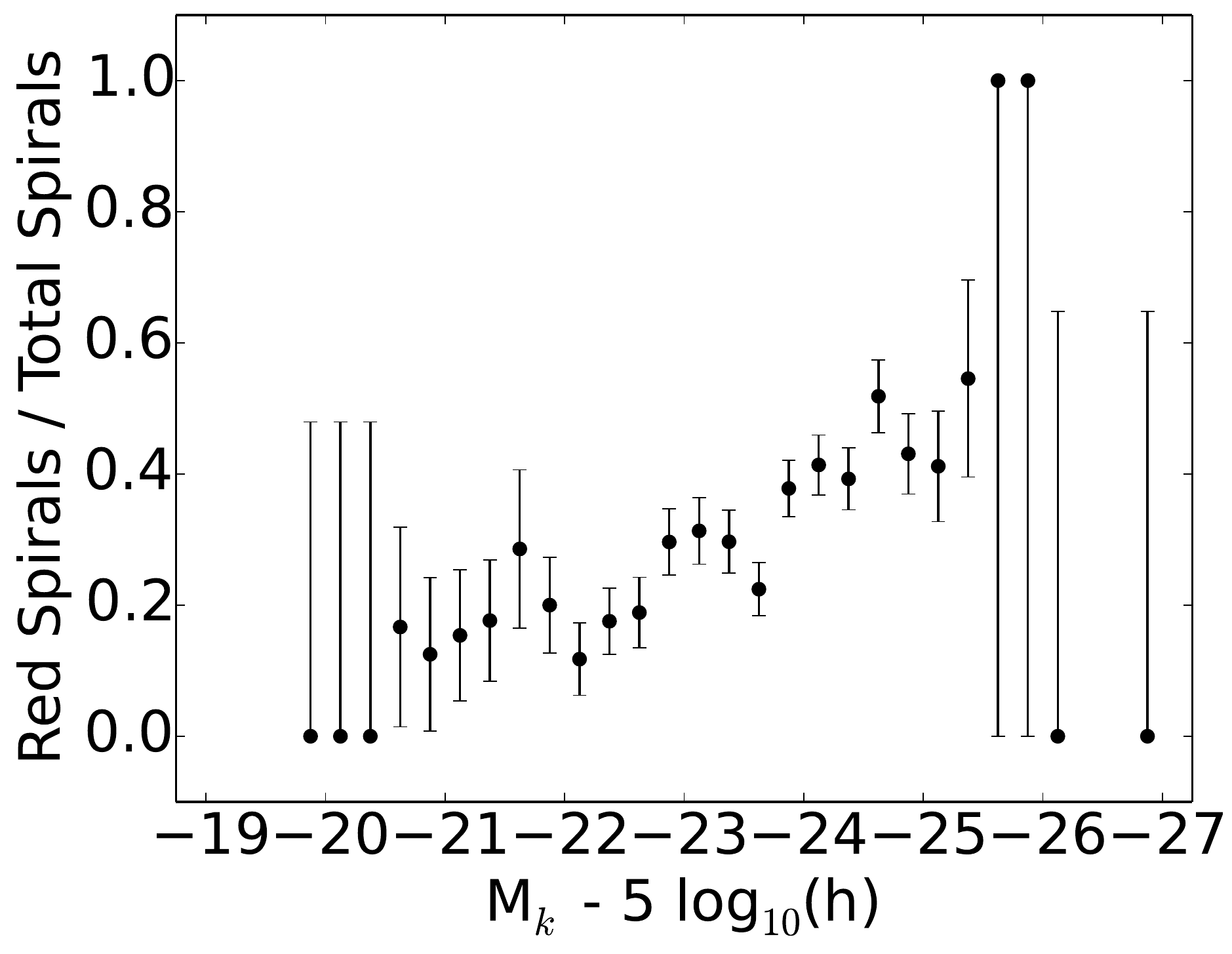}{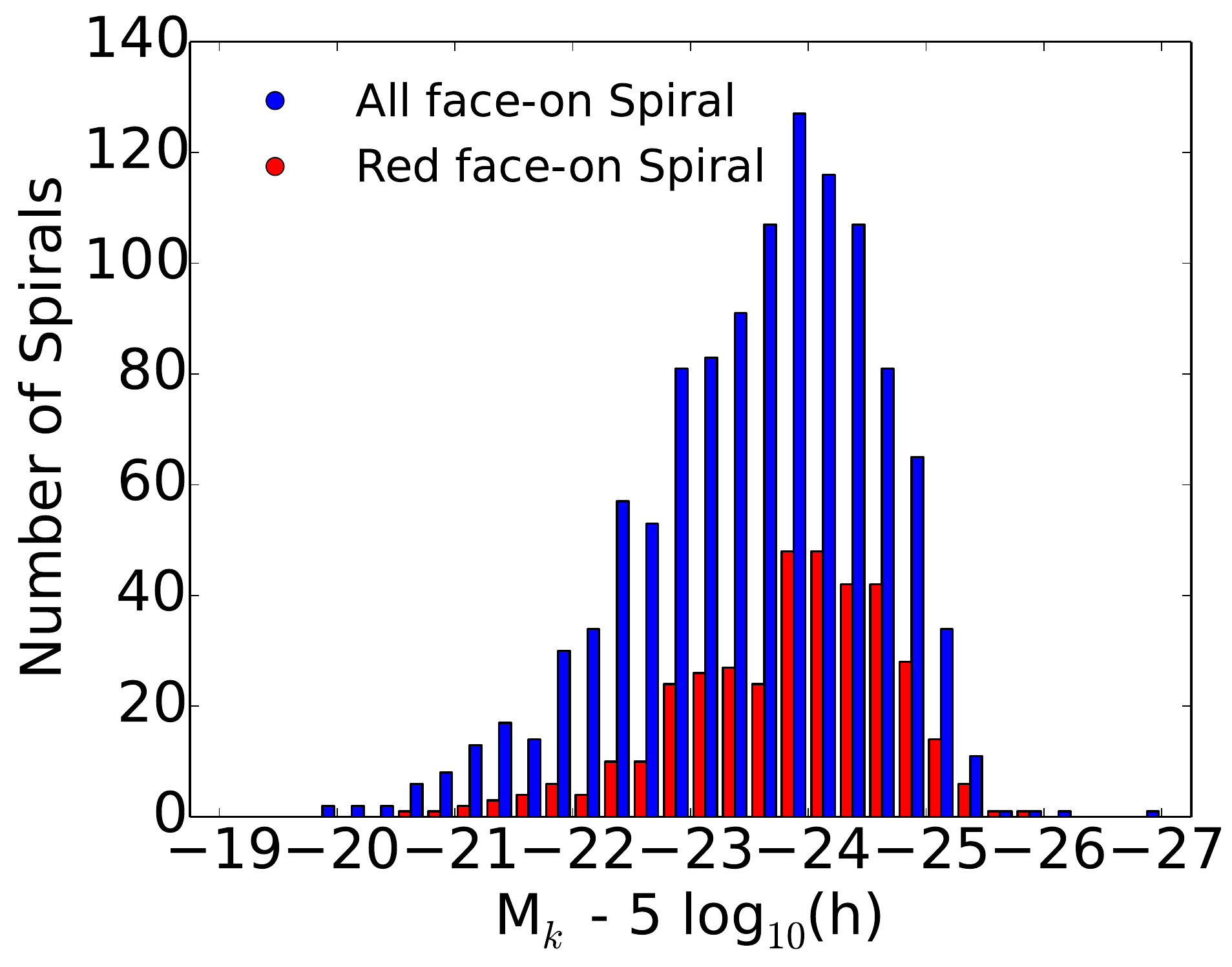}
\caption{Fraction of red to total spiral population (\textit{left}), and galaxy number counts  (\textit{right}) for spirals with 2MASS axis ratio $b/a > 0.6$. Both are a function of absolute K magnitude for NASA Sloan Atlas u-r colors. Fits and uncertainties are calculated as in Figure 11. This plot shows a similar but less extreme upward trend for red/total fraction with increasing stellar mass.}
\label{spiralfracface_fig}
\end{center}
\end{figure*}

Lastly, by visually inspecting the NASA Sloan Atlas galaxies in our sample, we verify that they appear predominantly red in color and are not dust contaminated blue spirals or misclassified early-types. A random selection of red spiral galaxies is shown in Figure \ref{egredgal_fig}. Most appear disk-like and possess some kind of internal structure. Some appear to have star forming regions, however their overall color is still red. The galaxy in the middle panel highlights the caution which must be exercised in relation to simple eyeball classification of galaxies. This galaxy is classified as late-type (T = 0) in PGC, but as early-type (T = -1) in RC3 and 2MRS. The object has a NASA Sloan Atlas Sersic index of 4.2, which is characteristic of an early-type galaxy. This is an example of a case where a galaxy on the border between early and late-type has been placed in one category by the morphology source selection hierarchy of this paper, but may intrinsically be some kind of intermediate case.

\begin{figure*}[]
\begin{center}$
\begin{array}{ccc}
\resizebox{2.2in}{!}{\includegraphics{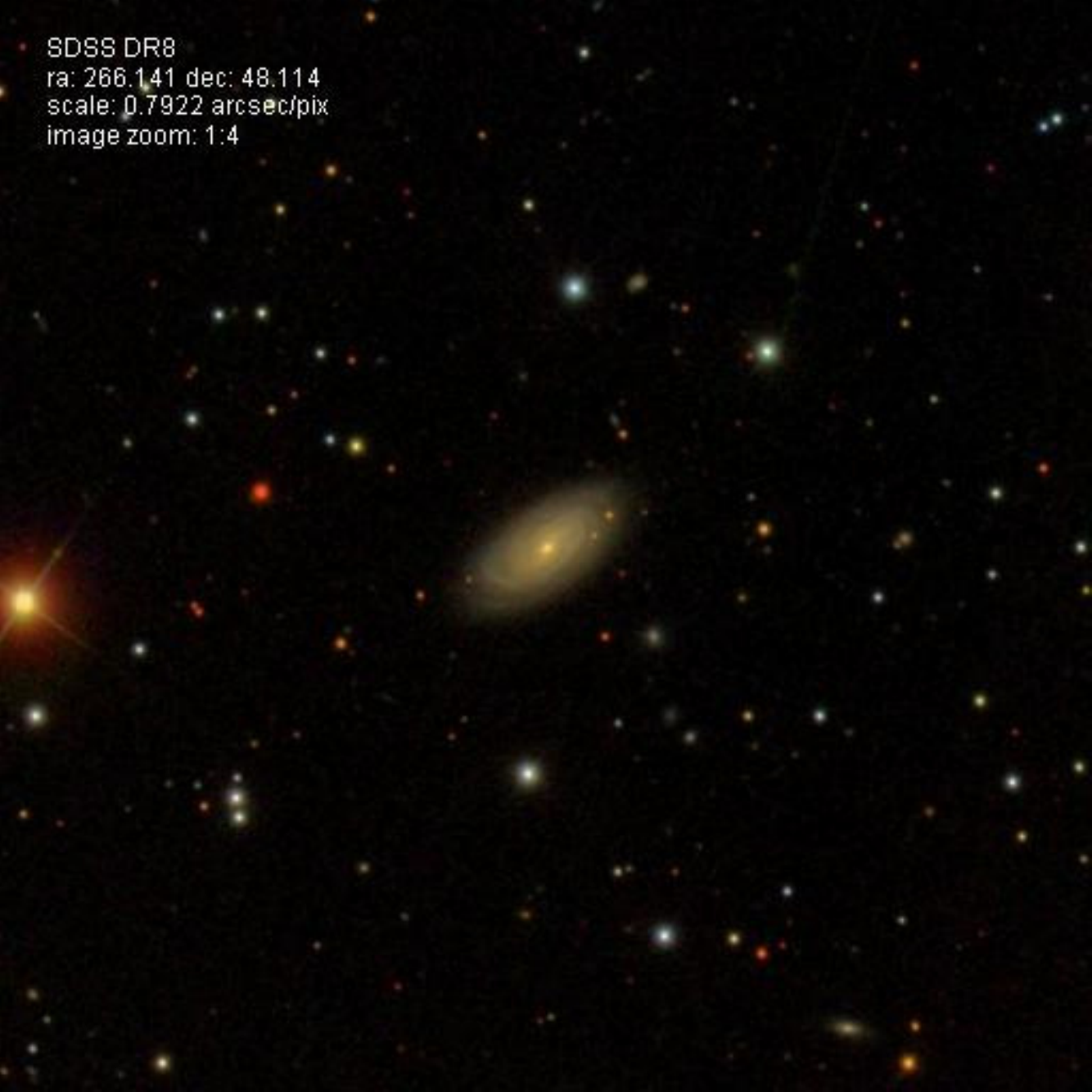}} &
\resizebox{2.2in}{!}{\includegraphics{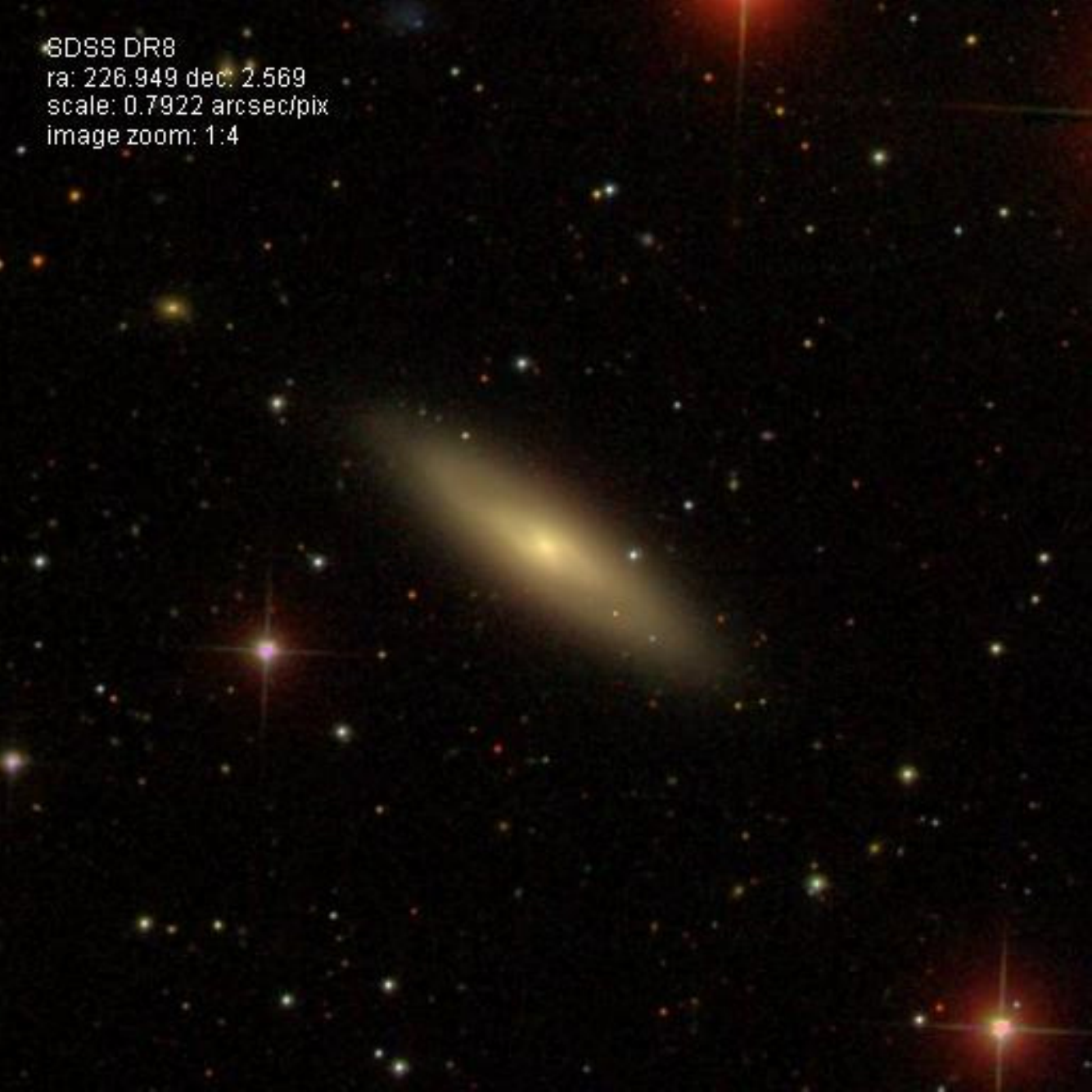}} &
\resizebox{2.2in}{!}{\includegraphics{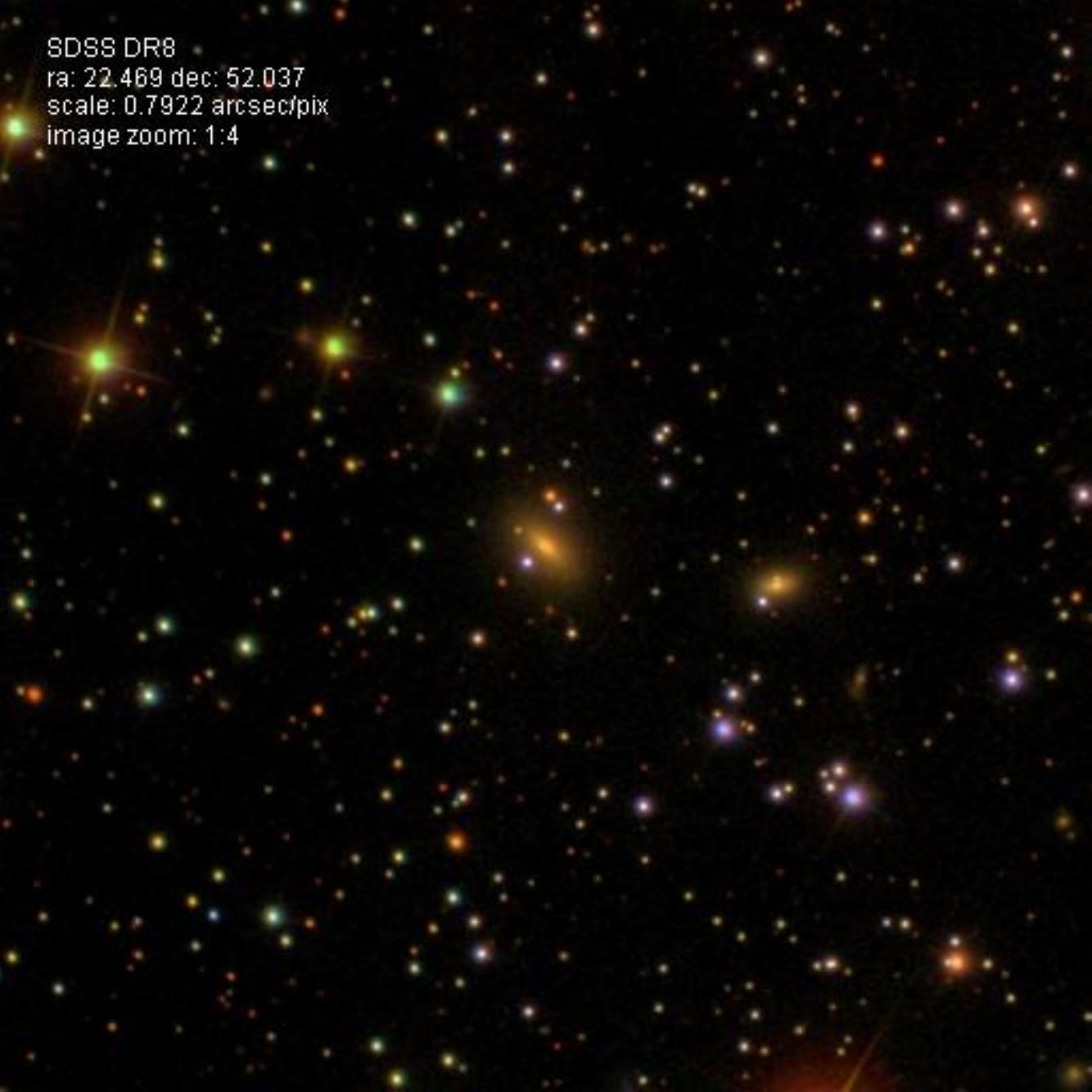}} \\
\resizebox{2.2in}{!}{\includegraphics{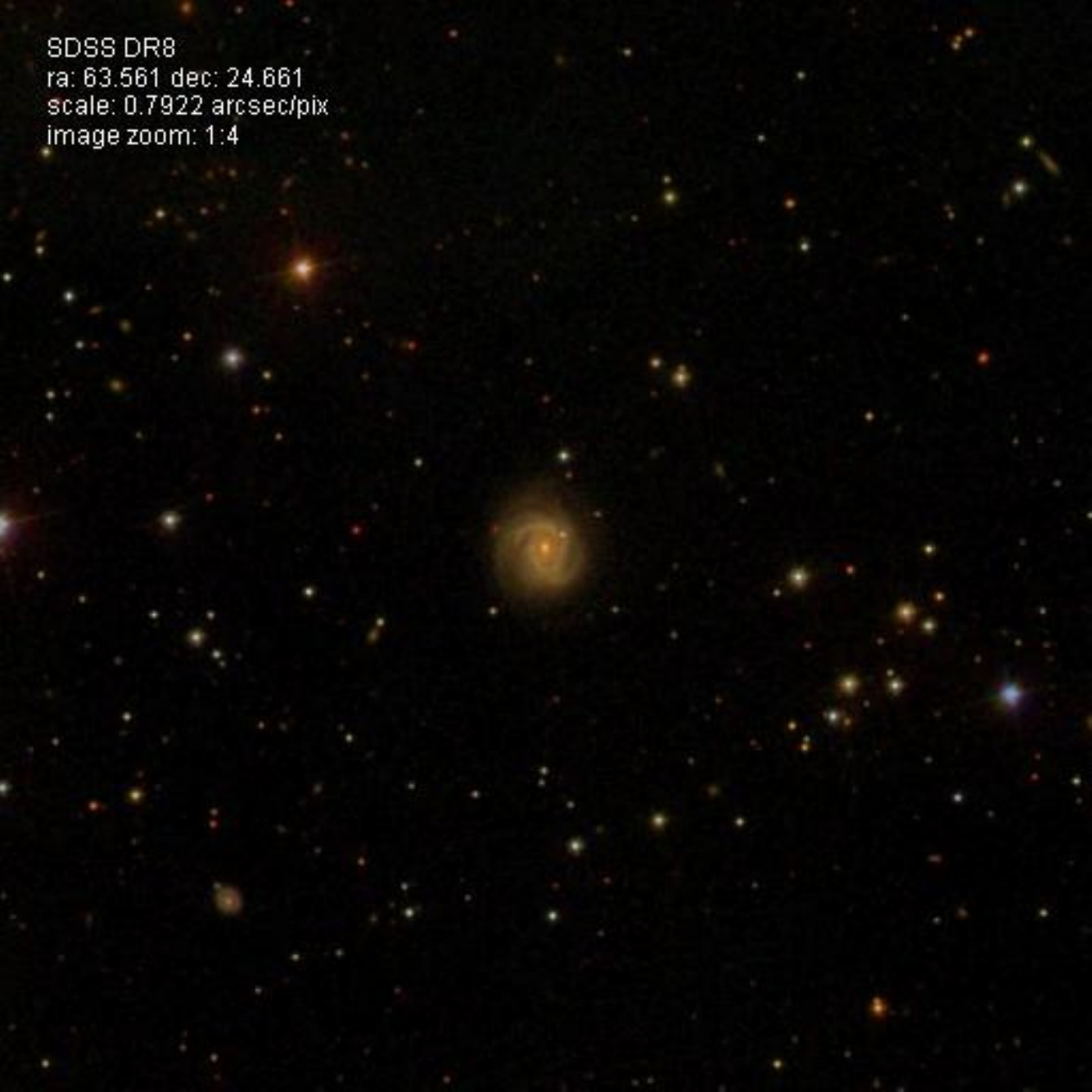}} &
\resizebox{2.2in}{!}{\includegraphics{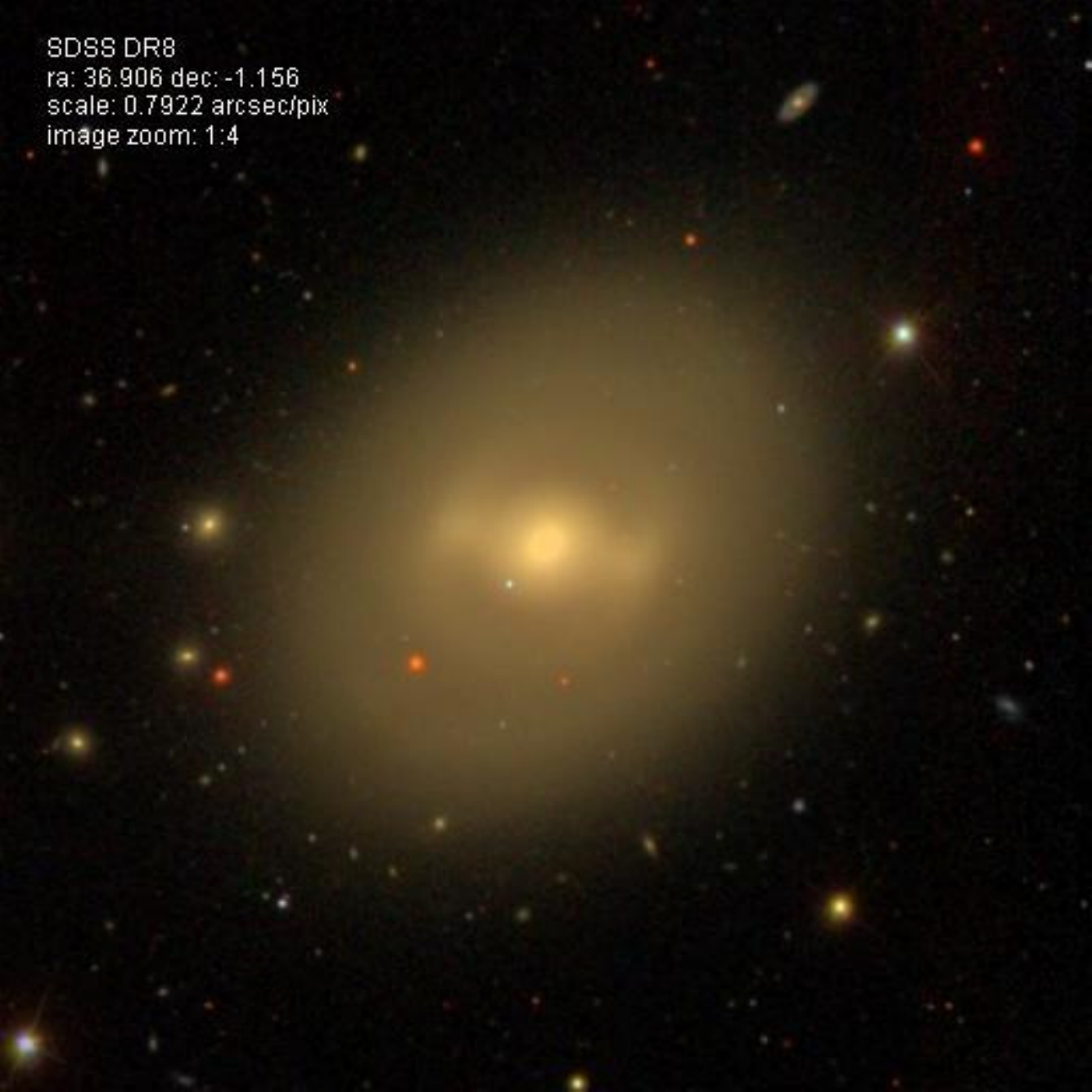}} &
\resizebox{2.2in}{!}{\includegraphics{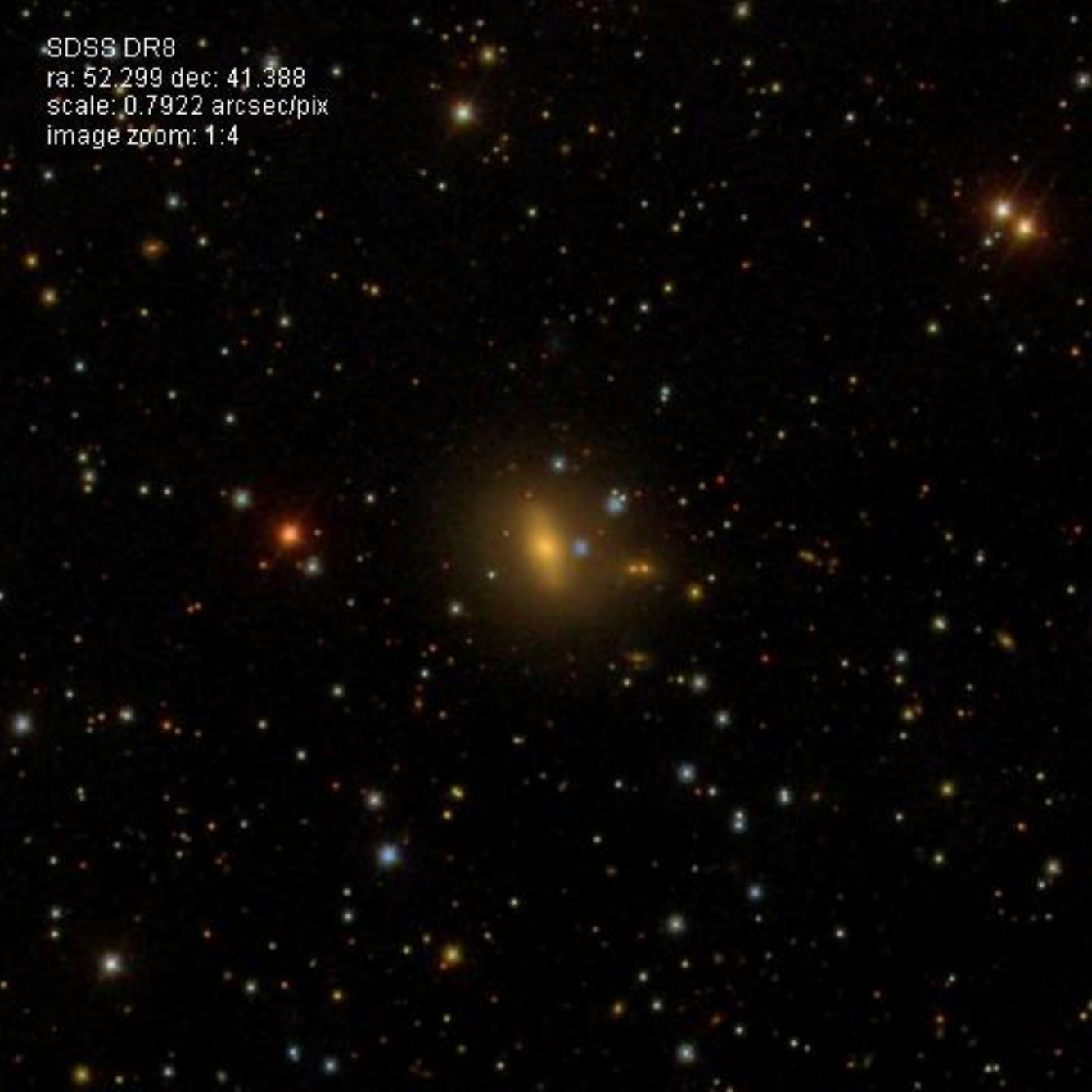}} \\
\resizebox{2.2in}{!}{\includegraphics{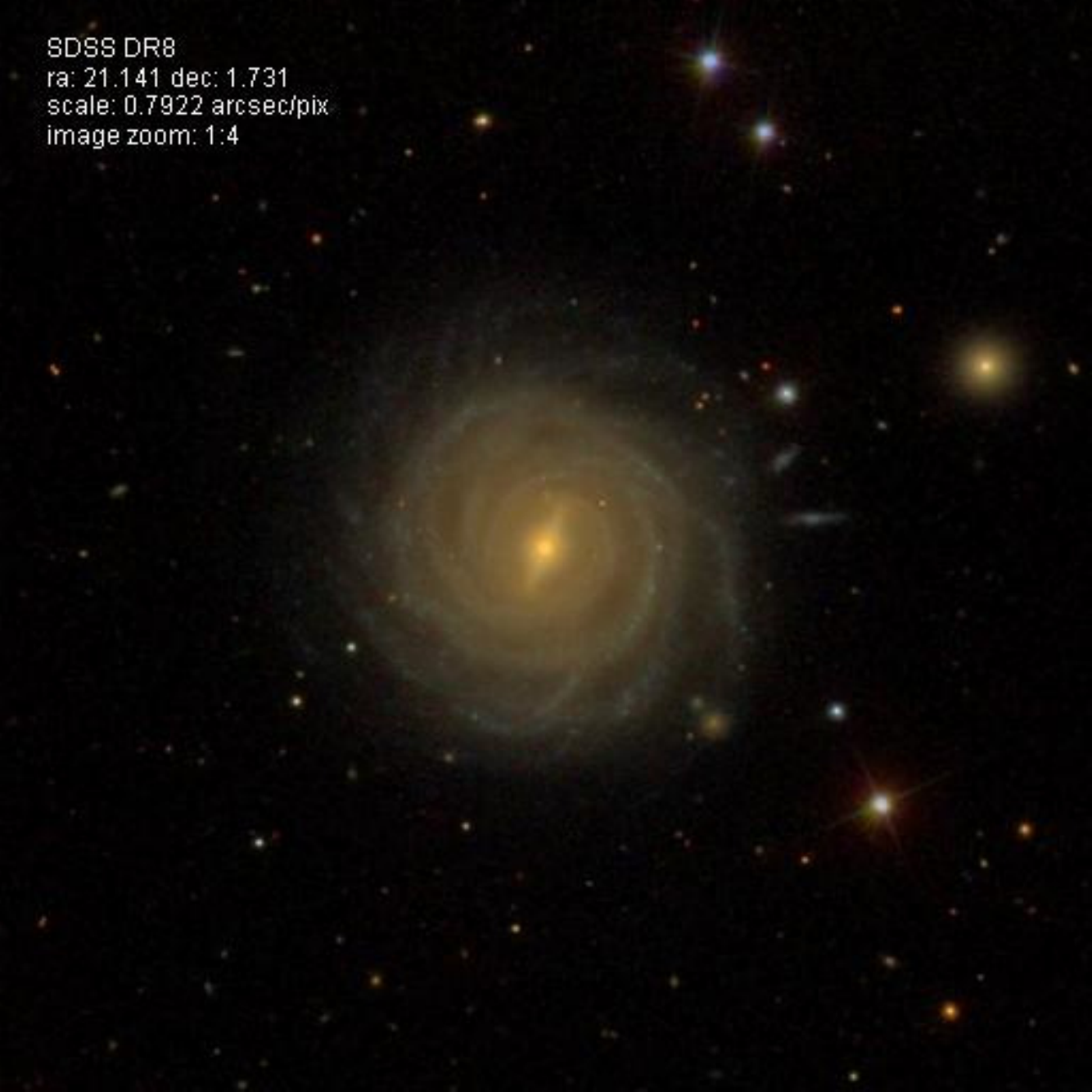}}&
\resizebox{2.2in}{!}{\includegraphics{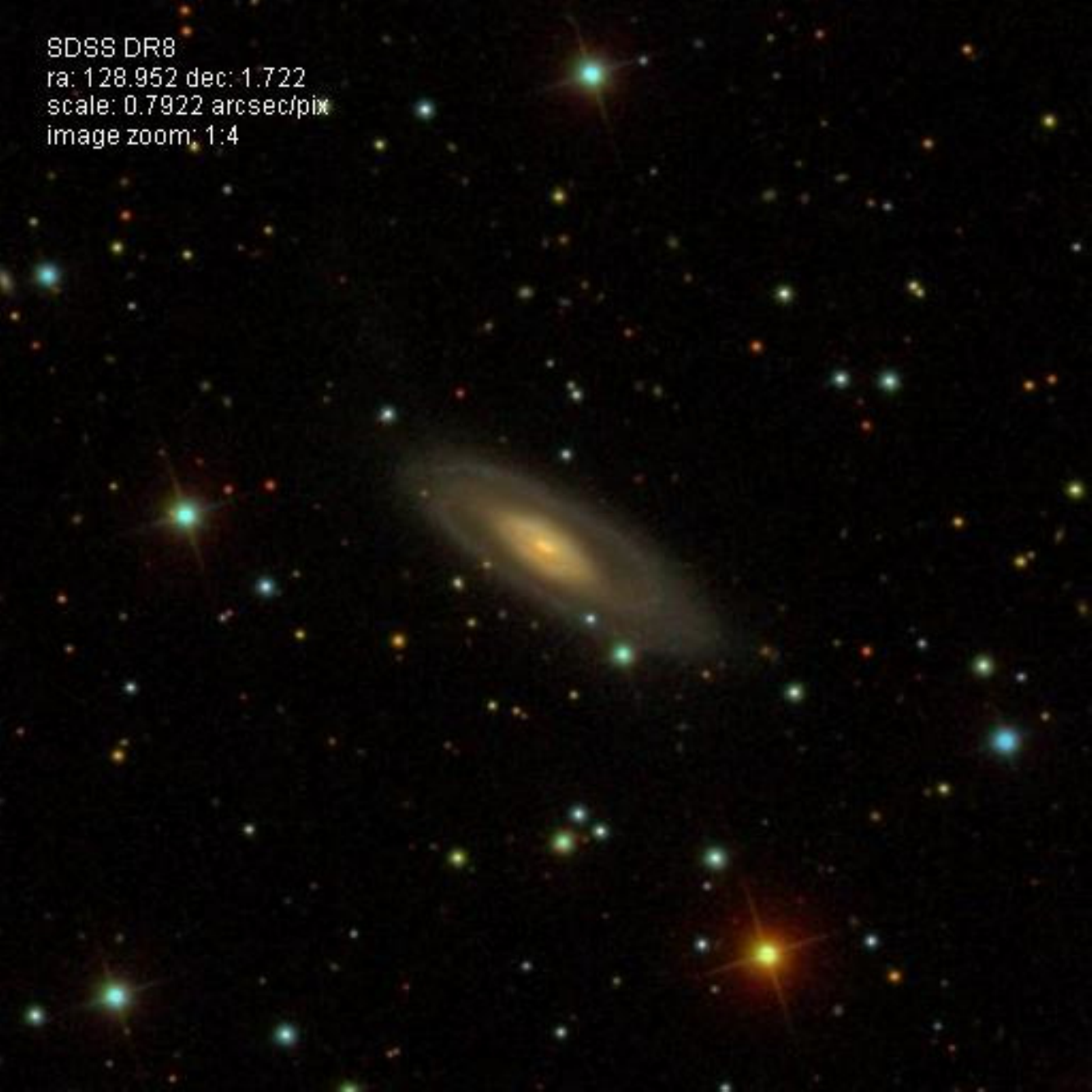}} &
\resizebox{2.2in}{!}{\includegraphics{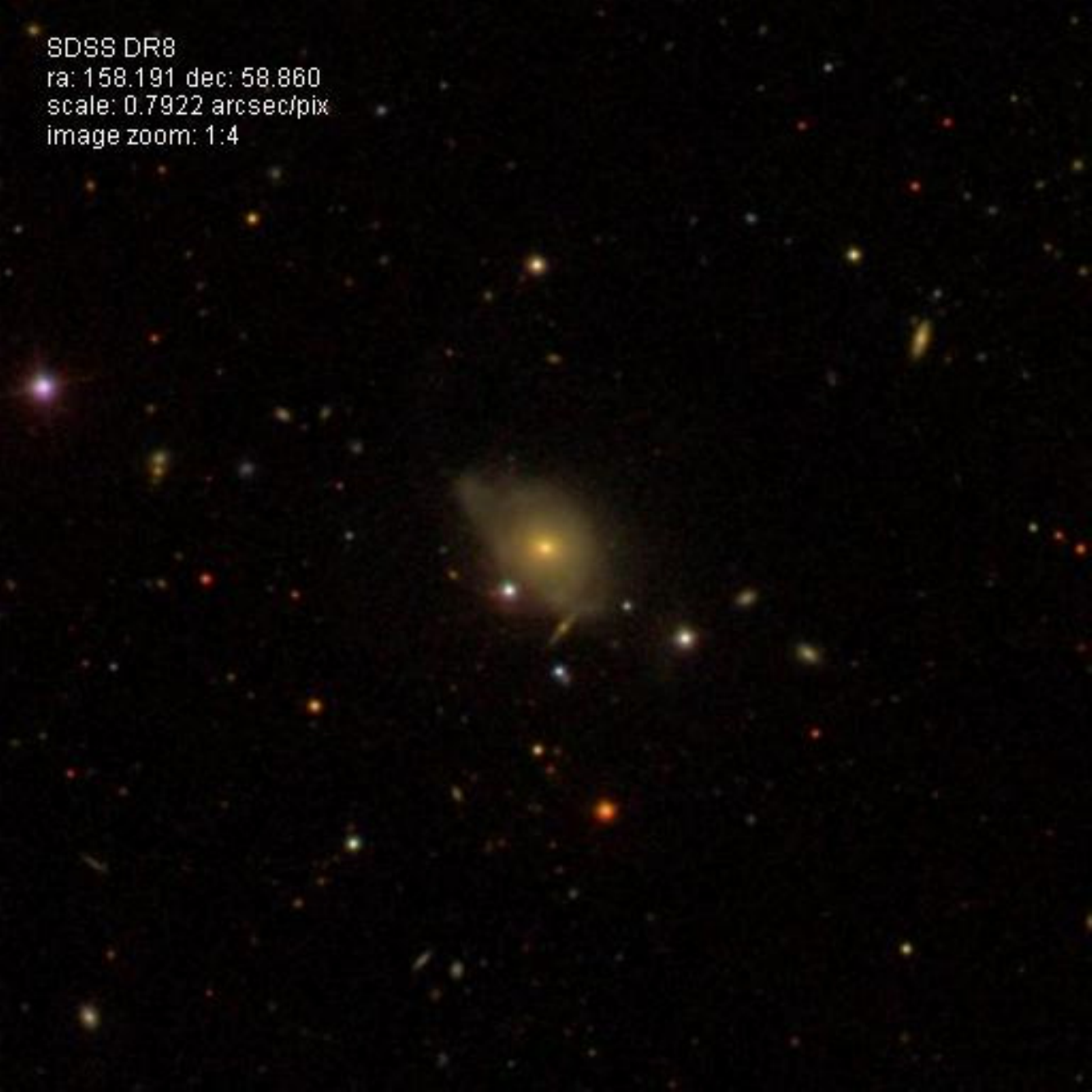}} \\
\end{array}$
\end{center}
\caption{A selection of red spiral galaxies from the NASA Sloan Atlas. A mixture of sizes, and inclinations are represented, however, most appear to be disk-like, with spiral structure and a predominantly red color. Of note is the galaxy in the middle panel. This galaxy is classified as late-type (T = 0) in PGC, but as early-type (T = -1) in RC3 and 2MRS. This is an example of a case where a galaxy on the border between early and late-type has been misclassified by the morphology source hierarchy of this paper. Also of note is the galaxy in the bottom right panel, which is classified as a late-type but appears to be the remnant of a minor merger/accretion. It could be argued that, in a traditional classification scheme, this is a misclassified early-type.}
\label{egredgal_fig}
\end{figure*}

If we change the selection hierarchy so that RC3 morphologies are selected in preference to PGC morphologies, we find that $\simeq$3\% of galaxies are shifted to the early-type population. The overall shape of the early-type and late-type LFs is unaffected and the trend showing an increase in fraction of red spirals with increasing stellar mass remains the same. We also test a scenario where we take the average of all available T-types for each galaxy. In this case, $\simeq$1.5\% of galaxies move from the late-type population to the early-type population, and we again note no discernible difference in the resulting luminosity functions or the properties of the red spiral population. 

\subsection{Red/Blue K-band Spiral Galaxy Luminosity Function and Possible Surface Brightness Incompleteness}

In Figure \ref{lfredbluespiral_fig} we plot luminosity functions, with maximum likelihood fitted Schechter functions for the red and blue spiral populations from the NASA Sloan Atlas, as defined in \S3.5. The resulting blue spiral luminosity function has a faint end slope defined by an $\alpha$ value of -1.21. The red spiral luminosity function has a considerably different shape, with an obvious turnover and power-law index defined by an $\alpha$ value of -0.69.

\begin{figure}[]
\begin{center}
\includegraphics[width = .49\textwidth]
{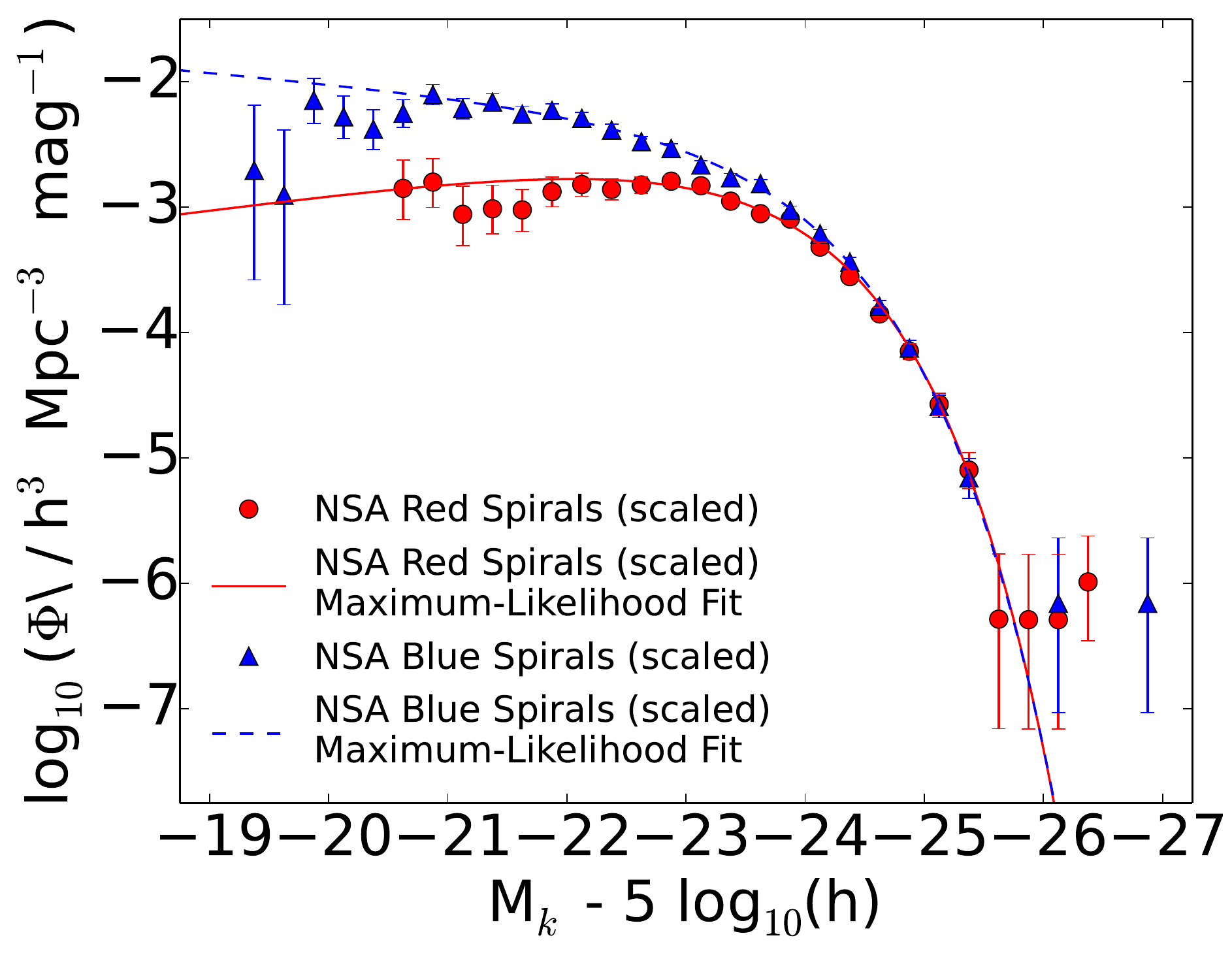}
\caption{Two LFs for blue spirals and for red spirals, scaled to correct for changes in density, using the function in Figure \ref{densitycorrect_fig}. The shapes of these two color-selected functions differ considerably at the faint end. The red and blue spiral populations contribute equally to the bright end of the function, with the red spiral function falling off considerably at the faint end whilst the blue function remains relatively positive. The slope of the faint end red spiral function is more similar in form to previously observed LFs of red galaxy populations at higher redshifts. Parameters for these fits may be seen in Table \ref{lfcompare_tab}.}
\label{lfredbluespiral_fig}
\end{center}
\end{figure}
In Figure \ref{lfredlate_fig} we provide color selected luminosity functions, with the blue sample comprising blue late-type galaxies and the red sample comprising of red early-type and red late-type galaxies (we assume blue early-type galaxies are negligible). When comparing this to our original late and early-type functions, we see red/early-type function dominate the overall LF even more at the bright end. The faint end slopes of both functions are largely unchanged, however the faint end of the red plus early-type function can be seen to be slightly flatter ($\alpha$ = -0.87) than our original early-type luminosity function ($\alpha$ = -1.00).

\begin{figure}[]
\begin{center}
\includegraphics[width = .49\textwidth]
{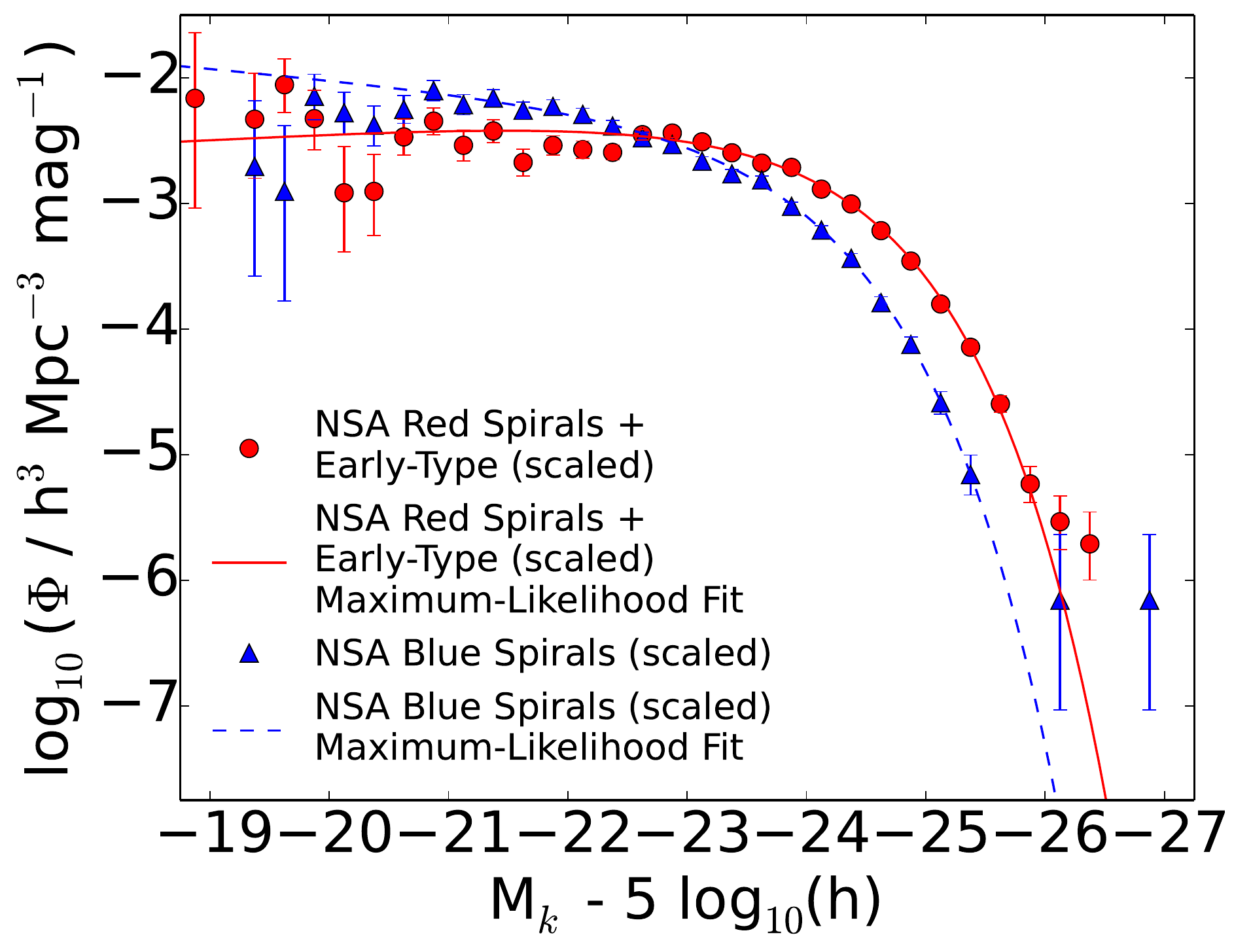}
\caption{Luminosity functions for blue spiral galaxies (late-types) and early-type galaxies with all red spirals added to that population. The functions are scaled for over-densities using the function in Figure \ref{densitycorrect_fig}. These corrections lower the faint end of both LFs very slightly. Adding the red spiral population to the early-type population raises the middle of the function and lowers the faint end, producing a shape closer to, but less extreme than that expected from a color-selected sample. Fit parameters are displayed in Table \ref{lfcompare_tab}.}
\label{lfredlate_fig}
\end{center}
\end{figure}
To further investigate the possible effects of low surface brightness incompleteness on our sample, particularly as a function of color, in Figure \ref{sbabsmag_fig} we plot 2MASS XSC mean surface brightness against M$_K$ for our total sample, red sample and blue sample. In Figure \ref{meansb_fig}, we indicated that our data was beginning to push up against a surface brightness limit of $\simeq$19 mag arcsec$^{-2}$ at our survey limit of K$_{tot}$ $\leq$ 10.75. This is slightly more obvious when we plot as a function of M$_K$, where galaxies with M$_K$ $>$ -21 in our blue sample again beginning pushing up against this surface brightness limit.

\begin{figure*}[]
\begin{center}$
\begin{array}{ccc}
\includegraphics[width=0.66\columnwidth]{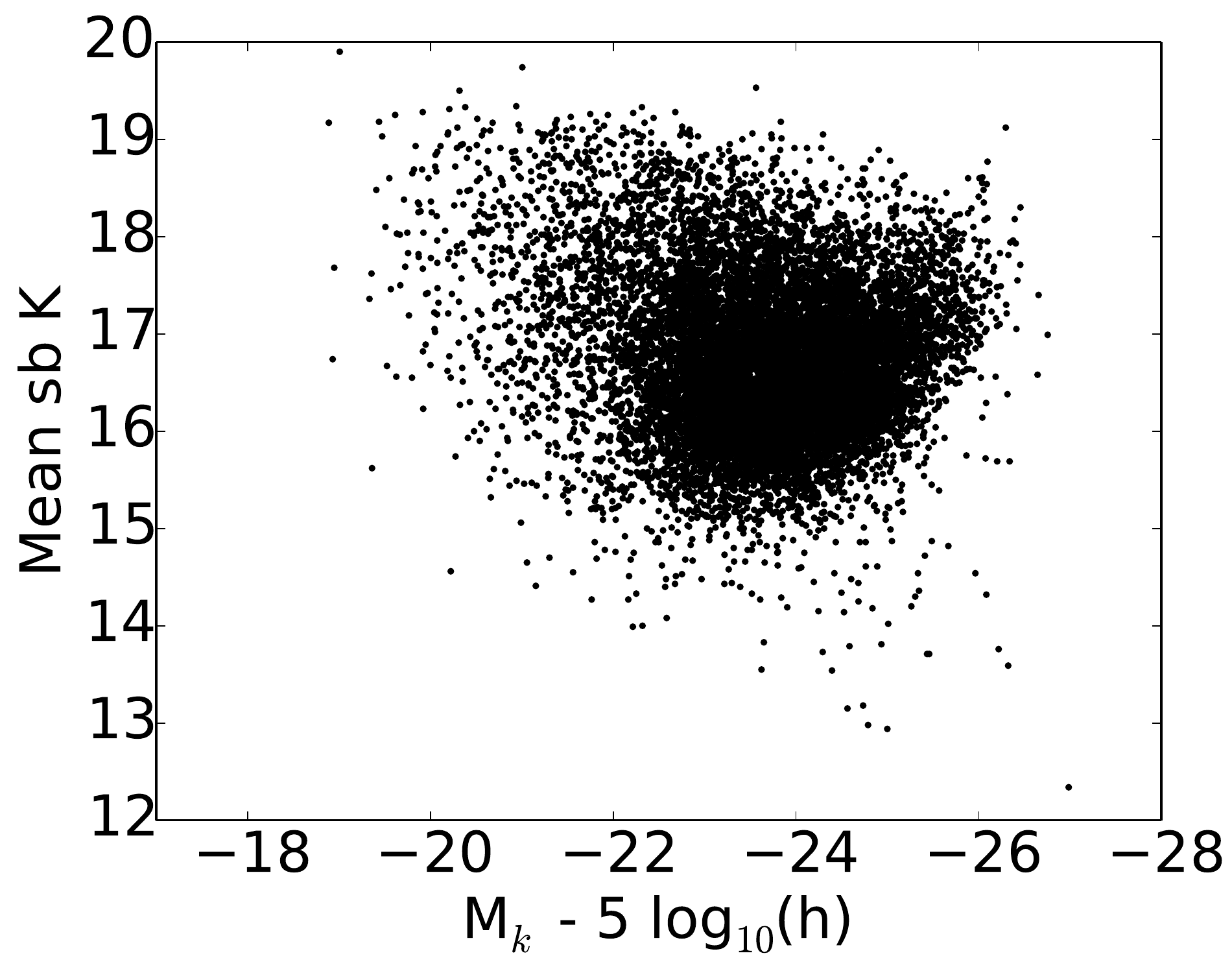} &
\includegraphics[width=0.66\columnwidth]{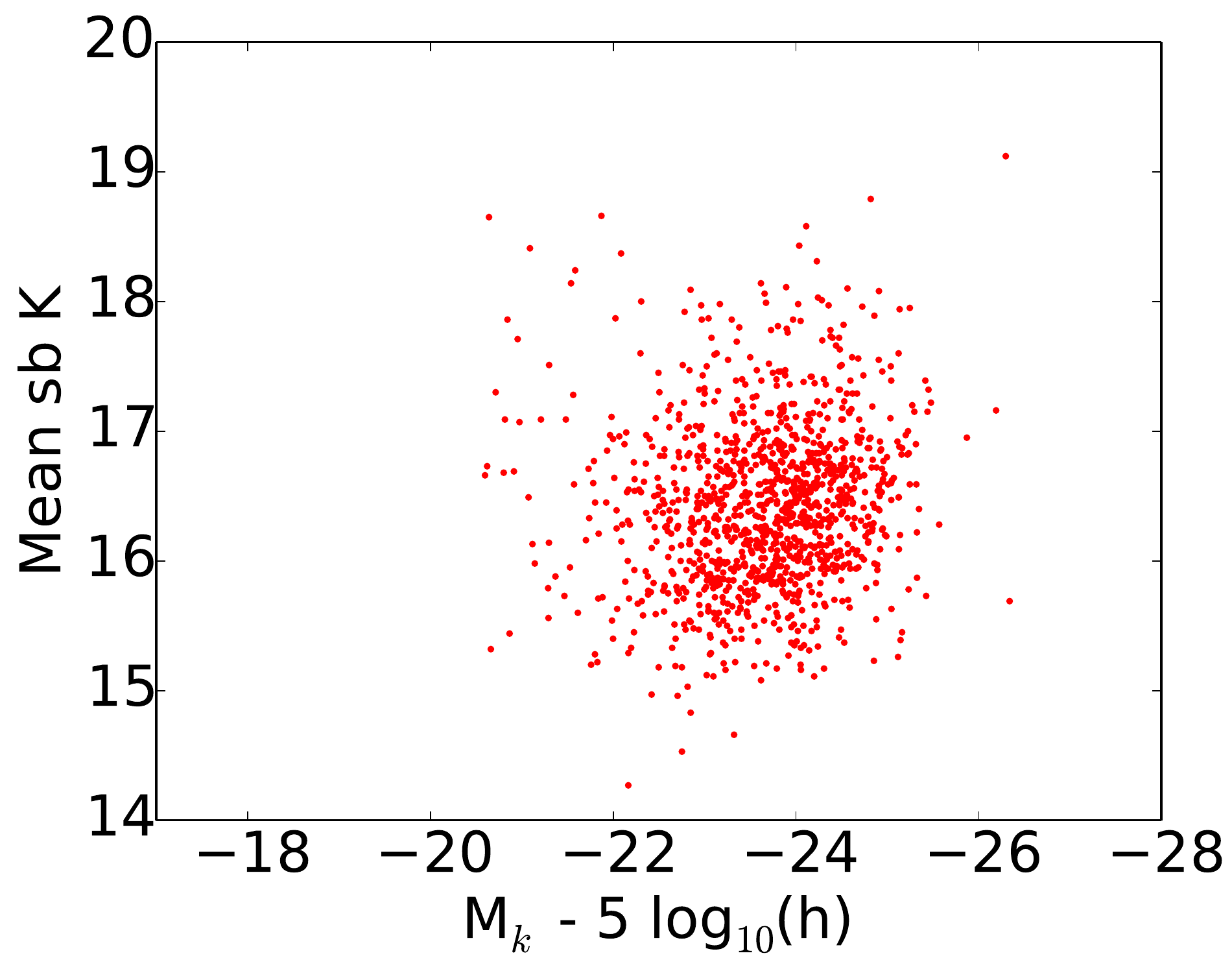} &
\includegraphics[width=0.66\columnwidth]{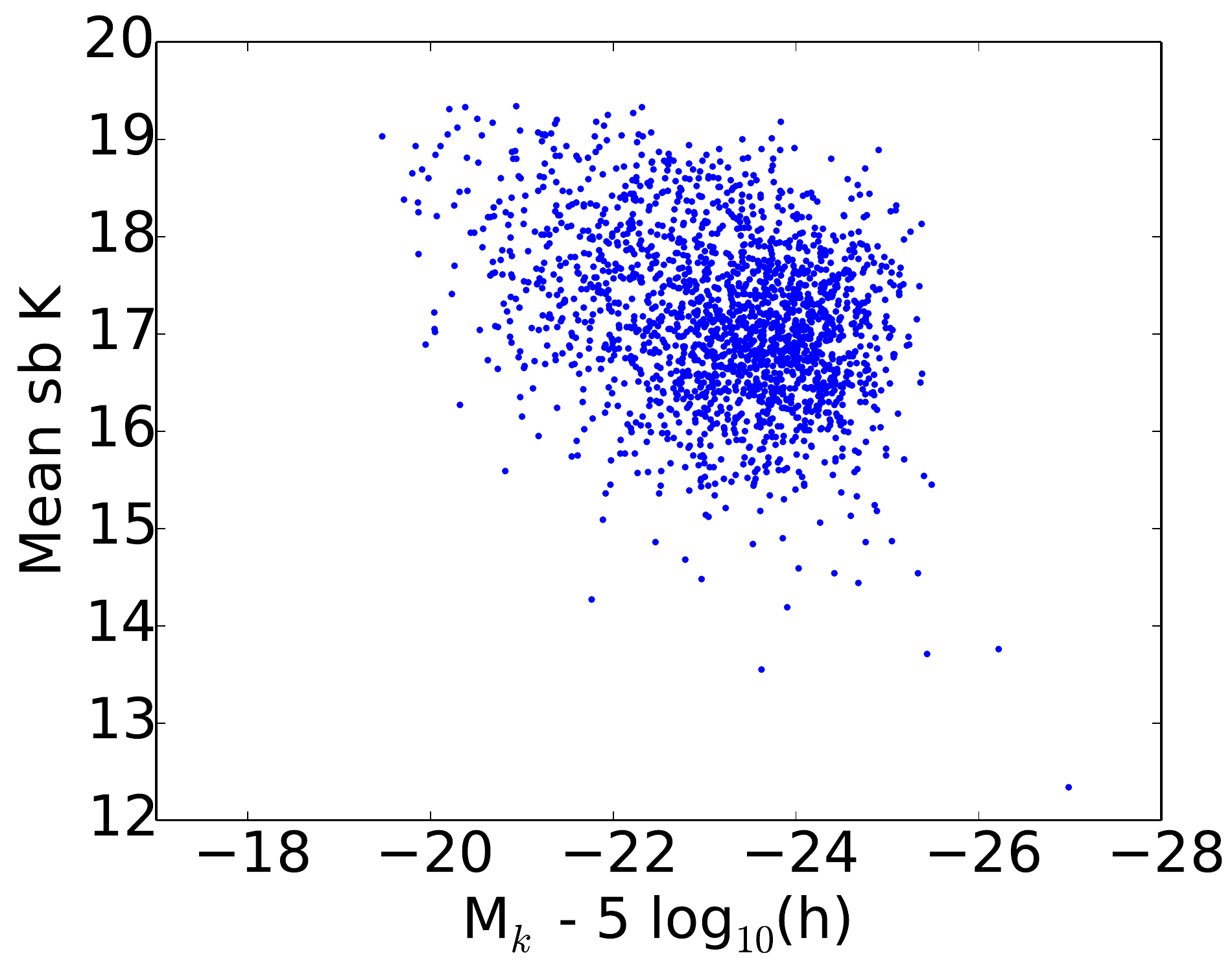}
\end{array}$
\end{center}
\caption{Plots of 2MASS XSC mean surface brightness as a function of K-band absolute magnitude for our total sample (\textit{left}), NSA red galaxies (\textit{middle}) and NSA blue galaxies (\textit{right}). For our total sample, and for the blue sample, galaxies with M$_K$ $>$ 21 appear to be pushing up against the 2MASS XSC surface brightness limit of 19 mag arcsec$^{-2}$ discussed in \S 2.4.}
\label{sbabsmag_fig}
\end{figure*}
Figures \ref{lfredbluespiral_fig} and \ref{lfredlate_fig} illustrate that the Schechter form fits the faint end of the blue late-type function poorly, and this may be due to our surface brightness limits. To investigate this we fitted a Schechter function to the blue spiral subsample galaxies with M$_K$ $<$ 21. For the density scaled fit, limiting the function to brighter absolute magnitude steepens the faint end slope from $\alpha$ = -1.21 to $\alpha$ = -1.28. 

\subsection{Comparison with Literature Luminosity Functions}

Maximum likelihood fit parameters for this paper as well as a number of previous LF papers are provided in Table \ref{lfcompare_tab}. For our early-type and late-type luminosity functions, we compare only to \citet{kocha01} as their methodology and selection criteria are the most similar to ours. Though \citet{bell03} also select for morphology, they use concentration indexes rather than eyeball morphologies, which are not directly comparable. 

Though the different functions agree to an extent, there are obvious differences, notably at the bright end of the LF. This is demonstrated in Figure \ref{lftotlitdiv_fig}, where we plot the Schechter functions from past literature, divided by the Schechter functions from this paper. It can be seen from Figure \ref{lftotlitdiv_fig} that Schechter function fits that look good on log-log plots can have discrepancies of tens of percent relative to 1/V$_{\rm{MAX}}$ luminosity functions \citep[and this is also seen in Figure 10 of ][]{jones06}.

\begin{figure}[]
\begin{center}
\includegraphics[width = .49\textwidth]{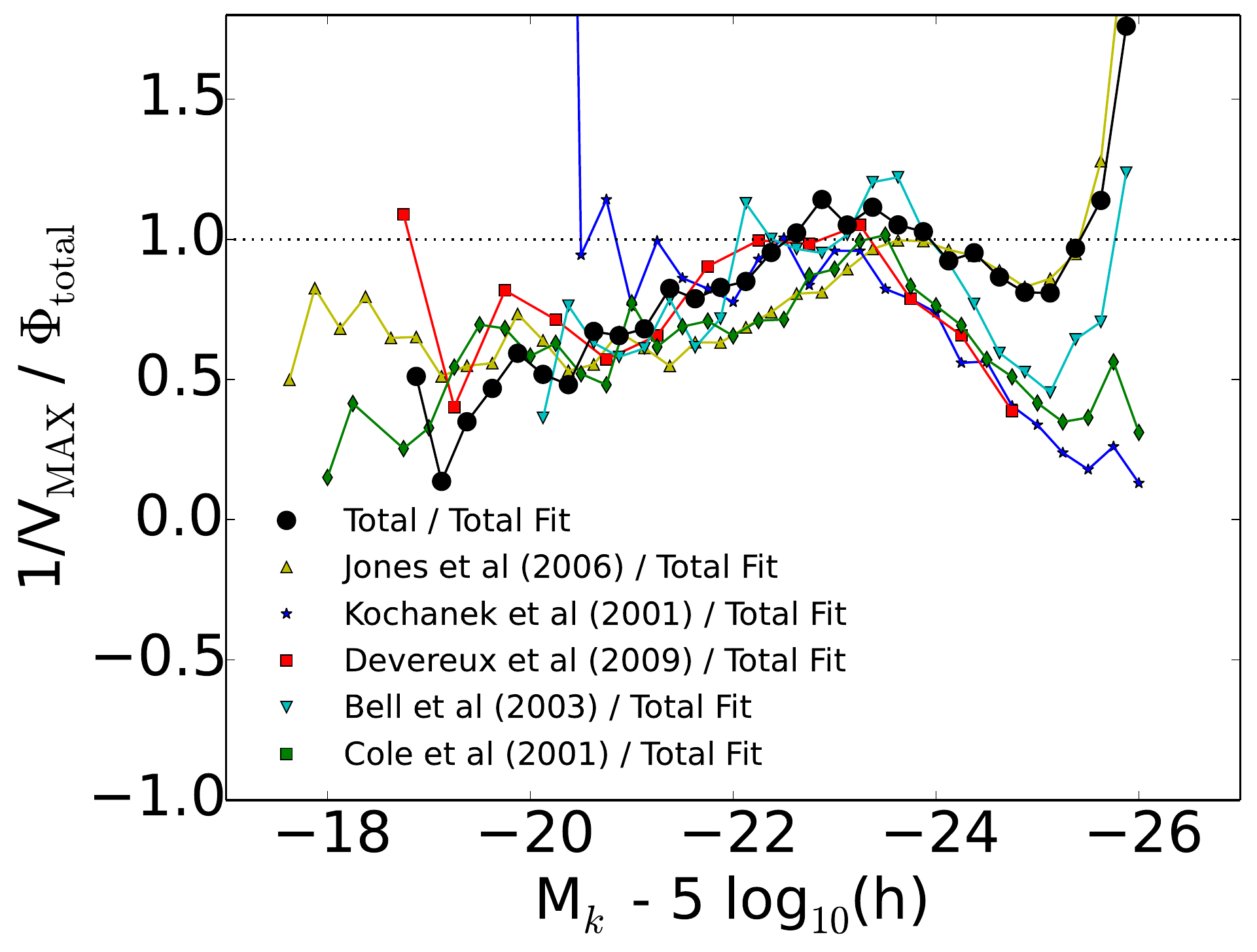}
\caption{1/V$_{\rm{MAX}}$ method data points from \citet{kocha01}, \citet{cole01}, \citet{bell03}, \citet{jones06} and \citet{dever09} divided by this paper's best density scaled Schechter function fit. This highlights the difficulty of fitting the bright and faint ends of the 1/V$_{\rm{MAX}}$ function with the Schechter form. For this paper, \citet{jones06} and \citet{bell03}, to a lesser extend, the Schechter form underestimates the bright end, whereas for \citet{dever09}, \citet{cole01} and \citet{kocha01} it overestimates. The faint end data points are comparably overestimated by the Schechter form in all cases. There is good agreement around M$^*$ for all functions.}
\label{lftotlitdiv_fig}
\end{center}
\end{figure}
The best agreement is with the work of \citet{jones06}, however this is to be expected as there is much overlap in the data used to produce both luminosity functions. One difference of note is the smoothness of the faint end of this function when compared to the others. This is due to the use of redshift-independent distances as well as flow corrected redshift derived distances in this project, rather than just the latter. The density correction made to the functions smooths the faint end even more. In terms of differences to studies other than \citet{jones06}, there are a number of factors that could account for these. 

We expect differences between luminosity functions calculated using different types of photometry. For example, our luminosity functions will be systematically offset from luminosity functions that were calculated using 2MASS isophotal photometry \citep[eg.][]{dever09}. The difference between isophotal and (brighter) total magnitudes is a function of surface brightness, with the two sets of magnitudes being offset by 0.2 magnitudes or less for galaxies with a mean surface brightness of $\mu_{\rm{K}}$ $<$ 18.5 \citep{jones04}. We also see offsets between our function and the other functions we compare to due to their choice of different photometry. While we and \citet{jones06} use total magnitudes, \citet{bell03} and \citet{cole01} used Kron magnitudes, \citet{dever09} used isophotal magnitudes, and \citet{kocha01} used isophotal magnitudes with a correction to approximate total magnitudes.

Differences can also be attributed to the use of small galaxy sample sizes resulting in poor estimates of spatial density due to higher density in large scale structure. Compared to our sample size of 13,325 galaxies, \citet{kocha01} had a much smaller sample of 3,878 galaxies, and \citet{dever09} had an even smaller sample of 1,345 galaxies. \citet{bell03} and \citet{cole01} had sample sizes of 6,282 and 5,683 respectively, though both pushed to far fainter apparent magnitudes. \citet{jones06} did not measure the luminosity functions of morphology-selected galaxies, and had a K-band magnitude limit of 12.75, so their sample of 60,869 galaxies is far larger than the others discussed here. Difficulties in obtaining accurate distances to objects due to galaxy peculiar motions would also result in incorrectly calculated absolute magnitudes. Though \citet{jones06}, \citet{kocha01} and \citet{dever09} have all made flow corrections for their galaxy samples, only \citet{dever09} also combined this with redshift-independent distance estimates. As we showed in Figure \ref{lfearlylate_fig}, the redshift limits imposed by each group affect the range of luminosities covered by each corresponding luminosity function. As \citet{kocha01} has a sample limited to $cz~>~2,000~{\rm km~s^{-1}}$, their sample does not extend to the faint absolute magnitudes that our sample does. \citet{dever09} has a limit of $cz~\leq~3,000~{\rm km~s^{-1}}$ and thus has less bright galaxies.

As illustrated by Figures \ref{lftotal_fig} and \ref{lftotlitdiv_fig}, the bright end of the \citet{kocha01} luminosity functions are considerably lower than our luminosity functions. The Schechter function fits to both the early-type and late-type functions of \citet{kocha01} have less negative $\alpha$ values, though the data points alone do not make this entirely obvious. Furthermore, the power law index of their early-type function is more negative than their late-type function, which is the opposite of what we see. If morphological type was truly a good proxy for color, we would expect an obvious downward slope at the faint end of the early-type curves and a steeper faint end slope for the late-type function.
\begin{table*}[]\centering
\caption{Maximum Likelihood fit parameters}
\scriptsize \begin{tabular}{cccccc}
\hline
sample & objects & mag limit & M$^*$ $-$ $5$ $\rm{log_{10}(h)}$ & $\alpha$ & log$_{10}$($\Phi^*$) \\ 
 & & & (mag) & & ({h $^3$ \rm Mpc}$^{-3}\,$ mag$^{-1}$) \\
 & & & & &  \\ 
\hline
 & & & & &  \\ 
\bf{Total} & & & & & \\
 & & & & & \\
This Paper & 13,325 & K$_{\rm{tot}}$ $\leq$ 10.75 & -23.87 $\pm$ 0.06 & -1.23 $\pm$ 0.08 & -2.13 $\pm$ 0.05 \\
This Paper (scaled) & 13,325 & K$_{\rm{tot}}$ $\leq$ 10.75 & -23.83 $\pm$ 0.06 & -1.17 $\pm$ 0.08 & -2.05 $\pm$ 0.05 \\
Devereux et al (2009) & 1,345 & K$_{\rm{iso}}$ $\leq$ 10.0  & -23.41 $\pm$ 0.46 & -0.94 $\pm$ 0.10 & -1.94 $\pm$ 0.10 \\
Jones et al (2006) & 60,869 & K$_{\rm{tot}}$ < 12.75  & -23.83 $\pm$ 0.03 & -1.16 $\pm$ 0.04 & -2.13 $\pm$ 0.01 \\
Bell et al (2003) & 6,282 & K$_{\rm{Kron}}$ $\leq$ 13.57 & -23.29 $\pm$ 0.05 & -0.77 $\pm$ 0.04 &  -1.84 $\pm$ 0.02 \\
Kochanek et al (2001) & 3,878 & K$_{20}$ < 11.25 & -23.39 $\pm$ 0.05 & -1.09 $\pm$ 0.06 & -2.06 $\pm$ 0.04 \\
Cole et al (2001) & 5,683 & K$_{\rm{Kron}}$ $<$ 13.2 & -23.44 $\pm$ 0.03 & -0.96 $\pm$ 0.05 & -1.97 $\pm$ 0.06 \\
 & & & & & \\
\bf{Early-Type} & & & & & \\
 & & & & & \\
This Paper & 5,640 & K$_{\rm{tot}}$ $\leq$ 10.75  & -24.03 $\pm$  0.06 &  -1.02 $\pm$ 0.10 & -2.58 $\pm$ 0.05  \\
This Paper (scaled) & 5,640 & K$_{\rm{tot}}$ $\leq$ 10.75  & -24.01 $\pm$  0.05 &  -1.00 $\pm$ 0.09 & -2.53 $\pm$ 0.04  \\
Kochanek et al (2001) & 1,781 & K$_{20}$ < 11.25  & -23.53 $\pm$ 0.06 & -0.92 $\pm$ 0.10 & -2.34 $\pm$ 0.05 \\
 & & & & & \\
\bf{Late-Type} & & & & & \\
 & & & & & \\
This Paper & 7,685 & K$_{\rm{tot}}$ $\leq$ 10.75  & -23.49 $\pm$ 0.06 & -1.13 $\pm$ 0.10 & -2.13 $\pm$ 0.04 \\
This Paper (scaled) & 7,685 & K$_{\rm{tot}}$ $\leq$ 10.75  & -23.43 $\pm$ 0.07 & -1.06 $\pm$ 0.10 & -2.05 $\pm$ 0.04 \\
Kochanek et al (2001) & 2,097 & K$_{20}$ < 11.25 & -22.98 $\pm$ 0.06 & -0.87 $\pm$ 0.09 & -2.0 $\pm$ 0.06 \\
 & & & & & \\
\bf{NASA Sloan Atlas Blue Late-Type} & & & & & \\
 & & & & & \\
 This Paper & 1,981 & K$_{\rm{tot}}$ < 10.75 & -23.47 $\pm$ 0.16 & -1.31 $\pm$ 0.23 & -2.35 $\pm$ 0.11 \\
 This Paper (scaled) & 1,981 & K$_{\rm{tot}}$ < 10.75 & -23.39 $\pm$ 0.14 & -1.21 $\pm$ 0.22 & -2.26 $\pm$ 0.10 \\
 This Paper (high surface brightness) & 1,899 & K$_{\rm{tot}}$ < 10.75 & -23.46 $\pm$ 0.15 & -1.28 $\pm$ 0.21 & -2.34 $\pm$ 0.10 \\
 This Paper (scaled, high surface brightness) & 1,899 & K$_{\rm{tot}}$ < 10.75 & -23.44 $\pm$ 0.16 & -1.28 $\pm$ 0.21 & -2.27 $\pm$ 0.10 \\
  & & & & & \\
\bf{NASA Sloan Atlas Red Late-Type} & & & & & \\
 & & & & & \\
 This Paper & 1,174 & K$_{\rm{tot}}$ < 10.75 & -23.33 $\pm$ 0.09 & -0.70 $\pm$ 0.11 & -2.50 $\pm$ 0.13  \\
 This Paper (scaled) & 1,174 & K$_{\rm{tot}}$ < 10.75 & -23.31 $\pm$ 0.09 & -0.69 $\pm$ 0.11 & -2.45 $\pm$ 0.11 \\
  & & & & & \\
\bf{NASA Sloan Atlas Early-Type + Red Late-Type} & & & & & \\
 & & & & & \\
 This Paper & 3,436 & K$_{\rm{tot}}$ < 10.75 & -23.76 $\pm$ 0.11 & -0.90 $\pm$ 0.19 & -2.27 $\pm$ 0.11 \\
 This Paper (scaled) & 3,436 & K$_{\rm{tot}}$ < 10.75 & -23.73 $\pm$ 0.10 & -0.87 $\pm$ 0.18 & -2.22 $\pm$ 0.10 \\
  & & & & & \\ & & & & & \\
\hline
\label{lfcompare_tab}
\end{tabular}
\end{table*}
If we apply our density correction to our luminosity function, the changes are very slight. The points at the faint end of all functions are slightly lowered by the density correction, and this is most significantly reflected in the change to $\alpha$ in all cases.

\section{Discussion}

The presence of a large population of red spirals in our sample, at face value, presents a simple solution to the observed difference in shape between color and morphology-selected luminosity functions. By moving the red spiral population into our early-type sample, we would expect the shape of the resulting function to change. The most noticeable changes are in the bright/high mass end of the function where most of the red spirals reside. The faint end changes from a value of $\alpha$ = -1.0 $\pm$ 0.09 to a value of $\alpha$ = -0.87 $\pm$ 0.18, which is in good agreement with the value of -0.83 $\pm$ 0.2 for $\alpha$ from \citet{baldr04}. 

While we can reproduce the faint end slope of the local galaxy luminosity function, we still cannot replicate the steep faint end slope of $\simeq$-0.5 seen in red sequence luminosity functions at higher $z$ \citep[e.g.,][]{brown07, bel04} but it is possible that this is due to selection effects which are only apparent at $z$ larger than our sample limits. Studies of evolving red galaxy luminosity functions have been made in the past \citep{bel04, brown07}, however, studies of evolving red/blue subsets of early/late-type galaxy luminosity functions are less common \citep[e.g.,][]{ilb05}.

The faint end slope of our blue spiral function is also similar to the blue function of \citet{baldr04}, with our $\alpha$ value of -1.21 $\pm$ 0.25 within the uncertainties of their blue function value of $\alpha$ = -1.18 $\pm$ 0.02. This faint end slope of the blue function is almost certainly affected by surface brightness incompleteness and should be steeper \citep[e.g.,][]{bell03, bla05}. We test this by imposing a faint end limit of M$_K$ $>$ -21 to the Schechter function fit, which produces a function with faint end slope of -1.28 $\pm$ 0.21. This is in good agreement with the faint end slope of $\simeq$ -1.3 found by \citet{bla05} for their sample of low surface brightness galaxies.

 We find that when comparing blue and red late-type populations, red spirals are preferentially found amongst the most massive spiral galaxies (supported by the findings of  \citet{mas10}, \citet{pimb12} and \citet{bell03_2}. In a simple toy model, where red spirals are faded blue spirals, we would expect red spirals to be fainter than the blue spiral population. Instead we find that red spirals are amongst the most luminous spiral galaxies and that these galaxies often show evidence for star formation (as shown in Figure \ref{egredgal_fig}), which is consistent with several previous studies that find continuing star formation within red spiral galaxies \citep[e.g.,][]{toj13, mas10, maha09, wolf09, cros13}.

We find that the fraction of spirals that are red increases with luminosity, that does not mean that red spirals dominate the bright end of the red galaxy luminosity function. This is not unexpected, as elliptical galaxies are expected to be the most massive members of any galaxy population. To illustrate this, we plot luminosity functions for NASA Sloan Atlas galaxies that we have classified as early-type, against those that we have classified as red spirals in Figure \ref{redearlylf_fig}. We see that the fraction of red galaxies that are spirals increases with decreasing luminosity, until $M_K \simeq -22$, at which point early-type galaxies may increase again. 
 
\begin{figure}[]
\begin{center}
\includegraphics[width = .49\textwidth]
{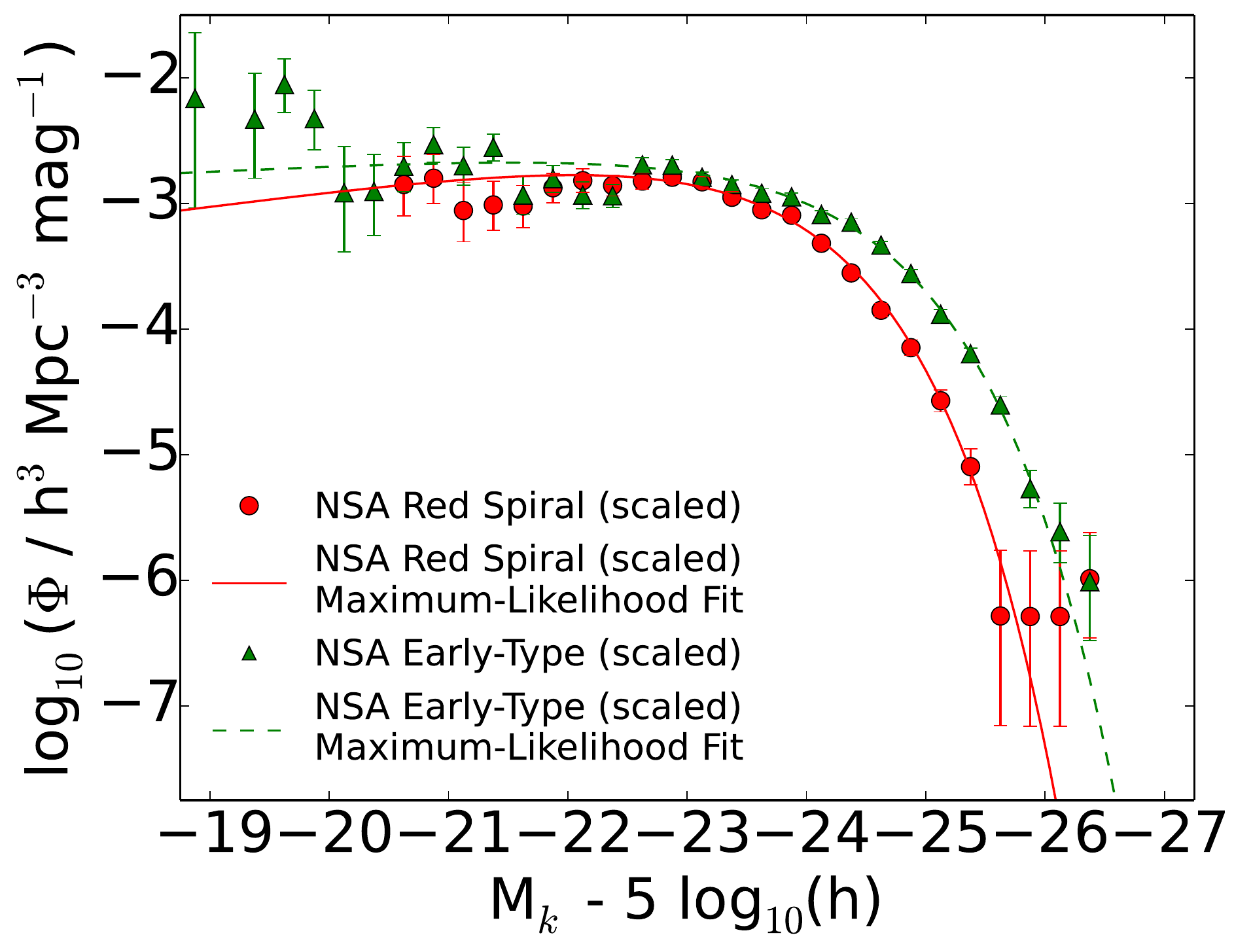}
\caption{Red-spiral LF plotted against an early-type LF for NASA Sloan Atlas galaxies. This shows that at brighter K absolute magnitudes (higher stellar masses) the red sequence should be dominated by massive elliptical or lenticular galaxies. At medium to low stellar masses, the two populations are roughly proportional.}
\label{redearlylf_fig}
\end{center}
\end{figure}
In addition, we note that disk galaxies without star formation exist in our sample, but they have largely been classified as early-type galaxies (using traditional classifications). \citet{bundy10} conclude that $\simeq$50\% of red sequence galaxies are disk-like, generally with large bulges, and dominate at lower masses. However, they differentiate disk and bulge galaxies using only axis ratios. \citet{vdwel09} also come to a similar conclusion, however they use a more sophisticated classification scheme called ZEST which ``combines the power of a principle component (PC) analysis of nonparametric measures of galaxy structure with information from a parametric fit". We have confirmed this by inspecting the less luminous optically red galaxies in our sample. If we consider only red galaxies with $M_K - 5log_{10}(h)$ > -20, we see a mixture of galaxies classified as elliptical and lenticular morphological types. Of the 7 galaxies shown in Figure \ref{lowmagred_fig}, 2 galaxies have NASA Sloan Atlas Sersic indices less than 2, 2 have Sersic indices between 2 and 3, and 3 have Sersic Indices higher than 3. It is plausible that these 4 galaxies with lower Sersic indices could be misclassified disk galaxies.

\begin{figure*}[]
\begin{center}$
\begin{array}{ccc}
\resizebox{2.2in}{!}{\includegraphics{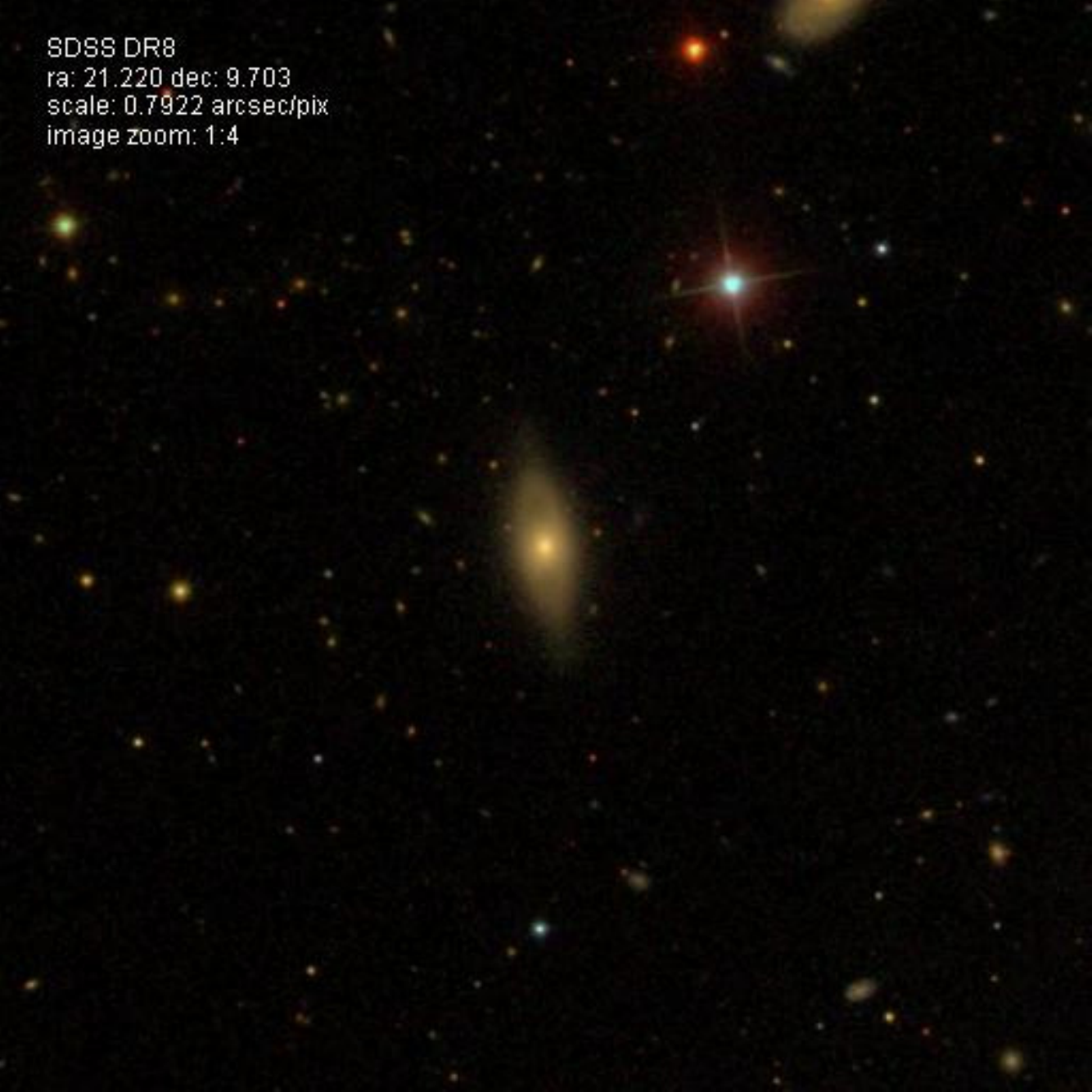}} &
\resizebox{2.2in}{!}{\includegraphics{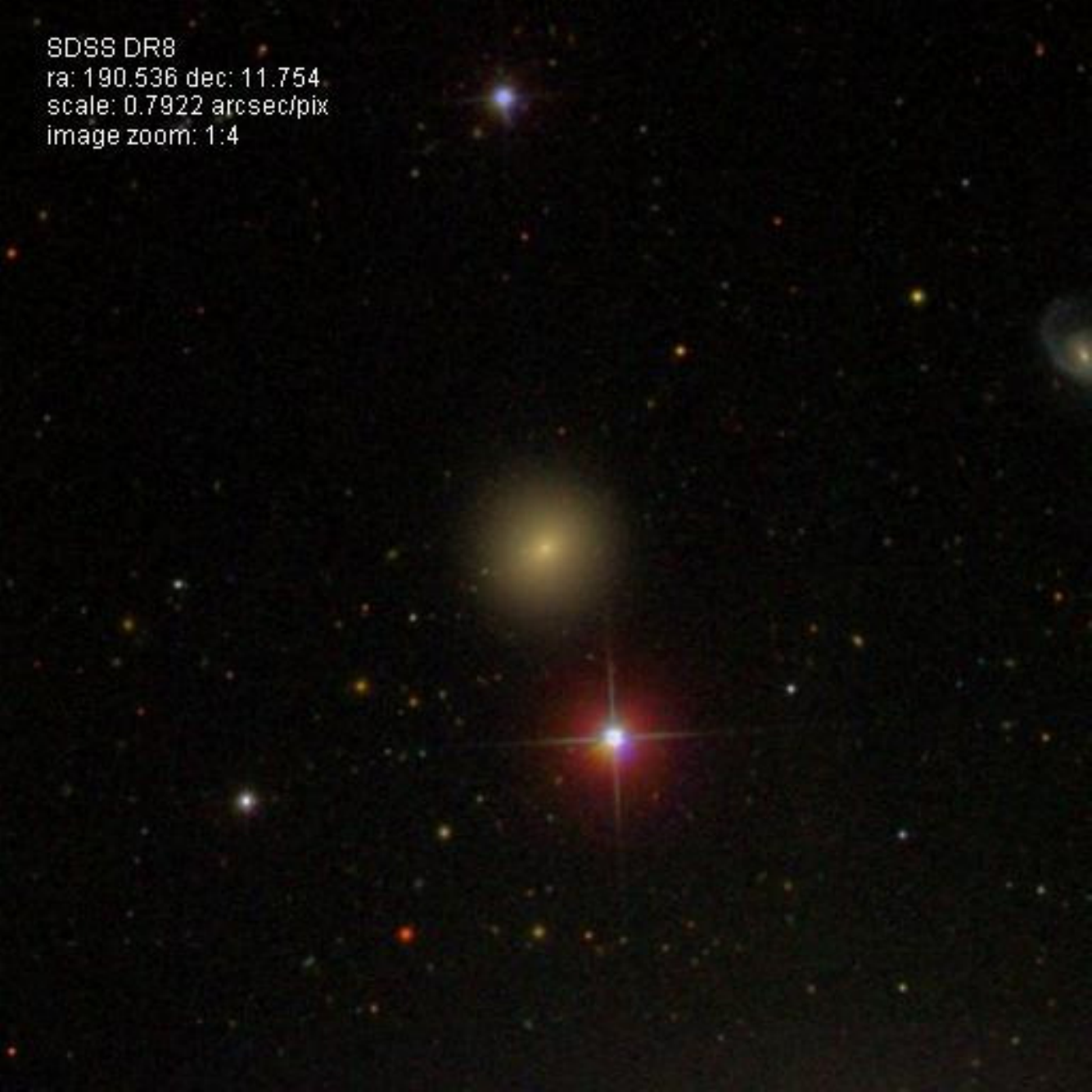}}&
\resizebox{2.2in}{!}{\includegraphics{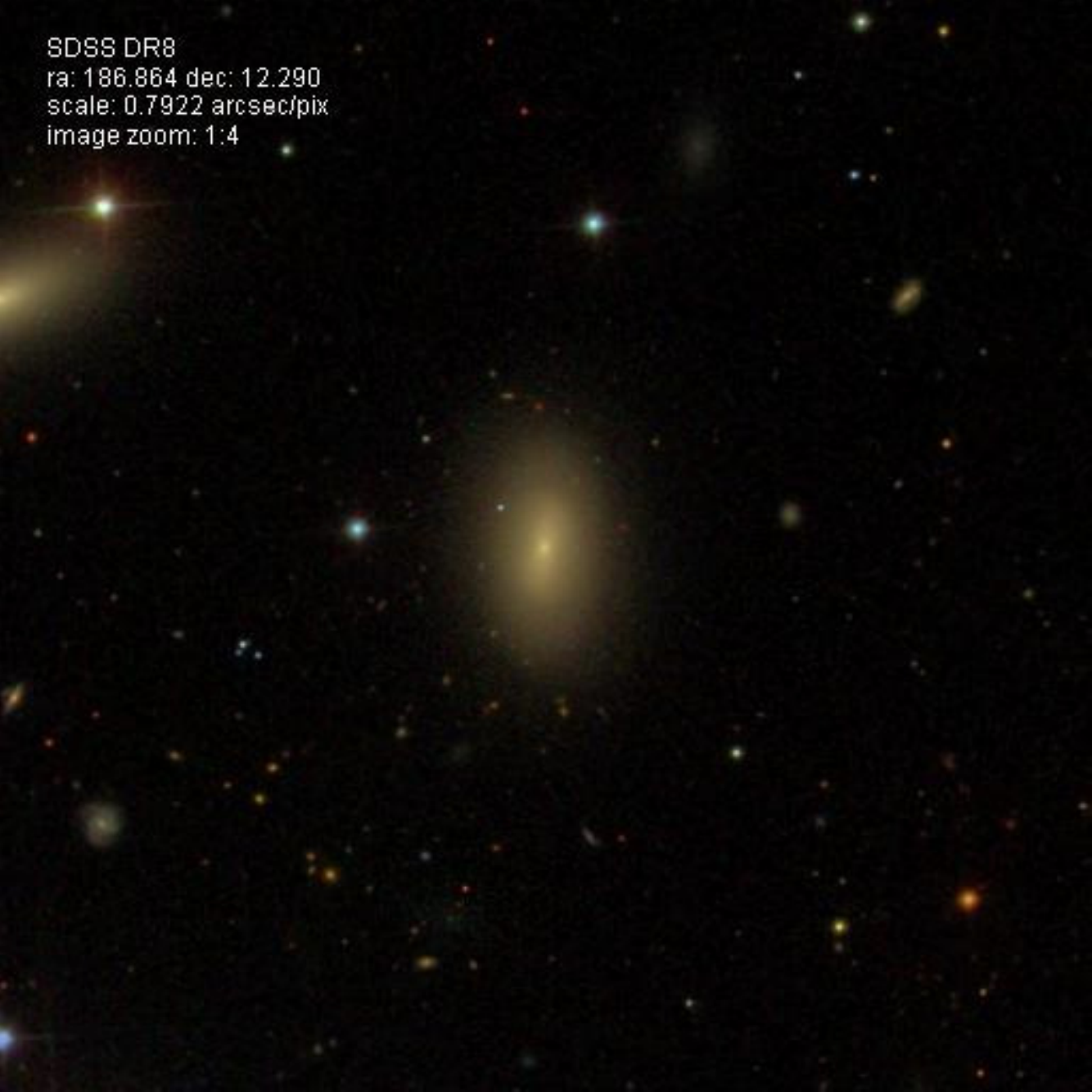}} \\
\resizebox{2.2in}{!}{\includegraphics{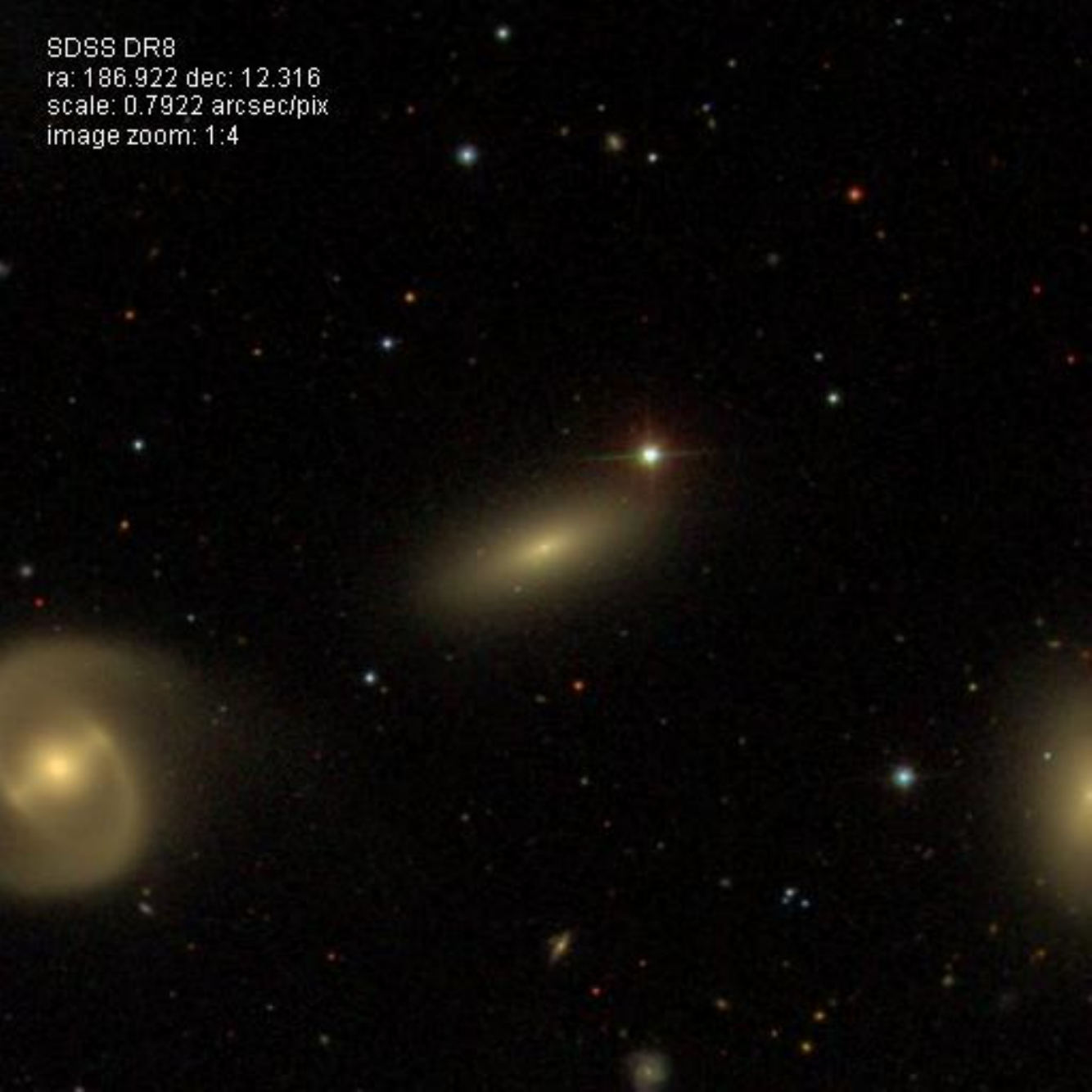}} &
\resizebox{2.2in}{!}{\includegraphics{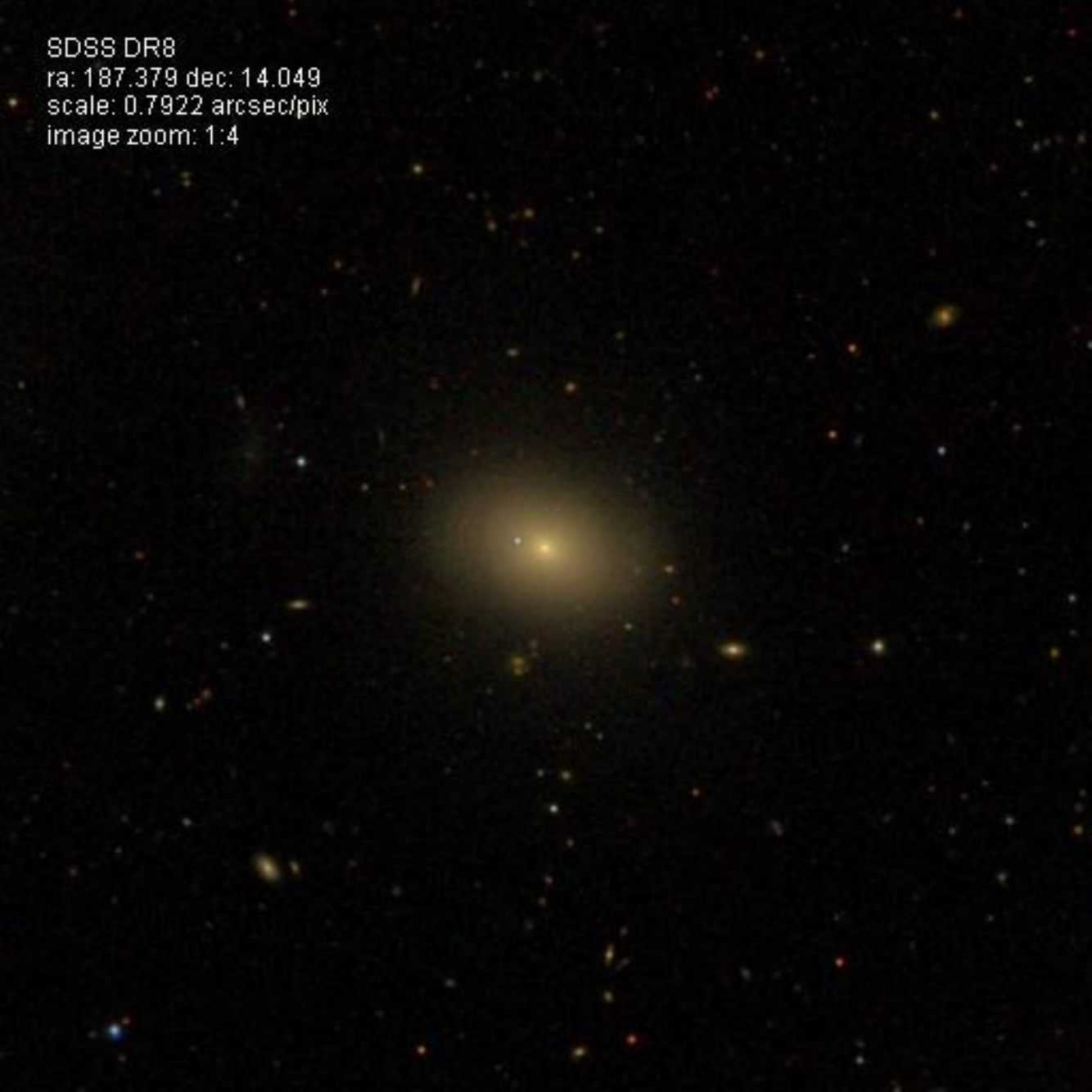}} &
\resizebox{2.2in}{!}{\includegraphics{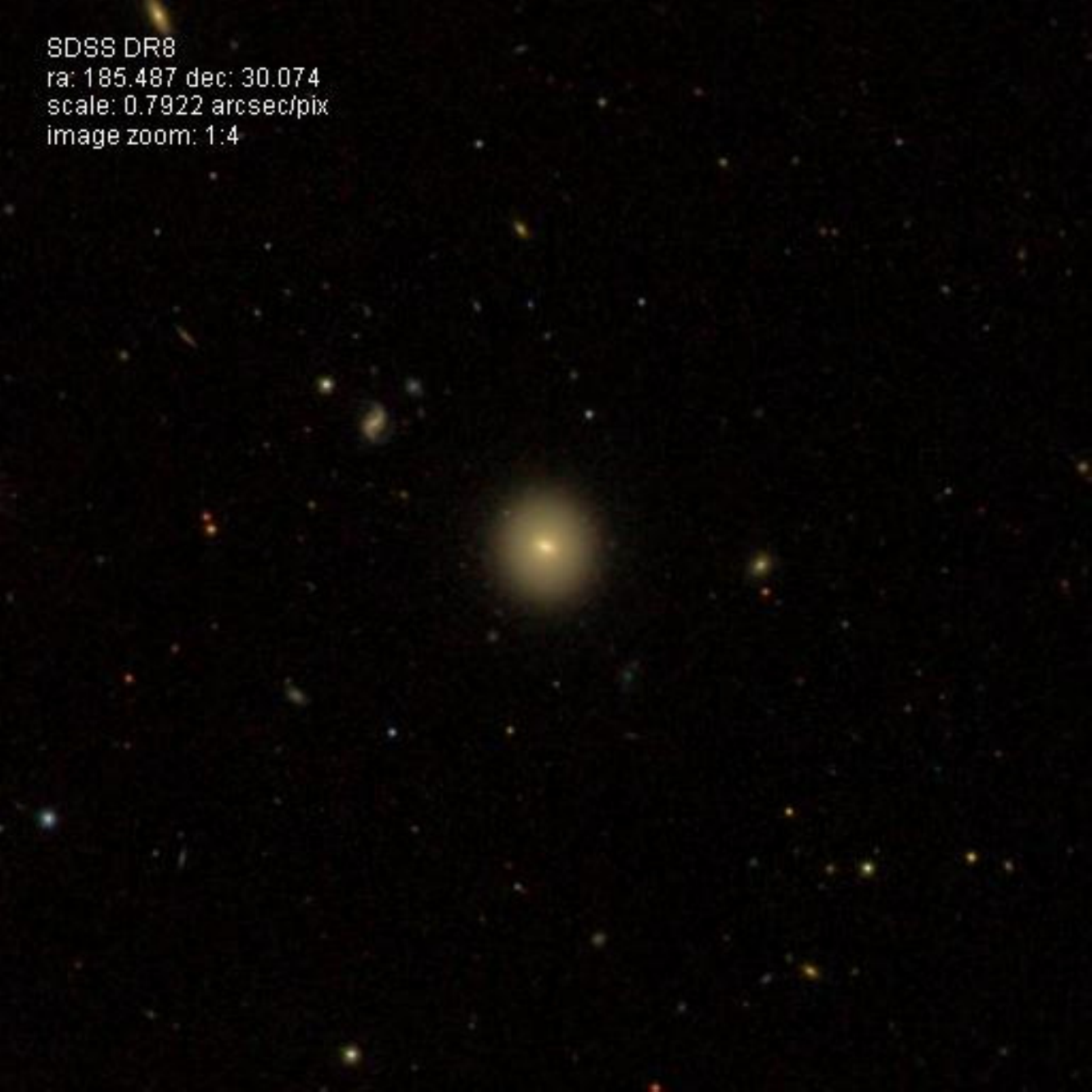}} \\
\resizebox{2.2in}{!}{\includegraphics{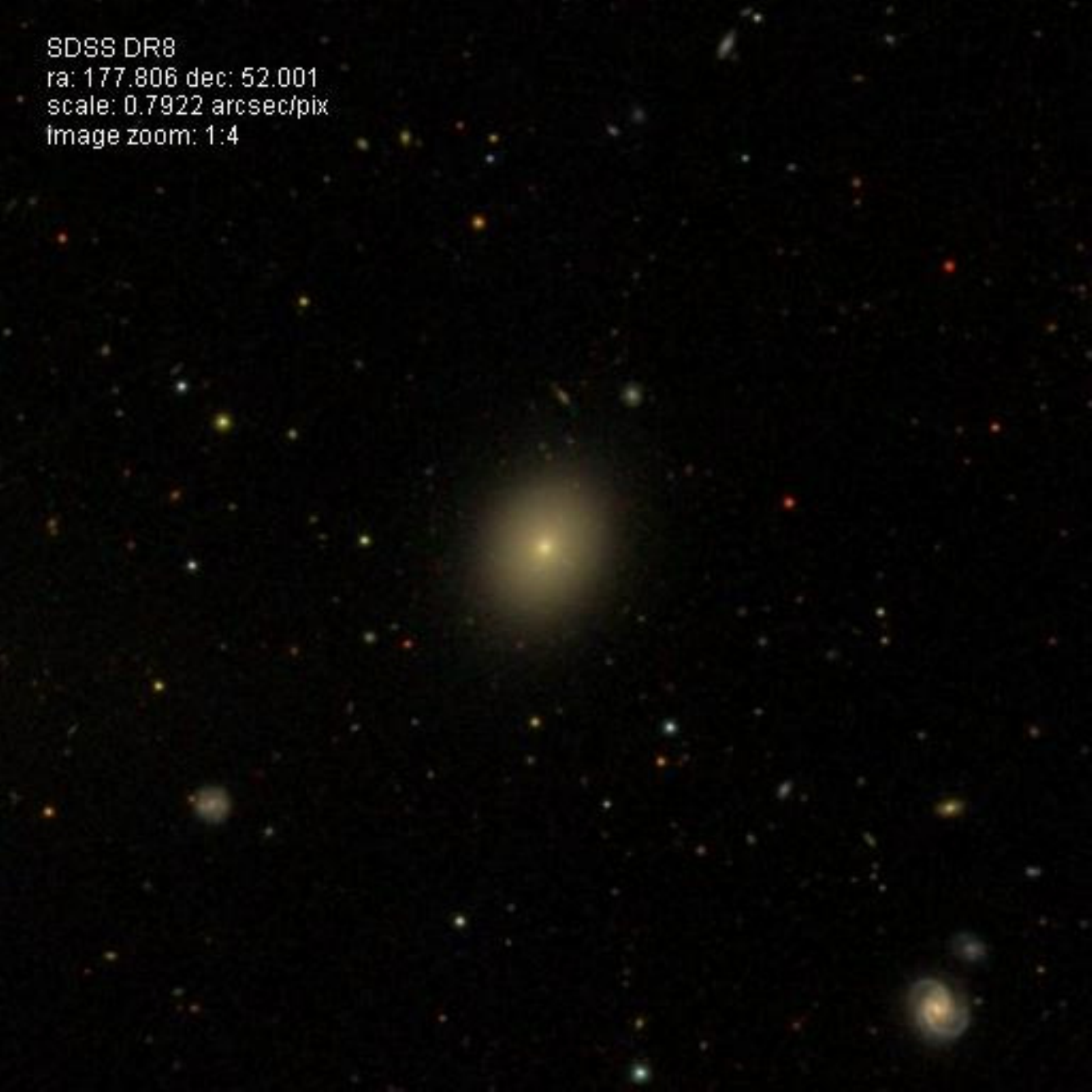}} & & \\
\end{array}$
\end{center}
\caption{Galaxies above the optically red color cut in Figure \ref{color_fig} with $M_K - 5log(h)$ > -20. These galaxies are a mix of disks and bulges. Though all are classified as early-type in available catalogs, many of the more disky objects, especially the first panel could arguably be classified as late-type. This is supported by their Sersic indicies, with the top 3 and left middle panel all below values of 3, and the last three larger than 3.}
\label{lowmagred_fig}
\end{figure*}

\section{Conclusion}

We have determined near infrared K-band luminosity functions using 13,325 $K_{\rm{tot}}$ $\leq$ 10.75 galaxies in the local Universe with known morphologies and redshifts. There are small differences in the shape of our LFs relative to prior literature, in part resulting from sample sizes, different photometric methods, the availability of redshift-independent distances, bulk flow corrections and corrections for over-density.

In this paper we have investigated the discrepancy between the shapes of color-selected and morphology-selected luminosity functions. The difference in shape is explained (in part) by massive red spiral galaxies. Our red galaxy LF is in agreement with the shape of other local Universe LFs, however, our red galaxy LF does not exhibit the same faint end slope as some red LFs at higher $z$. Our blue galaxy function also agrees well with previous local Universe LFs, and when corrected for surface brightness incompleteness, exhibits the same steep faint end shape as comparable low surface brightness optical functions.

There are comparable numbers of red and blue spiral galaxies at high stellar masses, but red spirals do not dominate the overall red galaxy population at its bright end. Higher mass red spirals are still forming stars, and disk galaxies without star formation have generally been classified as early-type galaxies (using traditional classifications), though their visual appearance and axis ratios indicate a far more disk like morphology. Sersic indices and visual inspection of the faint end of the red luminosity function indicates that a large fraction of these galaxies may be faded disks rather than true bulge dominated galaxies. 

\section{Acknowledgements}

The authors would like to note that this research has made use of the NASA/IPAC Extragalactic Database (NED) that is operated by the Jet Propulsion Laboratory, California Institute of Technology, under contract with the National Aeronautics and Space Administration. This publication makes use of data products from the Two Micron All Sky Survey, which is a joint project of the University of Massachusetts and the Infrared Processing and Analysis Center/California Institute of Technology, funded by the National Aeronautics and Space Administration and the National Science Foundation. We also acknowledge the usage of the HyperLeda database (http://leda.univ-lyon1.fr) as well as the many public datasets that were incorporated into this galaxy sample.

MJIB acknowledges the financial support of the Australian Research Council Future Fellowship FT100100280.

Though the majority of the work on this manuscript was completed at Monash University, NJB would like to thank the ICG, Portsmouth University, for their hospitality while the final touches were being made.

Lastly, we sincerely thank the referee, Eric Bell, with his constructive criticism of the manuscript. With it, this manuscript was much improved and expanded.

\bibliographystyle{apj}
\bibliography{article.bib}

\end{document}